\documentclass[useAMS, usenatbib]{mnras}
\usepackage{graphicx,amsmath,color,amssymb}

\usepackage[pdftitle={}]{hyperref}

\newcommand{\beq}{\begin{equation}}
\newcommand{\eeq}{\end{equation}}
\newcommand{\barr}{\begin{eqnarray}}
\newcommand{\earr}{\end{eqnarray}}

\newcommand{\Ly}{\textrm{Ly}}
\newcommand{\Lya}{Lyman-$\alpha$}
\newcommand{\Lyb}{Lyman-$\beta$}

\newcommand{\Data}{\mathcal{D}}
\newcommand{\model}{\mathcal{M}}

\newcommand{\Prob}{\textrm{Pr}}

\newcommand{\diag}{\textrm{diag\;}}
\newcommand{\nan}{\textrm{NaN}}
\newcommand{\normal}{\mathcal{N}}

\newcommand{\dd}{\textrm{d}}

\newcommand{\mdla}{\mathcal{M}_{\textrm{DLA}}}
\newcommand{\mnodla}{\mathcal{M}_{\neg\textrm{DLA}}}
\newcommand{\mkdla}{\mathcal{M}_{\textrm{DLA(k)}}}
\newcommand{\msubdla}{\mathcal{M}_{\textrm{sub}}}

\newcommand{\midlak}{\{\mathcal{M}_{\textrm{DLA}(i)}\}_{i=1}^{k}}

\newcommand{\taumf}{\tau_{0,\mathrm{MF}}}
\newcommand{\betamf}{\beta_\mathrm{MF}}

\newcommand{\lambdarest}{\lambda_{\textrm{rest}}}
\newcommand{\lambdaobs}{\lambda_{\textrm{obs}}}
\newcommand{\zqso}{z_{\textrm{QSO}}}

\newcommand{\lambdavec}{\boldsymbol{\lambda}}
\newcommand{\yvec}{\boldsymbol{y}}
\newcommand{\muvec}{\boldsymbol{\mu}}

\newcommand{\Kvec}{\boldsymbol{\mathrm{K}}}
\newcommand{\nuvec}{\boldsymbol{\nu}}

\newcommand{\omegavec}{\boldsymbol{\omega}}
\newcommand{\Omegavec}{\boldsymbol{\Omega}}
\newcommand{\AAtext}{\textrm{\AA}}

\newcommand{\effectivetau}{\tau_{\textrm{eff,HI}}}
\newcommand{\effectivetauvec}{\boldsymbol{\tau}_{\textrm{eff,HI}}}
\newcommand{\zvec}{\boldsymbol{z}}

\newcommand{\avec}{\boldsymbol{a}}
\newcommand{\Alyavec}{\boldsymbol{\mathrm{A}}_{\textrm{F}}}
\newcommand{\Avec}{\boldsymbol{\mathrm{A}}}

\newcommand{\Sigmavec}{\boldsymbol{\mathrm{\Sigma}}}

\newcommand{\siv}{\textsc{Siv}}

\newcommand{\omegadla}{\Omega_{\textrm{DLA}}}
\newcommand{\lognhi}{\log_{10}{N_{\textrm{HI}}}}
\newcommand{\nhi}{N_{\textrm{HI}}}
\newcommand{\zdla}{z_{\textrm{DLA}}}
\newcommand{\kms}{\,\textrm{km\,s}^{-1}}
\newcommand{\cm}{\,\textrm{cm}}

\newcommand{\msun}{\mathrm{M}_{\odot}}

\begin{document}

\title[GP DLAs from SDSS DR16]{Damped Lyman-alpha Absorbers from Sloan Digital Sky Survey DR16Q with Gaussian processes}
\author[ M.-F. Ho et al.]{Ming-Feng Ho$^1$\thanks{E-mail: mho026@ucr.edu}, Simeon Bird$^1$\thanks{E-mail: sbird@ucr.edu}, Roman Garnett$^2$\\
$^1$Department of Physics and Astronomy, University of California, Riverside, CA\\
$^2$Department of Computer Science and Engineering, Washington University in St. Louis, One Brookings Drive, St. Louis, MO\\
}

\date{\today}

\pagerange{\pageref{firstpage}--\pageref{lastpage}} \pubyear{2020}
\pagenumbering{arabic}
\label{firstpage}

\maketitle

\begin{abstract}
We present a new catalogue of Damped {\Lya} absorbers from SDSS DR16Q, as well as new estimates of their statistical properties.
Our estimates are computed with the Gaussian process models presented in \cite{Garnett17,Ho:2020} with an improved model for marginalising uncertainty in the mean optical depth of each quasar.
We compute the column density distribution function (CDDF) at $2 < z < 5$, the line density ($\dd N/ \dd X$), and the neutral hydrogen density ($\omegadla$).
Our Gaussian process model provides a posterior probability distribution of the number of DLAs per spectrum, thus allowing unbiased probabilistic predictions of the statistics of DLA populations even with the noisiest data.
We measure a non-zero column density distribution function for $\nhi < 3 \times 10^{22} \cm^{-2}$ with $95\%$ confidence limits, and $\nhi \lesssim 10^{22} \cm^{-2}$ for spectra with signal-to-noise ratios $> 4$.
Our results for DLA line density and total hydrogen density are consistent with previous measurements.
Despite a small bias due to the poorly measured blue edges of the spectra,
we demonstrate that our new model can measure the DLA population statistics when the DLA is in the {\Lyb} forest region.
We verify our results are not sensitive to the signal-to-noise ratios and redshifts of the background quasars although a residual correlation remains for detections from $\zqso < 2.5$, indicating some residual systematics when applying our models on very short spectra, where the SDSS spectral observing window only covers part of the {\Lya} forest.
\end{abstract}


\section{Introduction}
\label{sec:introduction}
Damped {\Lya} absorbers (DLAs) are strong {\Lya} absorption features discovered in quasar spectral sightlines.
At the densities required to produce neutral hydrogen column densities above the DLA threshold, $\nhi > 10^{20.3}\, \cm^{-2}$ \citep{Wolfe1986},
the gas of DLAs is self-shielded from the ionising effect of the ultra-violet background (UVB) \citep{Cen2012} but diffuse enough to have a low star formation rate \citep{Fumagalli:2013}.
DLAs contain a large fraction of the neutral hydrogen budget after reionisation \citep{Gardner1997,Noterdaeme12,Zafar2013,Crighton2015},
which make them a direct probe of the distribution of neutral gas.

Numerical simulations tell us DLAs are associated with a wide range of halo masses, with a peak value in the range of $10^{10} - 10^{11}\,\msun$ \citep{Haehnelt1998,Prochaska1997,Pontzen2008,Rahmati:2014}.
Through cross-correlating the DLAs with the {\Lya} forest, \cite{FontRibera:2012} measured a DLA bias factor $b_\mathrm{DLA} = 2.17 \pm 0.2$. This implies a median host halo mass of $\sim 10^{12}\,\msun$, assuming all DLAs arise from halos of the same mass and. However, a model which assumes a power-law distribution function of DLA cross-section as a function of halo mass is only in marginal tension with the data \citep{Bird2015}. 
Furthermore, a later measurement from SDSS-DR12 \citep{PerezRafols:2018} found a bias factor $b_\mathrm{DLA} = 1.99 \pm 0.11$, and a median host halo mass $\sim 4 \times 10^{11}\, \msun$, in good agreement with simulations.
Alternative measurements by cross-correlating with CMB lensing data are broadly consistent with both simulated DLAs and {\Lya} clustering \citep{Alonso:2018,Lin:2020}. 


In the cosmology context, the
{\Lya} forest is a successful probe of matter clustering between $2 < z < 6$ \citep{Croft:1998,McDonald:2000,Viel:2004,McDonald:2005b,Irvic:2017,Chabanier:2019}.
However, high column density absorbers such as DLAs will bias cosmological parameter estimates from {\Lya} and thus need to be masked out \citep{McDonald:2005a}.
Simulations have been performed to study the effect of damped absorbers on the {\Lya} 1-D and 3-D flux power spectrum \citep{Rogers:2018b,Rogers:2018a}, and a recent Bayesian fitting method has been proposed to better understand how DLA contaminants affect cosmological inference using the BAO peak \citep{Cuceu:2020}.


In this work,
we present new estimates for the column density distribution function (CDDF), the abundance of DLAs, and the average neutral hydrogen density at $z = 2 - 5$ for DLAs in the Sloan Digital Sky Survey IV quasar catalogue from Data Release 16 (SDSS-IV/eBOSS DR16) \citep{Dawson:2016,SDSSDR16Q:2020}.
We compute DLA population statistics using the Gaussian process (GP) model presented in \cite{Ho:2020},
a modified version of the machine learning framework from \cite{Garnett17}.
We retrain our model on SDSS DR12 \citep{Eisenstein:2011,Dawson:2013,Alam:2015,Paris2018}
and generate a DLA catalogue from DR16Q \citep{SDSSDR16Q:2020}.
We compute DLA population statistics from the DLA catalogue,
which update the estimates we made in \cite{Bird17,Ho:2020}.

The pipeline presented in \cite{Garnett17}
provided for the first time probabilistic detections of DLAs in each spectrum, which comes with a posterior distribution on putative DLAs for the column density and the absorber redshift.
With the aid of a full posterior probability distribution for the number of DLAs in each quasar spectrum, ``soft'' detections in noisy data become available. We propagate uncertainties from each individual spectrum into the global population, without setting any hard threshold on the minimum required probability for the presence of DLAs.
We are thus able to include even noisy spectra in our sample of DLAs.

\cite{Ho:2020} added an alternative model for sub-DLAs, which regularised excessive detections at low column density. We also included absorption from the mean optical depth in the {\Lya} forest in the GP mean function. This helped prevent the pipeline from using DLAs to compensate for {\Lya} forest absorption in the spectrum, essential at high redshift.
In this work, we further improve this aspect of our model. We marginalise out uncertainty in the effective optical depth in each spectrum using the measured mean optical depth as a prior when computing the evidence for the null, DLA, and sub-DLA models.

Several other DLA search methods for SDSS spectra have been implemented.
These range from visual inspection surveys \citep{Slosar11},
visually guided Voigt profile fitting \citep{Prochaska05,Prochaska2009}, and template fitting \citep{Noterdaeme:2009,Noterdaeme12}, to machine learning based methods such as a convolutional neural network (CNN) approach \citep{Parks18} and an unpublished Fisher discriminant analysis \citep{Concordance2012}.
The CNN method \citep{Parks18} was also run to identify DLAs as part of the SDSS DR16 quasar catalogue \citep{SDSSDR16Q:2020}. We compare the DLAs detected by our GP model and the DLAs in DR16Q in Section~\ref{sec:comparison}.

Machine learning methods have also been proposed to classify broad absorption lines (BALs), including a line-finder based convolutional neural network (CNN) \citep{QuasarNET:2018} and a hybrid of a CNN with a principal component analysis \cite{Guo:2019}.



Section~\ref{sec:method} will briefly outline our modelling decisions and the changes to the model made in this work.
Section~\ref{sec:data} describes the cuts we applied to SDSS DR16Q.
We recap our modelling details in Section~\ref{subsec:GP}.
We present our results in Section~\ref{sec:results}, including the CDDF in Section~\ref{subsec:cddf} and the incidence rate of DLAs and total HI density in Section~\ref{subsec:omega_dla}.
In Section~\ref{sec:systematics}, we discuss the possible remaining systematics in our method.
Section~\ref{sec:lymanlimit} shows population statistics for DLAs in Ly$\infty$ to Ly$\beta$.
In Section~\ref{sec:comparison}, we briefly compare our DLA catalogue to the DLAs presented in the SDSS DR16Q catalogue,
which implemented a CNN model \citep{Parks18} to classify DLAs.
We conclude in Section~\ref{sec:conclusion}.

\section{Methods}
\label{sec:method}
Here we briefly recap our Gaussian process (GP) based framework for detecting DLAs using \textit{Bayesian model selection}. We summarise the general approach, while more comprehensive mathematical details may be found in \cite{Garnett17,Ho:2020}. A quasar sightline has spectroscopic observations $\Data = (\lambdavec, \yvec)$, where $\lambdavec$ is a vector of rest wavelength bins, and $\yvec$ is a vector of observed flux at these wavelength bins.
Suppose we have built likelihood functions for a set of models $\{\model_i\}$.
We can evaluate the posterior probability of a model, $\model$, given a quasar observation, $\Data$, based on Bayes' rule:
\begin{equation}
   \Prob(\model \mid \Data) =
   \frac{p(\Data \mid \model)
      \Prob(\model)}{
      \sum_i p(\Data \mid \model_i)\Prob(\model_i)
   },
   \label{eq:model_selection}
\end{equation}
where $p(\Data \mid \model)$ is the model evidence of the quasar spectrum $\Data$ given model $\model$, $\Prob(\model)$ is the prior probability of model $\model$, and the denominator on the right-hand-side is the sum of posterior probabilities of all models in consideration.

Concretely, we have the model without DLAs ($\mnodla$), the model with $k$ DLAs ($\midlak$), and the model with sub-DLAs ($\msubdla$). We set $k=3$ here, allowing up to $3$ DLAs per spectrum.
We consider a posterior probability of a sub-DLA, $\msubdla$, not to be a DLA detection, as in \cite{Ho:2020}.
Section~\ref{subsec:GP} describes the details of how we compute the model evidence for each model.

Table~\ref{tab:notations} lists mathematical notation and definitions of parameters used throughout the paper.

\begin{table*}
   \caption{Mathematical notations and definitions}
   \begin{tabular}{|l|l|}
      \hline
      Notation & Description \\
      \hline\hline
       $\mnodla$ & Null model, model without DLAs or subDLAs \\
      $\mdla$ & Model with DLAs ($20 \leq \lognhi \leq 23$) \\
      $\msubdla$ & Model with subDLAs ($19.5 \leq \lognhi < 20$) \\
      $p(\Data \mid \model)$ & Model evidence, marginalised likelihood \\
      $\Prob(\model)$ & Model prior\\
      $(\betamf, \taumf)$ & Parameters of power-law relation of effective optical depth model \\
      $\effectivetau(z; \betamf, \taumf)$ & Power-law model of effective optical depth \\
      $p(\betamf)$ & Prior of $\betamf$, assumed to be a normal distribution \\
      $p(\taumf)$ & Prior of $\taumf$, assumed to be a normal distribution \\
      $p(\zdla \mid \zqso, \mdla)$ & Prior of redshift of DLAs, a uniform distribution \\
      $p(\nhi \mid \mdla)$ & Prior of column density of a DLA, a data-driven distribution \\
      $\yvec$ & Vector of normalised observed flux\\
      $\lambdavec$ & Vector of wavelength pixels in restframe \\
      $\nuvec$ & Vector of instrumental noise variance\\
      $\muvec$ & Vector of GP mean model \\
      $\Sigmavec$ & Matrix of GP covariance \\
      $\Alyavec$ & Matrix of mean flux suppression from the effective optical depth (diagonal matrix) \\
      $\Kvec$ & Matrix to describe covariance of quasar emission spectrum ($2281 \times 2281$ matrix, $20 \times 2281$ parameters)\\
      $\Omegavec$ & Matrix of Lyman series absorption noise (diagonal matrix)\\
      \hline
   \end{tabular}
   \label{tab:notations}
\end{table*}

\subsection{Data}
\label{sec:data}

Our GP model requires a training set \textit{without DLAs} for training the null model, $\mnodla$.
We use the DLAs in SDSS DR12Q detected by \cite{Ho:2020} as our true DLA labels.
Here we list the subset of DR12 quasars omitted from our training sample:
\begin{itemize}
   \item Quasars with $\zqso < 2.15$, which have almost no {\Lya} forest, are removed.
   \item \texttt{BAL}: quasars with a broad absorption line (BAL) probability larger than $0.75$ (\texttt{BAL\_PROB} $\geq 0.75$) are removed, as suggested by \cite{SDSSDR16Q:2020}. \texttt{BAL\_PROB} is derived from QuasarNET \citep{QuasarNET:2018}.
   \item \texttt{CLASS\_PERSON == 30}: quasars classified as BALs by human visual inspection are removed.
   \item \texttt{ZWARNING}: spectra flagged with \texttt{ZWARNING} for pipeline redshift estimation are removed, but extremely noisy spectra with \texttt{TOO\_MANY\_OUTLIERS} are kept.
\end{itemize}
We have in total $89,408$ spectra without DLAs for training the null model.

We also use the same above criteria to select the DR16Q spectra for applying our model.
In addition to the above criteria,
the DR16Q quasar sample to which our model is applied is a subset of the full DR16Q sample chosen following additional conditions:
\begin{itemize}
   \item \texttt{IS\_QSO\_FINAL == 1}: We require this flag in the quasar sample, specifying that a spectrum is robustly classified as a quasar.
   \item \texttt{CLASS\_PERSON == 3 or 0}: This flag specifies that the spectrum was classified by a human as a quasar (3) or was not visually identified (0).
   \item \texttt{SOURCE\_Z}: as suggested in Section 3.2 of  \citep{SDSSDR16Q:2020}, spectra with \texttt{Z > 5} and \texttt{SOURCE\_Z == PIPE} have a suspect redshift estimate and should not be used without a careful visual re-inspection.
   We thus remove these spectra from our analysis.
\end{itemize}

Integral to our method is a reliable quasar redshift estimate. It is not trivial to reliably estimate quasar redshifts in the large samples provided by DR16Q,\footnote{Indeed, we have extended our GP framework to provide a quasar redshift estimate \citep{Fauber:2020}.} and so we are careful to use the redshift estimates suggested by  \cite{SDSSDR16Q:2020}.
To ensure our quasar redshifts are as homogeneous as possible,
we use \texttt{Z\_PCA}, the recommended redshift estimate method for statistical analyses of a large ensemble of quasars.
We also remove the spectra where redshift measurements disagree with each other by more than $0.1$, which means we remove samples with $|z_i - z_j| > 0.1$ for $z_i, z_j \in$ \{\texttt{Z\_PIPE}, \texttt{Z\_PCA}, \texttt{Z}, \texttt{Z\_VI}\}. If \texttt{Z\_VI} is not present, we use only the other three redshift estimates.
Our final DR16Q sample size contains $159\,807$ {\Lya} quasar spectra.

\subsection{Gaussian process model}
\label{subsec:GP}
Consider a distant quasar with a known redshift, $\zqso$.
Each spectroscopic observation gives us the observed flux, $\yvec$, on a set of wavelength pixels in observed-frame wavelengths, $\lambdavec_\mathrm{obs}$.
Since the quasar redshift is assumed to be known, we shift into the rest frame, $\lambdavec = \lambdavec_\mathrm{obs} / (1 + \zqso)$.
Standard errors are provided with each observed flux pixel, $\sigma(\lambda_i)$, with $\lambda_i$ the $i$th pixel in $\lambdavec$, and we define the noise variance of each observed flux pixel as $\nu_i = \sigma(\lambda_i)^2$.
Given the observed flux of a quasar, we normalise all flux measurements by dividing the median flux observed between $[1425 \AAtext, 1475 \AAtext]$ in the rest-frame, a wavelength range redwards of the Ly$\alpha$ emission and avoiding major emission lines.

For each quasar observation, we have data $\Data = (\lambdavec, \yvec, \nuvec, \zqso)$.
We want to build a likelihood function to describe this data:
\begin{equation*}
   p(\yvec | \lambdavec, \nuvec, \zqso),
\end{equation*}
which is the likelihood of the flux $\yvec$ given all other observed quantities.
We model this likelihood as a Gaussian process:
\begin{equation*}
   p(\yvec | \lambdavec, \nuvec, \zqso) = \normal(\yvec; \muvec, \Sigmavec),
\end{equation*}
where $\muvec$ is the mean vector of the GP, and $\Sigmavec$ is the covariance matrix of the GP.
We will use bold lowercase italics for vectors and bold uppercase letters for matrices.

\subsubsection{Learning the GP null model}
\label{subsubsec:learning_GP}

A GP is fully specified by its first two central moments: the mean function, $\mu(\lambda)$, and the covariance kernel, $K(\lambda, \lambda')$, \citep{Rasmussen05}.
Our task now is to learn the mean function and the covariance function from the training set.
Suppose we have a set of quasar observations without any intervening DLAs, $\{\Data_1, \Data_2, \cdots, \Data_\mathrm{N_{spec}}\}$, where $\mathrm{N_{spec}}$ is the number of quasars in the training set.
We can then learn the mean function by taking a precision weighted average:
\begin{equation}
   \mu_j =
   \frac{
      \sum\limits_{i}{_{y_{ij} \neq \nan}} \, ({y_{ij}}/{ \nu_{ij})}
   }{
      \sum\limits_{i}{_{y_{ij} \neq \nan}}\, ({1}/{\nu_{ij})}
   },
   \label{eq:mean_model}
\end{equation}
where the summation is over $i$ index.
$j$ indicates $j$th pixel in the observed flux, $i$ represents $i$th spectrum, and we only average over the non-NaN values. Note this differs from \cite{Ho:2020}, where we used the mean rather than the precision weighted average. The precision weighted average can be viewed as a result of using an uninformative prior on $\mu_j$ and an independent Gaussian likelihood for each $y_{ij}$. If we have a set of normally disturbed flux pixels with each flux pixel follows $y_{ij} \sim \normal(\mu_j, \nu_{ij})$ with known variance $\nu_{ij}$ and an unknown $\mu_j$ with an uninformative prior, the posterior will be a normal distribution with a new mean equals a precision weighted average.

\begin{figure*}
   \includegraphics[width=2\columnwidth]{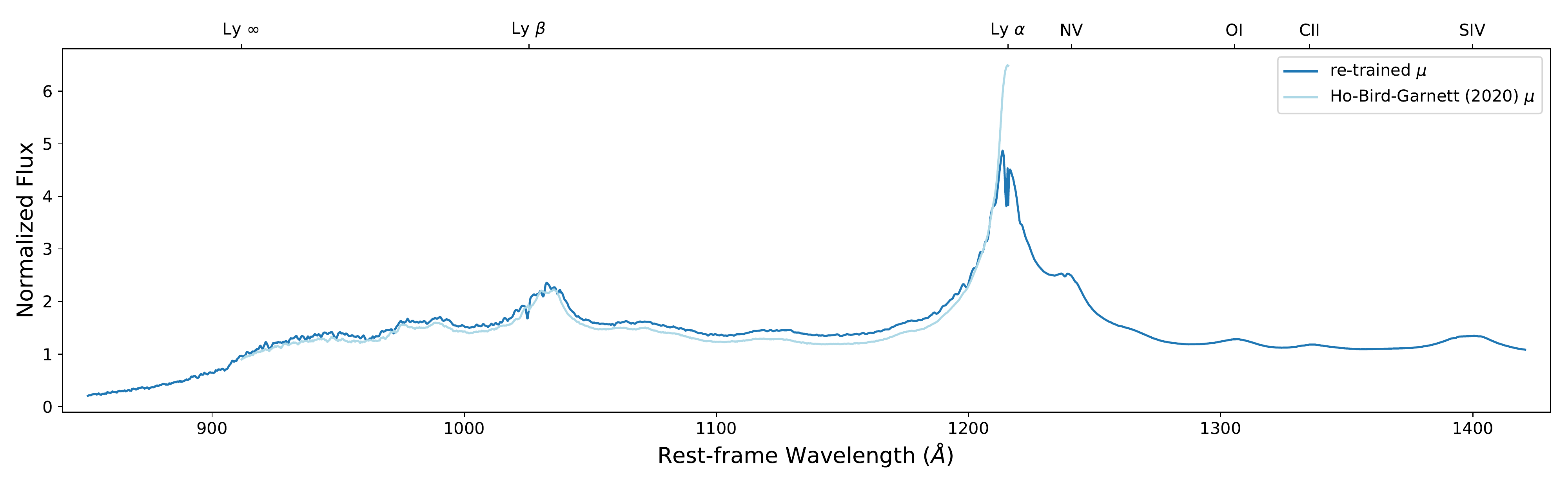}
   \caption{Our GP mean function using a precision weighted average of the rest-frame wavelengths.
   We extended our model compared \protect\cite{Ho:2020} (light blue), both bluewards past the Lyman break at $912${\AA}~and redwards past the {\siv} emission line.}
   \label{fig:mu}
\end{figure*}

Instead of training on the raw observed flux $\yvec$ directly, we follow \citep{Ho:2020} to train the mean function and the kernel on the flux after removing the average effect of the {\Lya} forest, the de-forest flux:
\begin{equation}
   \begin{split}
      y_{ij}   &\gets y_{ij} \cdot \exp( \effectivetau);\\
      \nu_{ij} &\gets \nu_{ij} \cdot \exp( 2 \cdot \effectivetau)\,,
   \end{split}
   \label{eq:de_forest}
\end{equation}
which means we replace observed flux and its variance with the flux and variance before the suppression of {\Lya} forest.
The effective optical depth is parameterised as:
\begin{equation}
   \begin{split}
      \effectivetau(z(&\lambdaobs); \betamf, \taumf) = \\
      &\sum_{i=2}^{N} \taumf \frac{\lambda_{1i} f_{1i}}{\lambda_{12} f_{12}}  (1 + z_{1i}(\lambdaobs))^{\betamf},
   \end{split}
   \label{eq:effective_tau_beta}
\end{equation}
where $\lambda_{1i}$ is the transition wavelength from {\Lya} to the $i$th member in the Lyman series, $f_{1i}$ represents the oscillator strength, $z_{1i}$ is the absorber redshift, and we set $N = 31$.
The absorber redshift is written as:
\begin{equation}
   \begin{split}
      1 + z_{1i}(\lambda, \zqso) &= \frac{\lambdaobs}{\lambda_{1i}}\\
      &= \frac{\lambda (1 + \zqso)}{\lambda_{1i}}.
   \end{split}
   \label{eq:abs_redshift}
\end{equation}
We parameterise the effective optical depth by a power-law relation with $\taumf$ and $\betamf$ parameters.
Here we specify a subscript ``MF'' to annotate the parameters modified by mean flux suppression. Fig~\ref{fig:mu} shows our new GP mean function, compared to \cite{Ho:2020}.

Taking this {\Lya} mean flux into account introduces a dependence on quasar redshift into the mean function of the GP for each quasar:
\begin{equation}
   \begin{split}
      \mu(\lambda, \zqso; \betamf, \taumf) &=
      \\\mu(\lambda) \cdot \exp(- \effectivetau(&z(\lambda, \zqso); \betamf, \taumf))\,.
   \end{split}
   \label{eq:mu_z}
\end{equation}
$\mu(\lambda)$ is the mean function we learned from Eq~\ref{eq:mean_model}.
We learn the mean function on a dense grid of wavelengths on a chosen rest-frame wavelength range:
\begin{equation}
   \lambda \in [850.75 \, \textrm{\AA}, 1420.75 \, \textrm{\AA}]
   \label{eq:modelling_range}
\end{equation}
with a linearly equal spacing of $\Delta \lambda = 0.25 \textrm{\AA}$.
\cite{Ho:2020} only modelled the null model in the {\Lya} region, $[911.75 \textrm{\AA}, 1215.75 \textrm{\AA}]$.
We extend the red end of our model to include a part of the metal line region until 1420.75 {\AA}.
This empirically improved the column density estimation of DLAs near the {\Lya} emission peak,
as otherwise part of the damping wing would go beyond 1215.75 {\AA} when a large DLA is very close to the quasar.

The mean function is thus written as a mean vector $\muvec(\zqso; \betamf, \taumf) = \mu(\lambdavec, \zqso; \taumf, \betamf)$ and the kernel is written as a matrix $\Sigma(\lambda, \lambda') = \Sigmavec$.
The covariance matrix's optimisation procedure is described in \cite{Garnett17,Ho:2020}.
We factorise the covariance matrix as in \cite{Ho:2020}:
\begin{equation}
   \Sigmavec_i = \Alyavec^{\top}(\Kvec + \Omegavec)\Alyavec + \diag{\nuvec_i}.
   \label{eq:Sigma}
\end{equation}
The $\Kvec$ matrix is a positive-definite symmetric matrix corresponding to the covariance between each quasar flux pixel.
$\Omegavec$ is a diagonal matrix describing the absorption noise:
\begin{equation}
   \diag{\Omegavec} = \omegavec \circ (1 - \exp(-\effectivetau(\zvec; \beta, \tau_0)) + c_0)^2\,.
   \label{eq:Omega}
\end{equation}
$\omegavec$ is freely optimisable while the {\Lya} flux term, $(1 - \exp(\effectivetau(\zvec; \beta, \tau_0)) + c_0)^2$, includes the redshift dependent noise variance with which we model the {\Lya} forest.
The optimised absorption noise parameters used here are:
\begin{equation}
   \begin{split}
      \tau_0 &= 0.000119\qquad
      \beta  = 5.15\qquad
      c_0    = 0.146.
   \end{split}
\end{equation}

The $\Alyavec$ is a diagonal matrix reflecting the mean vector suppression for each spectrum corresponding to the mean flux in the {\Lya} forest:
\begin{equation}
   \diag{\Alyavec} = \exp{(- \effectivetauvec(\zqso; \betamf, \taumf))}\,.
   \label{eq:Alya}
\end{equation}
The parameters of this matrix follow the values given in \cite{Kamble:2020}, which used a power-law relation to measure the effective optical depth in the {\Lya} forest in SDSS DR12:
\begin{equation}
   \begin{split}
      \taumf = 0.00554\qquad
      \betamf = 3.182,
   \end{split}
   \label{eq:meanflux_BOSS}
\end{equation}
with associated uncertainty for each parameter:
\begin{equation}
   \begin{split}
      \sigma_{\taumf} = 0.00064\qquad
      \sigma_{\betamf} = 0.074.
   \end{split}
   \label{eq:meanflux_BOSS_std}
\end{equation}

\begin{figure}
   \includegraphics[width=\columnwidth]{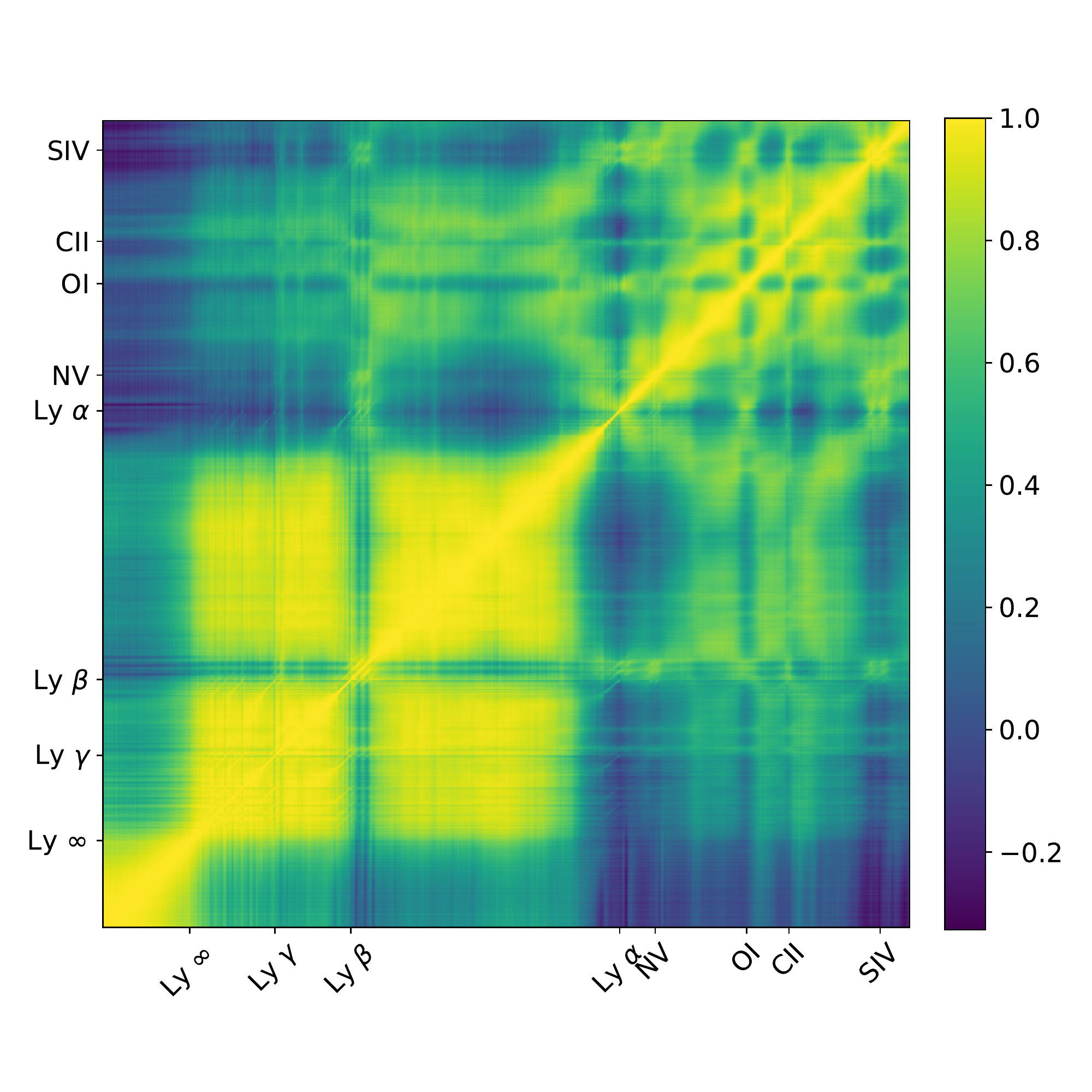}
   \caption{The correlation matrix learned from data, which is the covariance matrix $\Kvec$ normalised by the diagonal elements.
   Note that the correlation in the plot is pixel-by-pixel, and the matrix dimension is $2281\times2281$.
   Different emission lines and the Lyman break are visible in the plot.}
   \label{fig:covariance}
\end{figure}

The instrumental noise is encoded in the diagonal matrix $\diag{\nuvec_i}$, where $i$ simply denotes the $i$th quasar observation:
The final covariance matrix learned from our data is shown in Fig~\ref{fig:covariance}.
Comparing the kernel matrix we learned in this work to \cite{Ho:2020},
the current kernel is less noisy and contains several distinct features of emission lines.
The reduction in the noise is due to a larger training set, SDSS DR12Q catalogue, is used for optimising the kernel.

After having learned the GP null model, we can write down the null model likelihood function:
\begin{equation}
   \begin{split}
      p(\yvec \mid &\lambdavec, \nuvec, \zqso, \betamf, \taumf, \mnodla) =\\
      & \normal(\yvec; \muvec(\zqso; \betamf, \taumf),
      \Alyavec^{\top} (\Kvec + \Omegavec) \Alyavec + \diag{\nuvec_i}),
   \end{split}
   \label{eq:null_likelihood_function}
\end{equation}
where the notation $\mnodla$ specifies that our null GP model is conditioned on a training set \textit{without DLAs}.

\subsubsection{Model evidence for the null model}

Once we have trained our GP null model, $\mnodla$, according to Section~\ref{subsubsec:learning_GP},
we need to integrate out the nuisance parameters associated with {\Lya} forest absorption to get the model evidence.

In \cite{Ho:2020}, we only took the mean values of the meanflux parameters $(\betamf, \taumf)$ without their uncertainties, so the model evidence straightforwardly equals to Eq~\ref{eq:null_likelihood_function} without integration.
In this work, we take the uncertainties of meanflux suppression into account and integrate them out, according to \cite{Kamble:2020} prior.
The model evidence thus will be:
\begin{equation}
   p(\Data \mid \mnodla, \nuvec, \zqso) \propto
   p(\yvec \mid \lambdavec, \nuvec, \zqso, \mnodla),
\end{equation}
where we integrate out $(\betamf, \taumf)$
\begin{equation}
   \begin{split}
      p(\yvec \mid \lambdavec, \nuvec, \zqso, &\mnodla)
      =\\
      \int p(\yvec &\mid \lambdavec, \nuvec, \zqso, \betamf, \taumf, \mnodla)\\
      &p(\betamf)p(\taumf) \dd \betamf \dd \taumf
   \end{split}
   \label{eq:model_evidence_null}
\end{equation}
with
\begin{equation}
   \begin{split}
      p(\betamf) &= \normal(\betamf = 3.182, \sigma_{\betamf} = 0.074)\\
      p(\taumf) &= \normal(\taumf = 0.00554, \sigma_{\taumf} = 0.00064)\,.
   \end{split}
   \label{eq:boss_prior}
\end{equation}
We then use Quasi-Monte Carlo (\textsc{qmc}) to integrate out the meanflux parameters with $30\,000$ samples of $(\betamf, \taumf)$.
\textsc{qmc} takes samples from a so-called low-discrepancy sequence, leading to faster convergence.
Here we draw $30\,000$ samples generated from a scrambled Halton sequence, which gives samples approximately uniformly distributed on a unit square $[0, 1]^2$.
We then use inverse transform sampling to transform the Halton sequence to the distribution described in Eq~\ref{eq:boss_prior}.


\subsubsection{Model evidence for the DLA model}
Suppose we have a trained GP null model in Eq~\ref{eq:null_likelihood_function},
the DLA likelihood function will be the null model likelihood function multiplied by Voigt profiles for each line in the Lyman series of the absorber:
\begin{equation}
   \begin{split}
      p(\yvec \mid \lambdavec, \nuvec, &\zqso, \betamf, \taumf,
      \\ &\{{\zdla}_i\}_{i=1}^{k}, \{{\nhi}_i\}_{i=1}^{k}, \mkdla )
      \\
      = \normal (\yvec &; \avec_{(k)} \circ \muvec(\zqso; \betamf, \taumf),\\
      &\Avec_{(k)} (\Alyavec (\Kvec + \Omegavec) \Alyavec) \Avec_{(k)} + \diag{\nuvec_i})\,.
   \end{split}
   \label{eq:dla_likelihood_function}
\end{equation}
Here ${\Avec_{(k)}} = \diag{\avec_{(k)}}$ and $\avec_{(k)}$ is the function with $k$ voigt profiles, which represents $k$ DLAs:
\begin{equation}
   \begin{split}
      \avec_{(k)} = \prod_{i = 1}^{k} a(\lambdavec; {\zdla}_i, {\nhi}_i)\,.
   \end{split}
\end{equation}
$a(\lambdavec; {\zdla}, {\nhi})$ is a Voigt profile parameterised by the DLA's redshift, $\zdla$, and the column density of the DLA, $\nhi$.
The Voigt profile parameterisation used in this work is the same as \cite{Garnett17}.
We set the maximum number of DLAs per spectrum at $k = 3$ in this work, as there are rarely more than three DLAs per spectrum.
As described in \cite{Garnett17}, the default Voigt profile we use in this work includes {\Ly}$\alpha$, {\Ly}$\beta$, and {\Ly}$\gamma$ absorption, which allows us to constrain the DLA column density better when the {\Lyb} forest is in the observation window.

To get the model evidence, according to Eq~\ref{eq:dla_likelihood_function}, we need to integrate out the prior over the DLA parameters and the meanflux parameters $(\betamf, \taumf)$.
For convenience, we denote the parameters which need to be integrated out by $\theta = \{ \{{\zdla}_i\}_{i=1}^{k}, \{{\nhi}_i\}_{i=1}^{k}, \betamf, \taumf \}$.

For the model with a single DLA,
we have four parameters $\theta = \{ {\zdla}, {\nhi}, \betamf, \taumf \}$.
The model evidence is:
\begin{equation}
   \begin{split}
      p(\yvec \mid \lambdavec, &\nuvec, \zqso, \mdla)
      =\\
      \int p(\yvec &\mid \lambdavec, \nuvec, \zqso, \theta, \mdla)p(\theta \mid \zqso, \mdla) \dd \theta.
   \end{split}
   \label{eq:model_evidence_dla}
\end{equation}
By assuming each parameter is independent of each other,
we factorise the parameter prior as:
\begin{equation}
   \begin{split}
      p(&\theta \mid \zqso, \mdla)\\
      &= p(\zdla \mid \zqso, \mdla) p(\nhi \mid \mdla) p(\betamf) p(\taumf),
   \end{split}
   \label{eq:dla_prior}
\end{equation}
where we assign the \cite{Kamble:2020} prior for the meanflux parameters as in Eq~\ref{eq:boss_prior}.
We use the same prior for column density, $p(\nhi \mid \mdla)$, as \cite{Ho:2020}. This was trained using kernel density estimation on the $\lognhi$ distribution from \cite{Lee2013} DR9 DLAs with an addition of a $3\%$ uniform prior.

The $\zdla$ prior is uniform within the search range for DLAs.
We set this search range to be from {\Lyb} to {\Lya}.
Removing DLAs detected in the {\Lyb} forest ensures the purity of DLA samples in deriving the statistical properties of the DLA population. However, to generate a complete catalogue, we also consider a search range from the Lyman limit to {\Lyb}.

We used the same model and priors for the sub-DLA model as in \cite{Ho:2020}.
The sampling range of the redshifts of sub-DLAs is the same as for the DLA model.
Model priors are the same as \cite{Ho:2020}, based on the DLA catalogue in SDSS DR9 \citep{Concordance2012}.

\subsection{Example spectra}

\begin{figure*}
   \includegraphics[width=2\columnwidth]{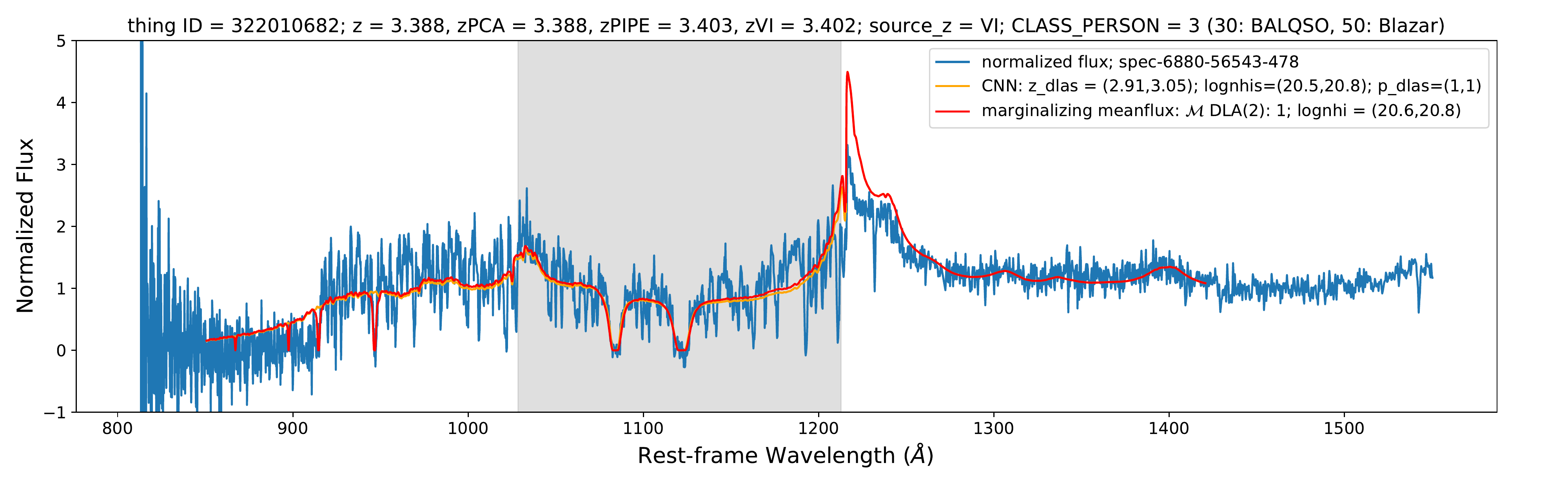}
   \includegraphics[width=2\columnwidth]{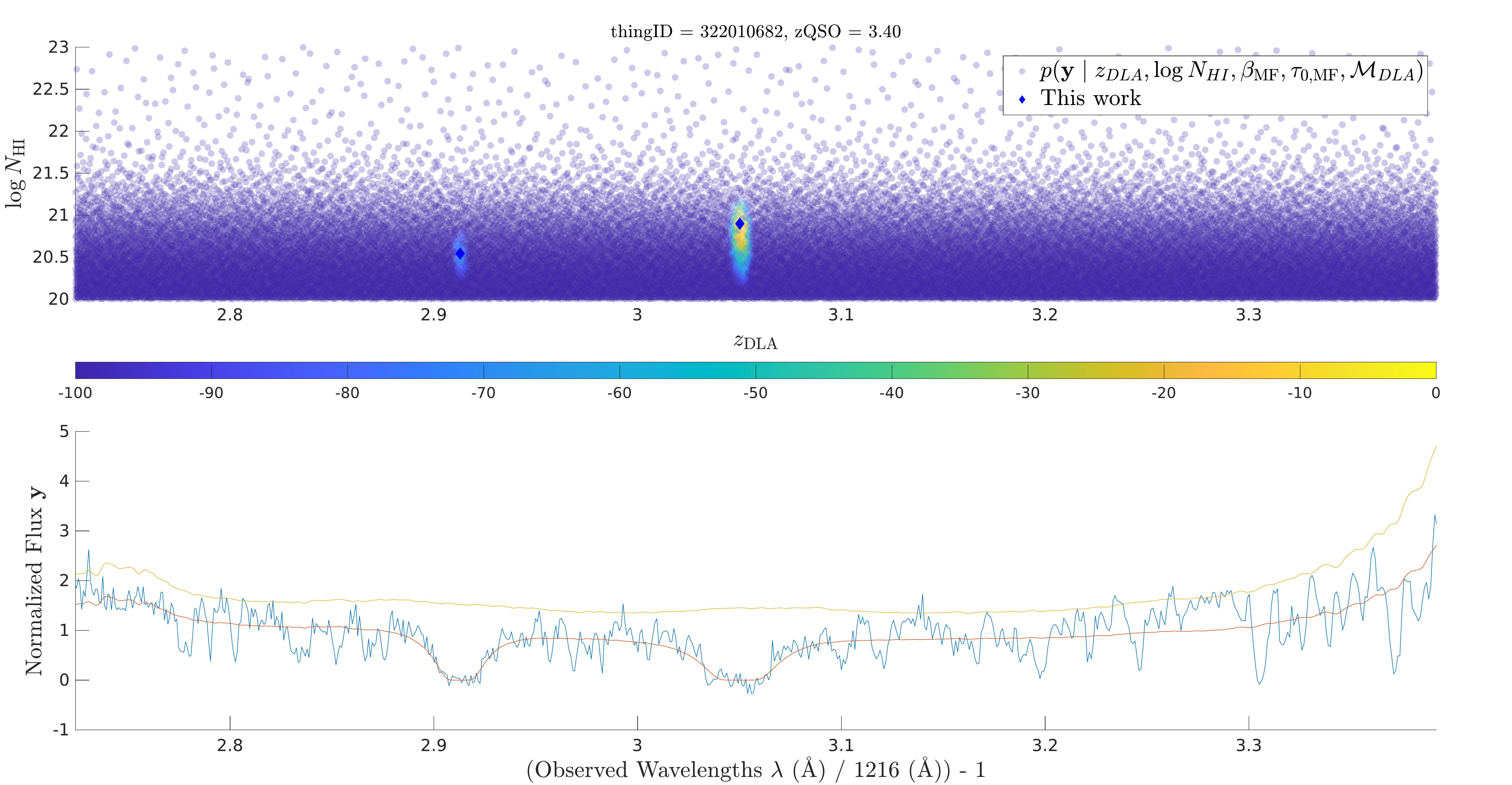}
   \caption{An example of a spectrum with distinct DLA features.
   \textbf{(Top):} The normalised observed spectrum in rest-frame wavelengths (blue) with the GP model (red) and the detection from the CNN model reported in DR16Q (orange).
   The title shows a series of column values in SDSS DR16Q catalogue, including SDSS identifier, best available redshift, PCA redshift, SDSS pipeline redshift, redshift from visual inspection, source for the best available redshift, and object classification from visual inspection ($0$: not inspected; $1$: star; $3$: quasar; $4$: galaxy; $30$: BAL quasar; $50$: Blazar(?)).
   Shaded area (grey) shows the sampling range of $\zdla$, which is from \Ly$\beta + 3\,000 \kms$ to $\zqso - 3\,000\kms$.
   The legend shows the spectrum is from  spec-6880-56543-478 (spec-\texttt{plate}-\texttt{mjd}-\texttt{fiber\_id}). The CNN model (orange) detected two DLAs, with redshifts of $\zdla = 2.91, 3.05$ and column densities of $\lognhi = 20.5, 20.8$, at DLA confidence $= 1$ for each DLA.
   Our GP model (red) also detected two DLAs with the model posterior $p(\mdla{_{(2)}} \mid \Data ) = 1$ and column densities $\lognhi = 20.6, 20.8$.
   \textbf{(Middle):} The sample likelihoods of detecting DLAs in the parameter space, $\theta \in (\zdla, \lognhi)$. Colour bar shows the normalised log likelihoods, $\log p(\yvec \mid \zdla, \lognhi, \taumf, \betamf, \zqso, \mdla)$, with the maximum log likelihood to be zero.
   We also show the maximum a posteriori estimates of DLAs in the blue squares.
   The posterior distribution sharply peaks at the parameter space,
   indicating the detection of these DLAs in high confidence.
   \textbf{(Bottom):} The observed flux (blue) as a function of absorber redshifts with the GP model (red) and the GP model before the meanflux suppression (yellow). The position on the x-axis directly corresponds to the x-axis in the middle plot.
   }
   \label{fig:example_accurate}
\end{figure*}

In this section, we show some example spectra to demonstrate our proposed model.
Figure~\ref{fig:example_accurate} shows an example with prominent DLA features.
As shown in the parameter space (middle plot), the posterior distribution is peaked at the \textit{maximum a posteriori} (MAP) values of those two DLAs.
Our GP model estimates the parameters of the DLAs with small uncertainties.
As shown in the top plot, our MAP values agree with the column densities measured by the CNN model reported in the DR16Q catalogue.

\begin{figure*}
   \includegraphics[width=2\columnwidth]{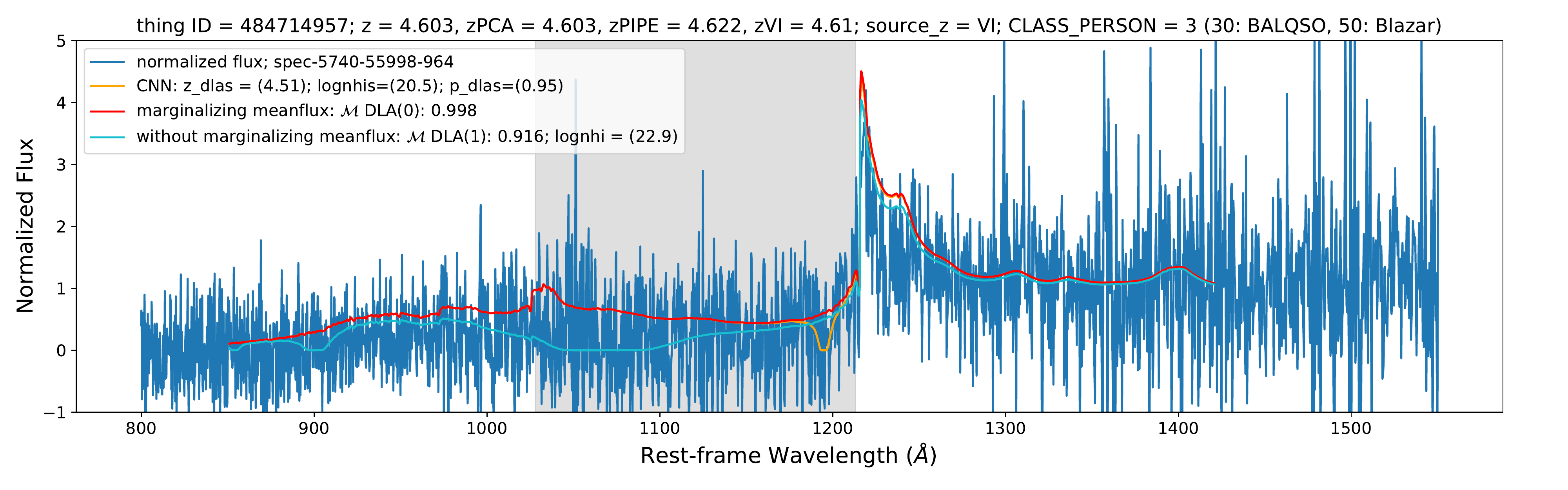}
   \caption{
   An example of a noisy spectrum with an uncertain meanflux.
   The normalised observed spectrum in rest-frame wavelengths (blue) with the GP model (red) and the detection from the CNN model reported in DR16Q (orange).
   We also plot the result without marginalising the uncertainty of meanflux prior (cyan).
   Shaded area (grey) shows the sampling range of $\zdla$, which is from \Ly$\beta + 3\,000 \kms$ to $\zqso - 3\,000\kms$.
   Our proposed model (red) indicates no DLA in the spectrum, with the null model posterior $p(\mnodla \mid \Data) = 0.998$.
   On the other hand, if our model ignores the uncertainty of $(\betamf, \taumf)$, it would falsely detect a DLA with $p(\mdla \mid \Data) = 0.916$ with $\lognhi = 22.9$ (cyan).
   When marginalising over the uncertainty in effective optical depth, our proposed model (red) avoids detecting a false-positive large DLA.
   }
   \label{fig:example_uncertain_taueff}
\end{figure*}

Figure~\ref{fig:example_uncertain_taueff} shows an extremely noisy spectrum, for
which our GP model is very uncertain about the effective {\Lya} absorption
in the spectrum.
The DLA models are degenerate with the absorption from the {\Lya} forest.
Without modelling the uncertainty in the mean flux,
the GP model does not know that the drop in the spectrum can be explained by {\Lya} forest absorption.
It instead fits a big DLA with $\nhi = 10^{22.9} \cm^{-2}$ as its preferred explanation for the drop in flux.

\subsection{Selection on the strength of Occam's razor}
\label{subsection:occams}

As we use more parameters to compute the DLA or sub-DLA model, the model selection will prefer to fit a Voigt profile to the GP if all candidate models are poorly fit.
Thus, the DLA or sub-DLA model's evidence is sometimes too strong compared to the null model.

The most common poor fit situations are quasar spectra with $\zqso < 2.5$ and with low signal-to-noise ratios (SNR).
As SDSS optical spectra have a fixed observing window, quasar spectra with $\zqso < 2.5$ have an incomplete {\Lya} forest.
The constraining power of the quasar becomes weaker as only part of the data fits into our modelling window, $[850.75\,\textrm{\AA}, 1420.75\,\textrm{\AA}]$.
Thus the DLA model and the null model are closer in likelihood space.

To avoid this situation, we introduced an additional Occam's razor in \cite{Ho:2020}, which is injected in the model selection as:
\begin{equation}
   \begin{split}
      \Prob(&\mdla \mid \Data) =\\
      &\frac{\Prob(\mdla)p(\Data \mid \mdla)\frac{1}{N}}{
         \left(\begin{matrix}
            \Prob(\mdla)p(\Data \mid \mdla)\\
            +\Prob(\msubdla)p(\Data \mid \msubdla)\\
         \end{matrix}\right)\frac{1}{N}
         +\Prob(\mnodla \mid \Data)
      },
   \end{split}
   \label{eq:occams}
\end{equation}
Here $N$ is the Occam's razor penalty, and we used $N =10\,000$ in \cite{Ho:2020}. We previously validated the Occam's razor strength by matching it to the DR9 concordance catalogue \citep{Concordance2012}.

In this work, however, we modify our null model to consider uncertainty from the mean flux measurement, which means it has more parameters. Thus, the null model gains more constraining power, so a weaker Occam's razor may be preferable.
To make our model posteriors more consistent with human identifications,
we decided to conduct a visual inspection on a small subset of the spectra.

We first train a model without Occam's razor and select at random from this model $239$ putative large DLAs with $\nhi > 10^{22}\cm^{-2}$ and $243$ putative small DLAs with $10^{20} \leq \nhi < 10^{21} \cm^{-2}$. We visually inspect each spectrum and compute the model posteriors with a range of strengths for Occam's razor, $N = \{1, 10, 100, 1\,000, 30\,000\}$.
We then treat each spectrum as a multiple-choice problem: if we think the model posterior of a given Occam's razor describes the given spectrum well, then we record one vote for this value of Occam's razor.
Multiple selections are allowed for each spectrum as the model posteriors are often very close. After collecting votes, the winning value of Occam's razor was $N = 1\,000$, a ten times reduction from our earlier value.

For quasar spectra with $\zqso > 2.5$ there are enough data points in the {\Lya} range that the strength of Occam's razor has a small effect. We will discuss the effect of Occam's razor in Section~\ref{sec:systematics}. We suggest incorporating variations due to Occam's razor into the uncertainty in population statistics for conservative usage.

\subsection{Summary of the modifications}

Here we summarise the modifications we made in this work, comparing to the model of \cite{Ho:2020}:
\begin{enumerate}
   \item Our training set is SDSS DR12 quasar spectra with DLAs detected by \cite{Ho:2020} removed. We considered a DLA to be detected if the posterior probability of a spectrum containing a DLA is larger than $0.9$, $P(\mdla \mid \Data) > 0.9$.
   \item The wavelength range modelled goes from $\lambdarest = 850.75$ {\AA} to $\lambdarest = 1\,420.75 $ {\AA}.
   \item The effective optical depth prior, $(\tau_0, \beta)$, is updated from \cite{Kim07} to \cite{Kamble:2020}.
   \item The uncertainty in the mean flux suppression parameters, $\tau_0$ and $\beta$, is marginalised while computing the model evidence.
\end{enumerate}
The first modification gives us a training set size containing $89,408$ spectra without DLAs.
The larger training set better learns the covariance structure of quasar emission lines, which allows the second modification: expanding the model to cover the Lyman break and {\siv} line.
The expansion enables the model to use the metal lines to constrain the correlations of the emission lines in the {\Lya} forest.
When using the previous modelling range, $[911.75 \AAtext, 1215.75 \AAtext]$, we found that we often detected DLAs with high $\nhi$ in the red end of the spectrum, where the code inserts a DLA at the quasar redshift to compensate for an oddly shaped {\Lya} emission line. This was possible because when we cut the spectrum at a rest frame wavelength $1215.75 \AAtext$, half of the damping wings were removed, allowing for more model freedom and dubious $\nhi$ estimation.

Third, to make the mean flux suppression prior for $(\tau_0, \beta)$ consistent with the DR12Q training set, we switched to the mean flux measurement based on BOSS DR12Q \citep{Kamble:2020}.
Our last modification is marginalising the uncertainty of \cite{Kamble:2020}'s parameters while marginalising the DLA parameters.

To compute the statistical properties of DLAs, we need to convert the posterior distribution of a DLA in each spectrum into the expected number of DLAs per redshift or column density bin, for which we use the method described in \cite{Bird17}.
We briefly summarise the modelling decisions we made to produce the DLA samples in the result section:
\begin{itemize}
   \item Search range: from {\Lyb} to {\Lya}.\footnote{For the CDDF in \cite{Ho:2020}, we used a sampling range from \Ly$\beta + 3\,000 \kms$ to \Ly$\alpha - 30\,000 \kms$ to avoid finding DLAs in the proximity zone.
   Here, we instead use \Ly$\beta + 3\,000 \kms$ to \Ly$\alpha - 3\,000 \kms$.
   This has a very moderate effect on our results, however,
   we provide a check of systematics due to removing DLAs near to the quasar redshift in Section~\ref{subsec:z_qsos}.}.
   \item Maximum number of DLAs: three.
   \item Maximum $\zqso$: quasar redshifts $< 7$.
   \item DLA redshift $2 < \zdla < 5$.
\end{itemize}

\section{Results}
\label{sec:results}

\subsection{Column density distribution function}
\label{subsec:cddf}

\begin{figure}
   \includegraphics[width=\columnwidth]{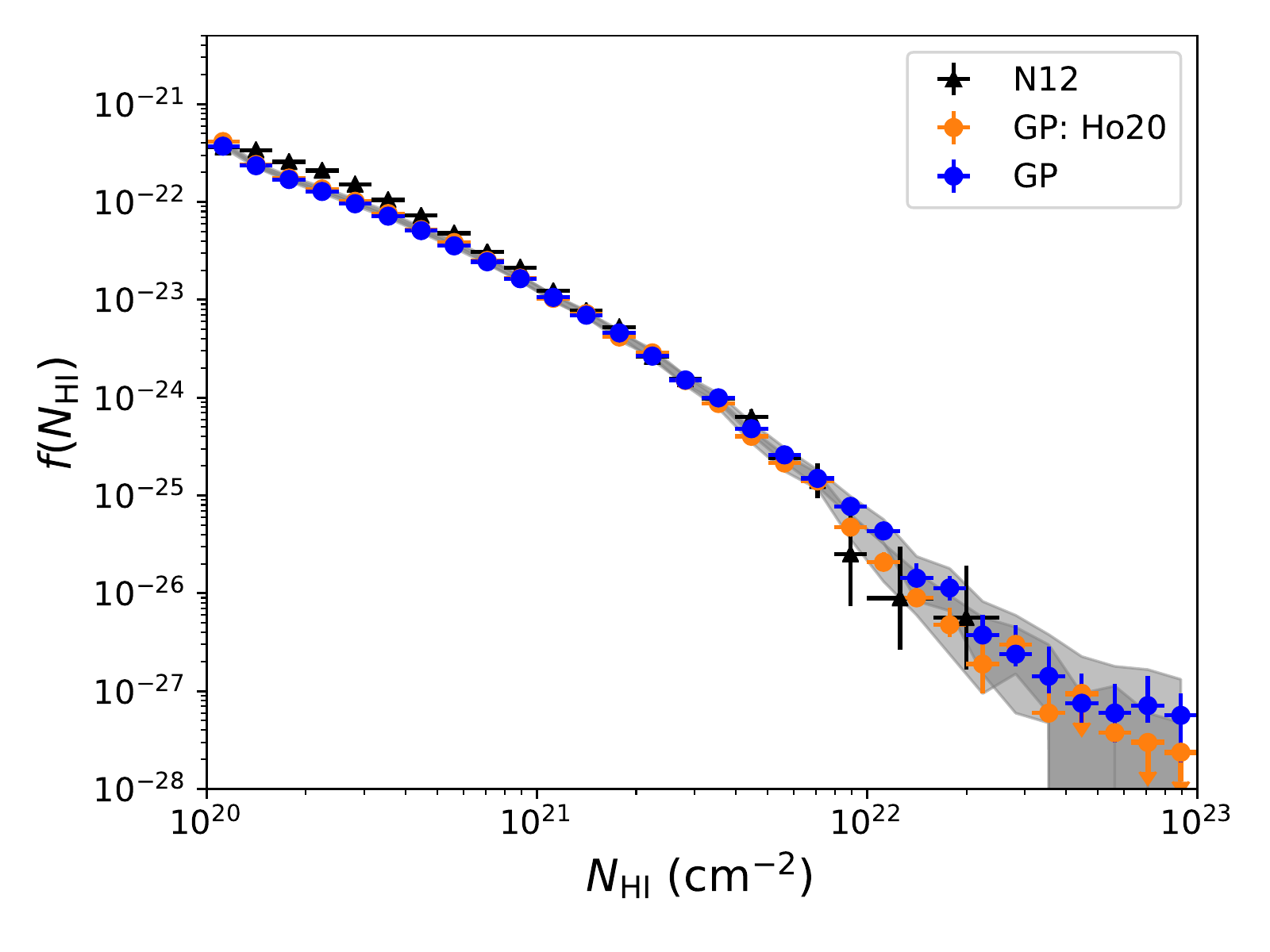}
   \caption{The CDDF, integrated over all $z < 5$ spectral path, derived from SDSS DR16Q spectra with our proposed Gaussian process models (GP; blue).
   The CDDF measurements from \protect\cite{Ho:2020} (GP: Ho20; orange) are plotted as a comparison.
   Error bars show the $68\%$ confidence limits, while grey areas show the $95\%$ confidence limits.
   Black dots are from \protect\cite{Noterdaeme12} (N12).
   }
   \label{fig:cddf}
\end{figure}

Figure~\ref{fig:cddf} shows the CDDF we estimate from DR16Q spectra.
In the following sections,
the CDDF is computed for $\nhi \in [10^{20}, 10^{23}]$, while the DLA incidence rate $\dd N /\dd X$ and the total HI density in DLAs $\omegadla$ are computed for $\nhi \in [10^{20.3}, 10^{23}]$.
Ho20 refers to \cite{Ho:2020}, a DR12 DLA catalogue that used a modified GP model from \cite{Garnett17}.

The CDDF is a histogram of column densities normalised by the effective spectral path that could contain DLAs. We count all spectral path with an absorber with $\zdla < 5$.
Error bars denote the $68 \%$ confidence limits, and the grey band represents the $95 \%$ confidence limits.
Note that the uncertainties here are the statistical uncertainties associated with the GP model.
They do not include uncertainty due to potential systematics.
Section~\ref{sec:systematics} will describe how possible systematics would affect the CDDF.

As shown in Figure~\ref{fig:cddf}, we observe non-zero column density until $3 \times 10^{22} \cm^{-2}$.
Our DR16 measurement is mostly consistent with our previous DR12 measurement until $\nhi \leq 9 \times 10^{21}$.
For $\nhi \geq 3 \times 10^{22}$,
both our DR12 and DR16 measurement are consistent with zero at $95 \%$ confidence level,
though there is one bin from DR16 not consistent with zero (see Table~\ref{tab:CDDF_analysis/dr16q_full_int_lyb_occam_zqso7_1_30_delta_z_0_1/cddf_all.txt}).
We also measure no turn over for the CDDF at the high column end, $\nhi \sim 10^{21.5} \cm^{-2}$. It was suggested in \cite{Schaye:2001} that molecular hydrogen sets a maximum ${\nhi}$ so that steepen the CDDF at the high end.
The latest simulated CDDF from \textsc{simba} \citep{Hassan:2020}, which included molecular hydrogen formation in their star formation recipe, predicts no turn over at the high end, consistent with our measurements.

\begin{figure}
   \includegraphics[width=\columnwidth]{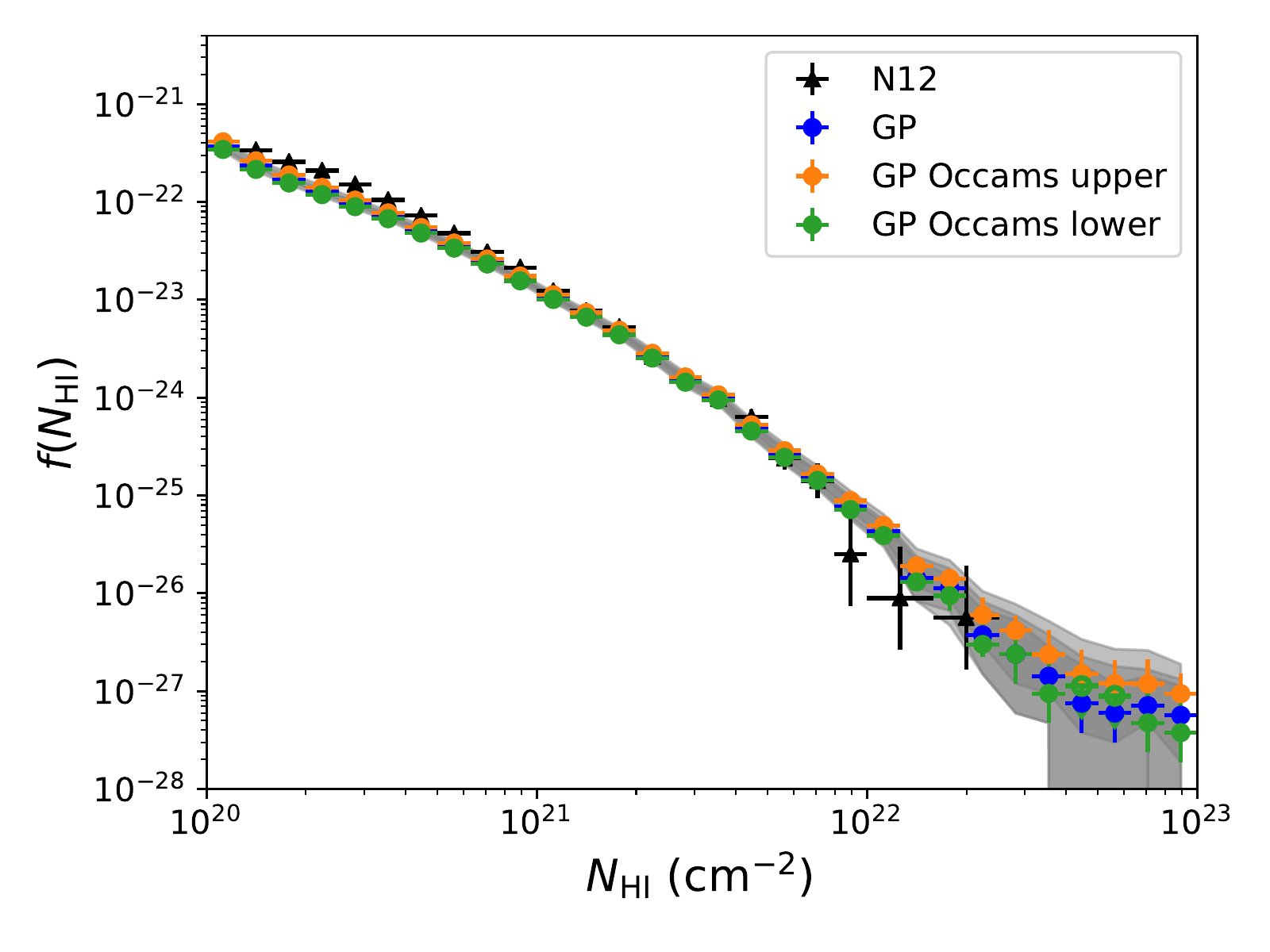}
   \caption{The CDDFs with different Occam's razor strengths, which discussed in Section~\ref{subsection:occams}.
   Occam's upper (orange) represents $N = 30\,000$ while Occam's lower (green) represents $N = 30$.
   We present our main result (GP; blue) with an optimal strength $N = 1\,000$, which we selected from visually inspecting a subset of the dataset.
   Note that the difference between different Occam's strengths is well within $95\%$ confidence limits.
   Black dots are from \protect\cite{Noterdaeme12} (N12).
   }
   \label{fig:cddf_occams}
\end{figure}

In Figure~\ref{fig:cddf_occams},
we plot the CDDF with different Occam's razor strengths.
When the Occam's razor strength is weak ($N=30$),
model selection will find DLAs even though the SNR is low,
so we get more absorbers at both high and low column density ends.
On the other hand,
if the razor strength is strong ($N = 30\,000$),
model selection will prefer to avoid finding DLAs at low SNR spectra,
which results in a decrease.

However, in general,
in Figure~\ref{fig:cddf_occams},
we observe the razor strength only marginally affects the CDDF.
Thus the small tension at the low end, $\nhi \in [10^{20}, 10^{20.3}]$, between our CDDF and N12 is more likely due to other reasons than Occam's razor.

\begin{figure}
   \includegraphics[width=\columnwidth]{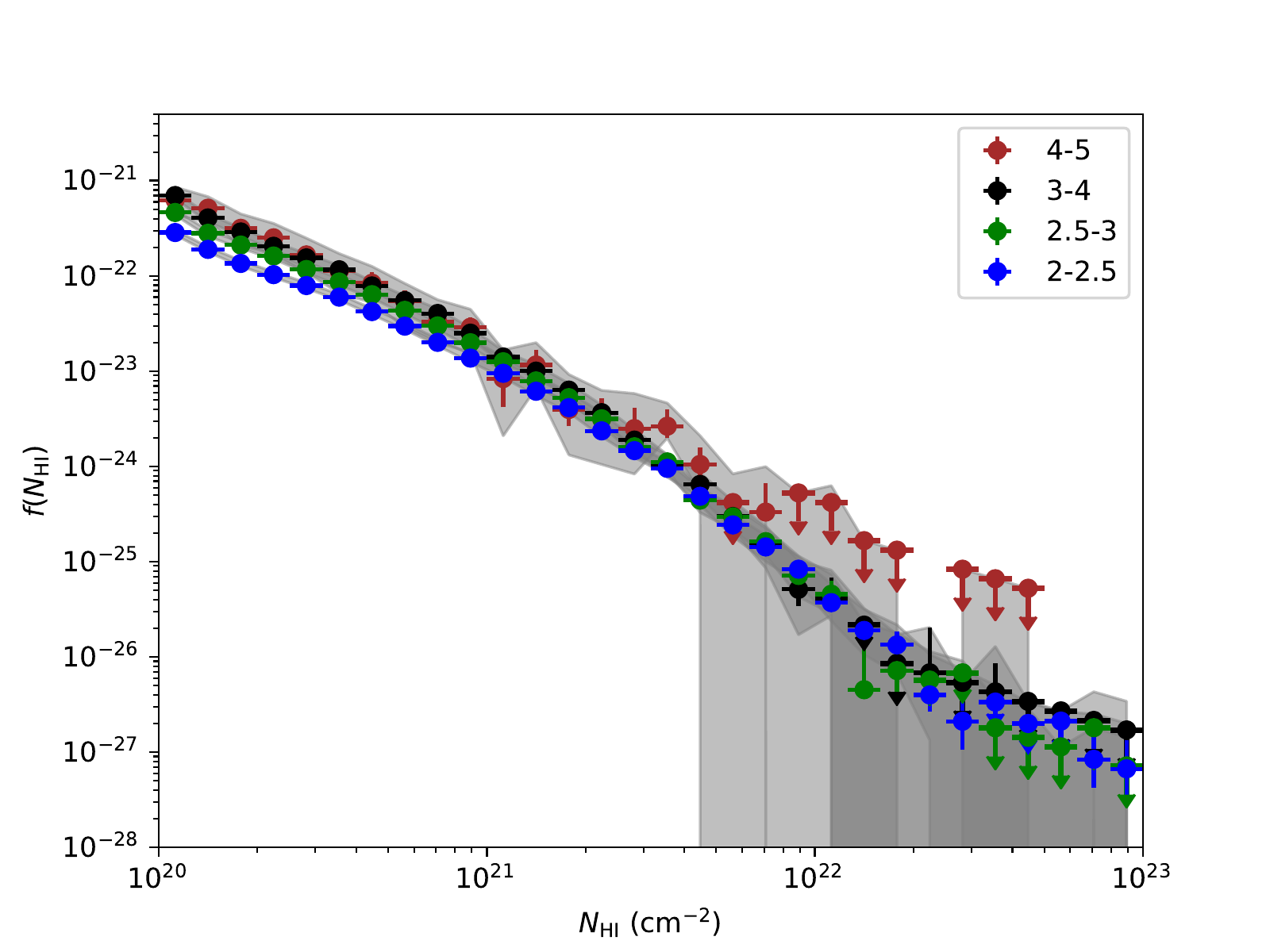}
   \caption{The CDDF derived from DLAs in a variety of redshift bins. Labels show the redshift bins in used.
   We show $68\%$ confidence limits in error bars and $95\%$ confidence limits in grey areas.
   If the bin is consistent with no detection at $68\%$ limits,
   we show a down-pointing arrow indicating the $68\%$ confidence upper limit.}
   \label{fig:cddf_zz}
\end{figure}

We show the redshift evolution of the CDDF in Figure~\ref{fig:cddf_zz}.
The downward pointing symbols indicate the $68\%$ upper confidence limit when the data is consistent with zero at $68\%$ confidence limits.
As we can anticipate, for high-redshift quasars with $\zqso > 4$, since the flux is highly absorbed,
we detect DLAs with larger uncertainties, and the number of large DLAs is consistent with zero.

In both \cite{Bird17} and \cite{Ho:2020},
we found that the CDDF is getting shallower at $z > 4$.
However, given our detection for $\nhi > 4 \times 10^{21}$ at $z > 4$ is highly uncertain and consistent with zero detection,
this trend is not significant in our current dataset.
Instead, the detection of DLAs with $\nhi < 4 \times 10^{21}$ at $z > 4$ is consistent with the measurements at $z \in [2.5, 4]$.
We find no strong evidence for an evolution of the slope of the CDDF at $z > 4$.

One possible reason why we found the CDDF was shallower at $z > 4$ in \cite{Bird17} and \cite{Ho:2020} is absorption due to the {\Lya} forest.
When the spectrum is highly absorbed, there is a degeneracy between a large DLA and the forest's absorption.
In \cite{Bird17}, we did not model the GP mean as a function of effective optical depth,
so it is possible the model was trying to use DLAs to compensate the excess absorption due to the forest,
which results in a shallower CDDF at $z > 4$.
In \cite{Ho:2020},
the slope of the CDDF is less shallow at $z > 4$, as we modelled the effective optical depth into our GP mean.
In this work, we integrated out the measurement uncertainty of the mean flux, and
the slope is almost indistinguishable from the CDDF at $z \in [2.5, 4]$.
This may indicate that, to understand the DLAs at $z > 4$ better,
we need a better measurement for the effective optical depth at $z > 4$.

One interesting feature in Figure~\ref{fig:cddf_zz} is the drop in the amplitude of the CDDF at $z \in [2, 2.5]$.
As we will discuss in Section~\ref{subsec:omega_dla},
the drop of CDDF at the low redshifts also shows in the DLA incident rate, $\dd N/ \dd X$.
We will discuss this in more detail in the next section.

\subsection{Redshift evolution of DLAs}
\label{subsec:omega_dla}

\begin{figure}
   \includegraphics[width=\columnwidth]{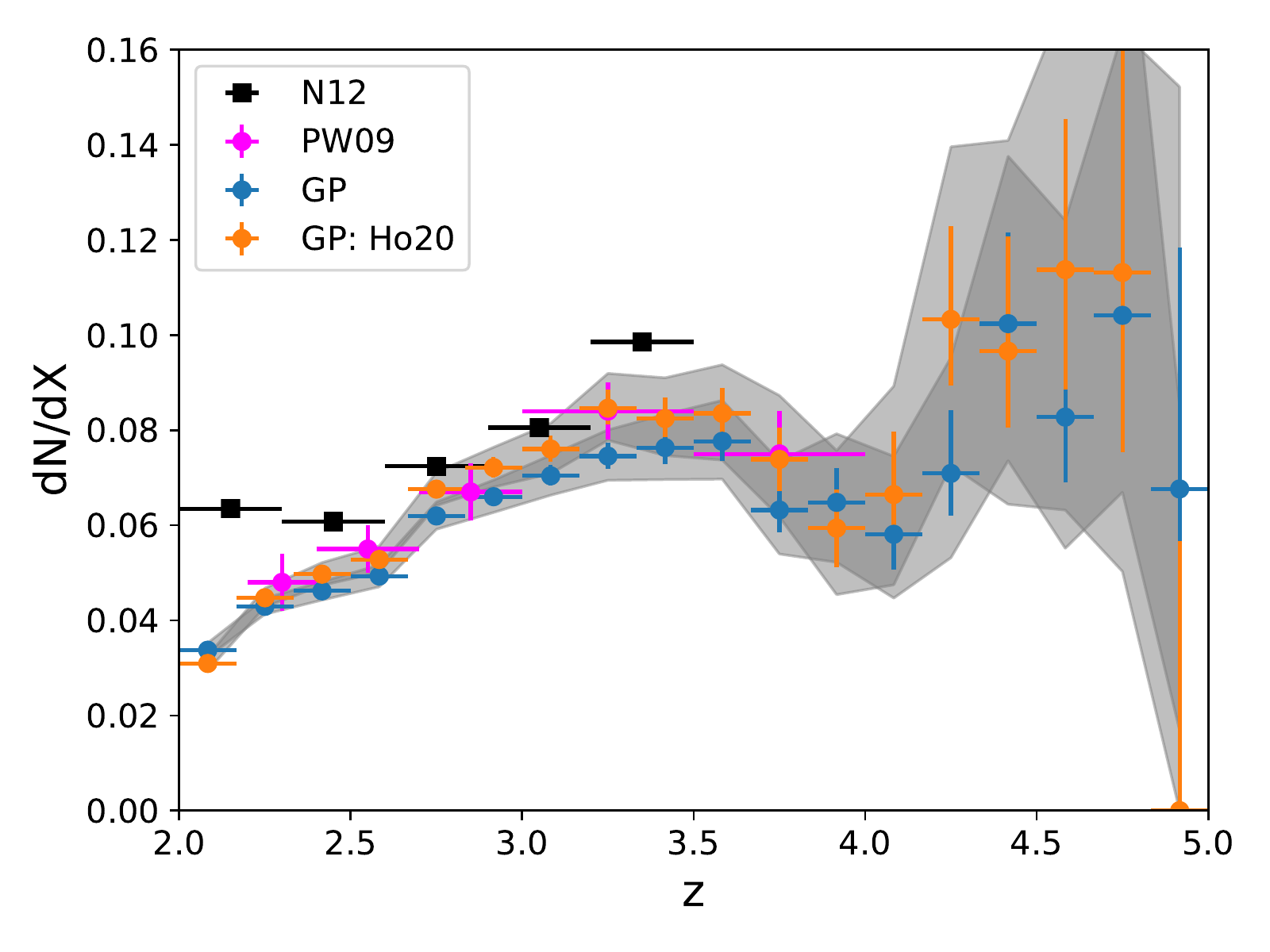}
   \caption{The incident rate of DLAs as a function of redshift,
   integrated over $\lognhi > 20.3$ spectra from our catalogue (GP; blue).
   We also plot the line densities from \protect\cite{Noterdaeme12} (N12; black) and \protect\cite{Prochaska2009} (PW09; pink),
   and \protect\cite{Ho:2020} (GP: Ho20; orange) as comparisons.
   }
   \label{fig:dndx}
\end{figure}

Figure~\ref{fig:dndx} shows the incident rate of DLAs, $\dd N/ \dd X$, as a function of absorber redshift.
Our results are consistent with \cite{Prochaska2009} and \cite{Ho:2020}
and are slightly lower than \cite{Noterdaeme12}.
$\dd N/ \dd X$ is sensitive to the weaker DLAs,
so the difference between \cite{Noterdaeme12} and \cite{Prochaska2009} is likely due to the false positive rate.

\cite{Prochaska2009} performed a visually-guided Voigt profile fitting on SDSS-DR5.
Though their sample size is smaller, with the help of the human eye, their method is likely less prone to
false positives than the automated template fitting used in \cite{Noterdaeme12}.
This difference may also explain the drop in amplitude of the CDDF at $z \in [2, 2.5]$.
\cite{Prochaska2009} and \cite{Noterdaeme12} have a larger discrepancy at $z \in [2, 2.5]$,
and \cite{Noterdaeme12} may overestimate the weak absorbers at this redshift range where the spectra are short.
Our measurement is consistent with \cite{Prochaska2009},
which implies that we detect fewer weak absorbers in low redshift bins
and explains the small tension in the CDDF at low-$\nhi$.

One noticeable feature in $\dd N/ \dd X$, which we have not discussed before, is the decrease of
the line density from $z = 3.5$ to $z = 4.0$ and another increase at $z > 4.0$.
This feature is also shown in our Ho20 measurement.
The drop of $\dd N/ \dd X$ at $z \in [3.5, 4]$ is consistent with \cite{Prochaska2009} at $95 \%$ confidence limits.
One interesting question is whether the increase from $z \in [4.0, 4.5]$ is real.
The measurements at $z > 4$ still have large error bars,
so it is hard to say whether $\dd N/\dd X$ at $z  > 4$ is an increase or a flat line.
More data, especially with high SNR, are needed to determine the trend of line density at $z > 4$.

\begin{figure}
   \includegraphics[width=\columnwidth]{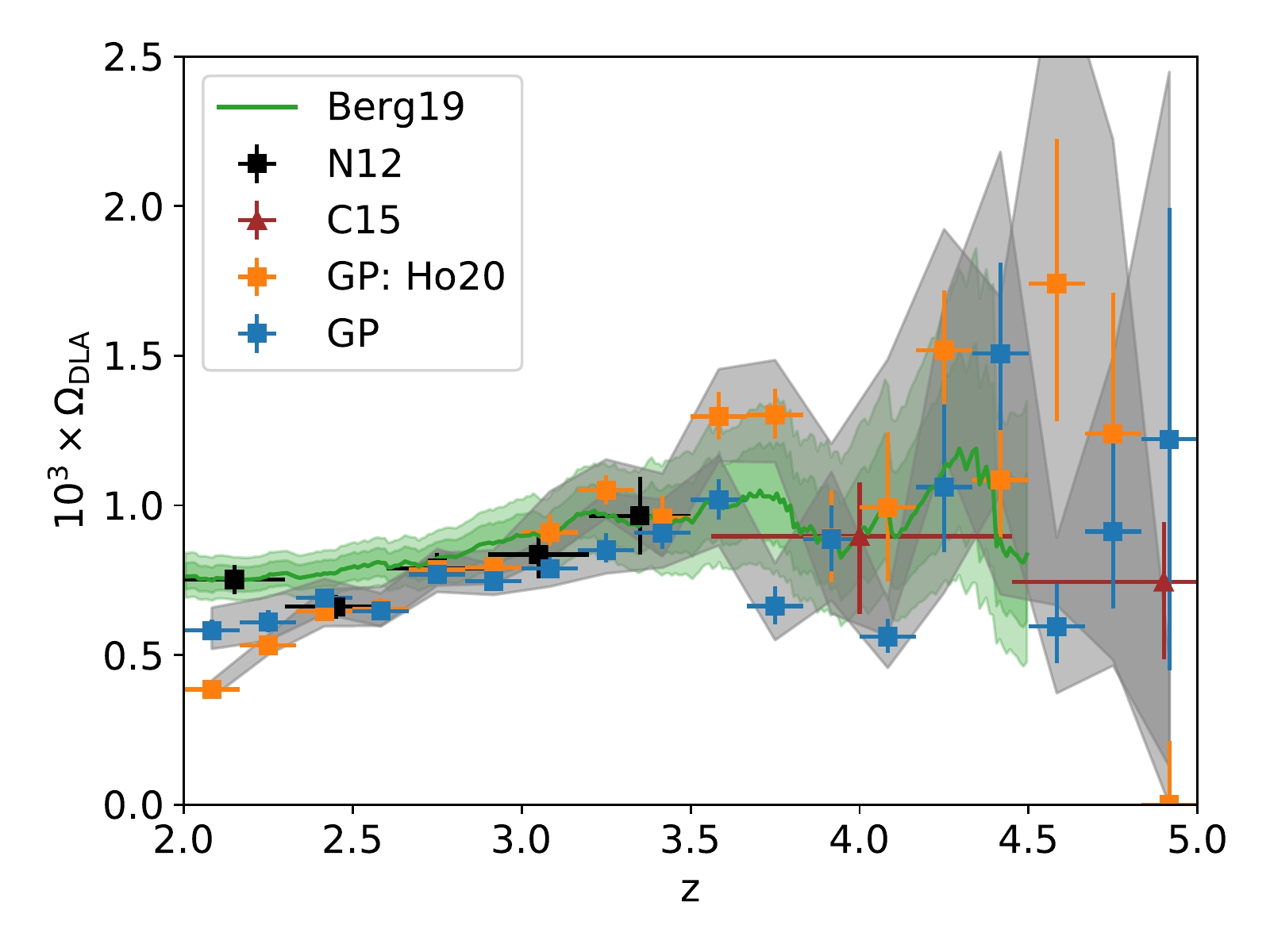}
   \caption{The total HI density in DLAs, integrated over DLAs with $\lognhi > 20.3$ in our catalogue (GP; blue).
   For comparison, we plot the measurements from
   \protect\cite{Berg:2019} (Berg19; green line and shaded area),
   \protect\cite{Noterdaeme12} (N12; black),
   \protect\cite{Crighton2015} (C15; red),
   and \protect\cite{Ho:2020} (GP: Ho20; orange).
   }
   \label{fig:omegadla}
\end{figure}

In Figure~\ref{fig:omegadla},
we show the total HI density in DLAs in terms of cosmic density.
Our results are mostly consistent with \cite{Noterdaeme12} at $z \in [2.5, 3.5]$.
At higher redshift bins, $z > 3.5$, our measurements are consistent with \cite{Prochaska2009} and \cite{Crighton2015}.
\cite{Crighton2015} used high signal-to-noise spectra from a smaller survey,
so they have larger error bars.
Comparing to Ho20,
our current $\omegadla$ has more mass at low-redshifts ($\zdla \sim 2$) and less mass at $\zdla \in [3.5, 4]$.
The trend of $\omegadla$ in DR16 is shallower than Ho20.

\begin{figure*}
   \includegraphics[width=\columnwidth]{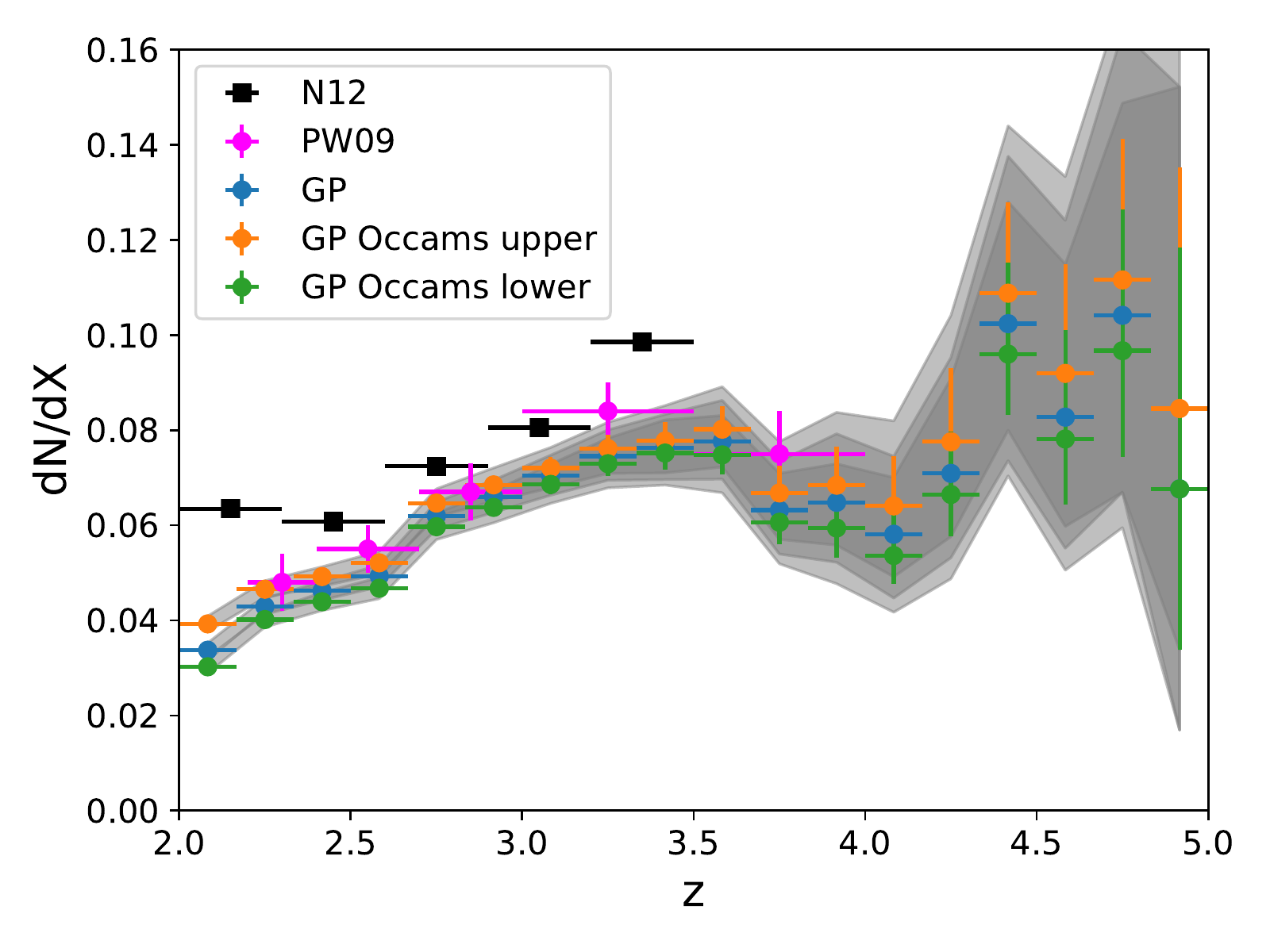}
   \includegraphics[width=\columnwidth]{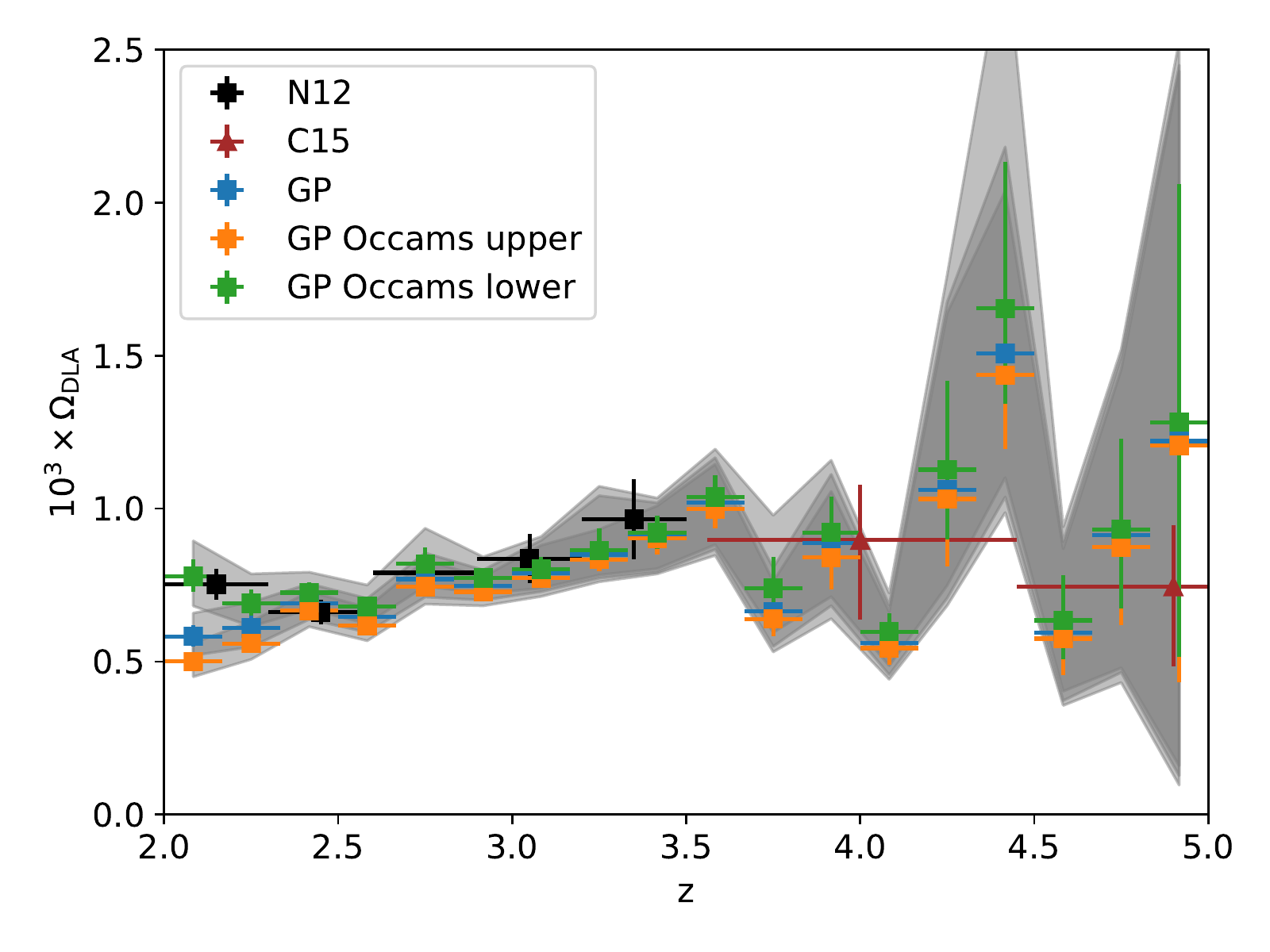}
   \caption{
   The line density \textbf{(left)} and $\omegadla$ \textbf{(right)} in DLAs as a function of redshifts with different Occam's razor strengths.
   Occam's upper (orange) represents $N = 30\,000$ while Occam's lower (green) represents $N = 30$.
   The main result (GP; blue) is computed with $N = 1\,000$.
   }
   \label{fig:omega_dla_occams}
\end{figure*}

We also plot the $\omegadla$ measured by \cite{Berg:2019} in Figure~\ref{fig:omegadla}.
We see our DR16 measurement is consistent with \cite{Berg:2019} even at $z > 3.5$.
There is a slight tension at $z < 2.5$, which may be because some low-redshift spectra are too short and noisy to measure column density confidently using our model.
SDSS spectra with $\zqso < 2.5$ only covers a region from the Ly$\alpha$ to Ly$\beta$ or shorter.
When the signal-to-noise is low, it is difficult to identify DLAs even using human eyes.
As shown in Fig~\ref{fig:omega_dla_occams},
a different selection of Occam's razor could moderately affect the two bins with $z < 2.33$.
The strength of Occam's penalty corresponds to a prior belief in detecting a DLA in a short and noisy spectrum, as discussed in Section~\ref{subsection:occams}.

Note that, from Figure~\ref{fig:cddf_zz} we see there are no solid detections for $\nhi > 3 \times 10^{21}$ DLAs at $z > 4$.
In \cite{Bird17}, $\omegadla$ was skewed towards high values at $z > 4$ even without real detections of large DLAs.
Our result in Figure~\ref{fig:omegadla} does not have this issue.
This may indicate our proposed method of integrating out the uncertainty on meanflux measurement helps us avoid the forest biasing the posterior density of $\nhi$ towards the high end.

In general, we observe an increase of $\omegadla$ from $z = 2$ to $z = 3.5$,
and a slight decrease from $z = 3.5$ to $z = 4$.
For $\zqso > 4$, the measurement error is larger and less correlated between redshift bins, as in
Ho20. This is reasonable given the lower quasar number density at high redshift.

Figure~\ref{fig:omega_dla_occams} shows the line density and $\omegadla$ with various Occam's razor strengths. As expected,
the razor strengths only moderately affect the statistics of low redshift spectra.
For $\dd N/ \dd X$, our results are consistent with \cite{Prochaska2009}, even with the weakest razor. N12 still detects somewhat more weak DLAs than we do, even though we only apply a small penalty for these short and noisy spectra.



\section{Checks for systematics}
\label{sec:systematics}

\subsection{Effect of signal to noise ratios}
\label{subsec:snrs}

\begin{figure}
   \includegraphics[width=\columnwidth]{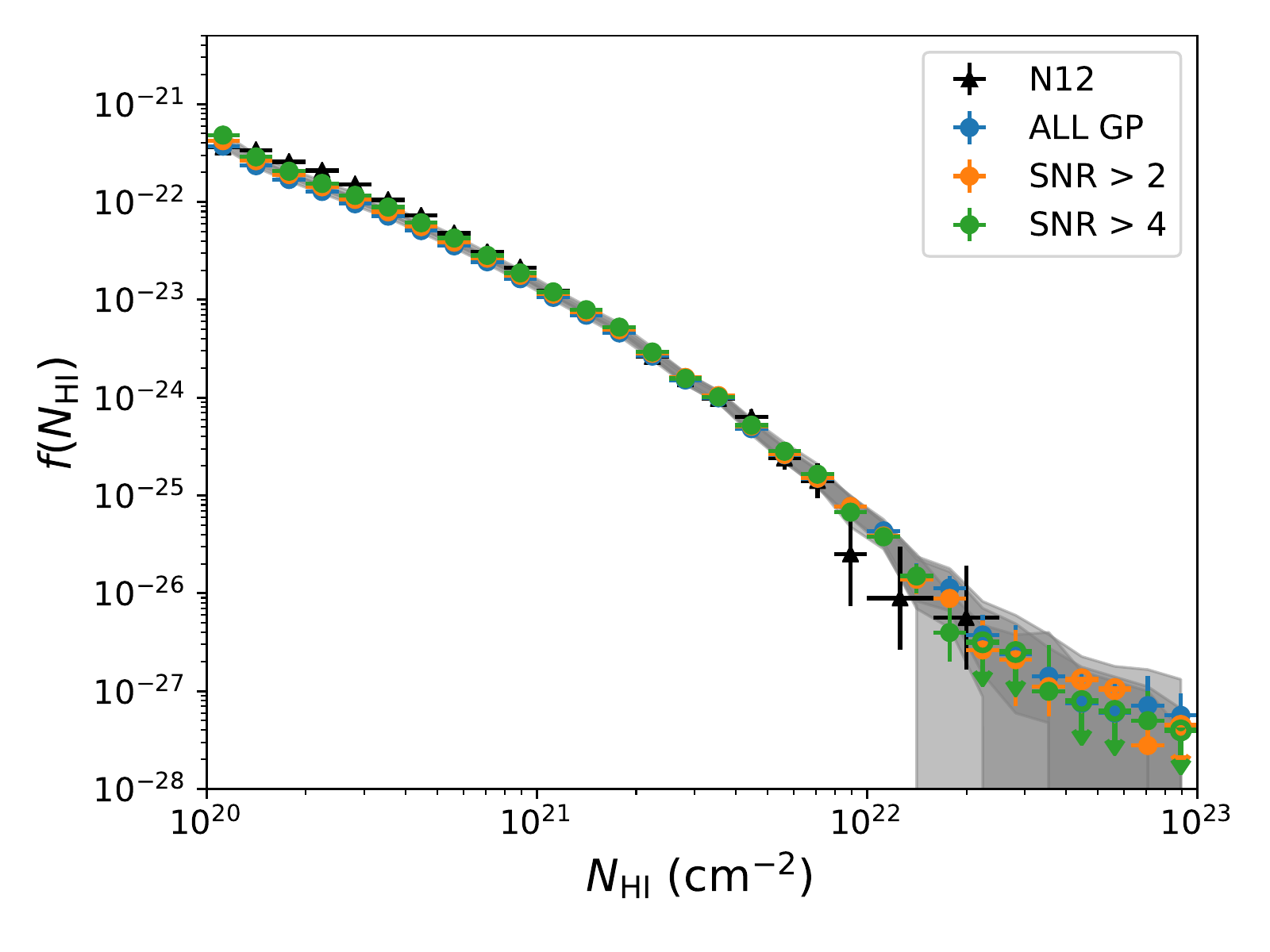}
   \caption{The CDDF of DLAs for a subset of samples with different minimal SNRs.
   SNR $> 2$ (orange) excludes $20\%$ of the noisiest spectra, and SNR $> 4$ (green) excludes $54\%$ of the spectra.
   68\% confidence limits are drawn as error bars, while 95\% confidence
   limits are shown as a grey filled band.
   }
   \label{fig:snrs_cddf}
\end{figure}

\begin{figure*}
   \includegraphics[width=\columnwidth]{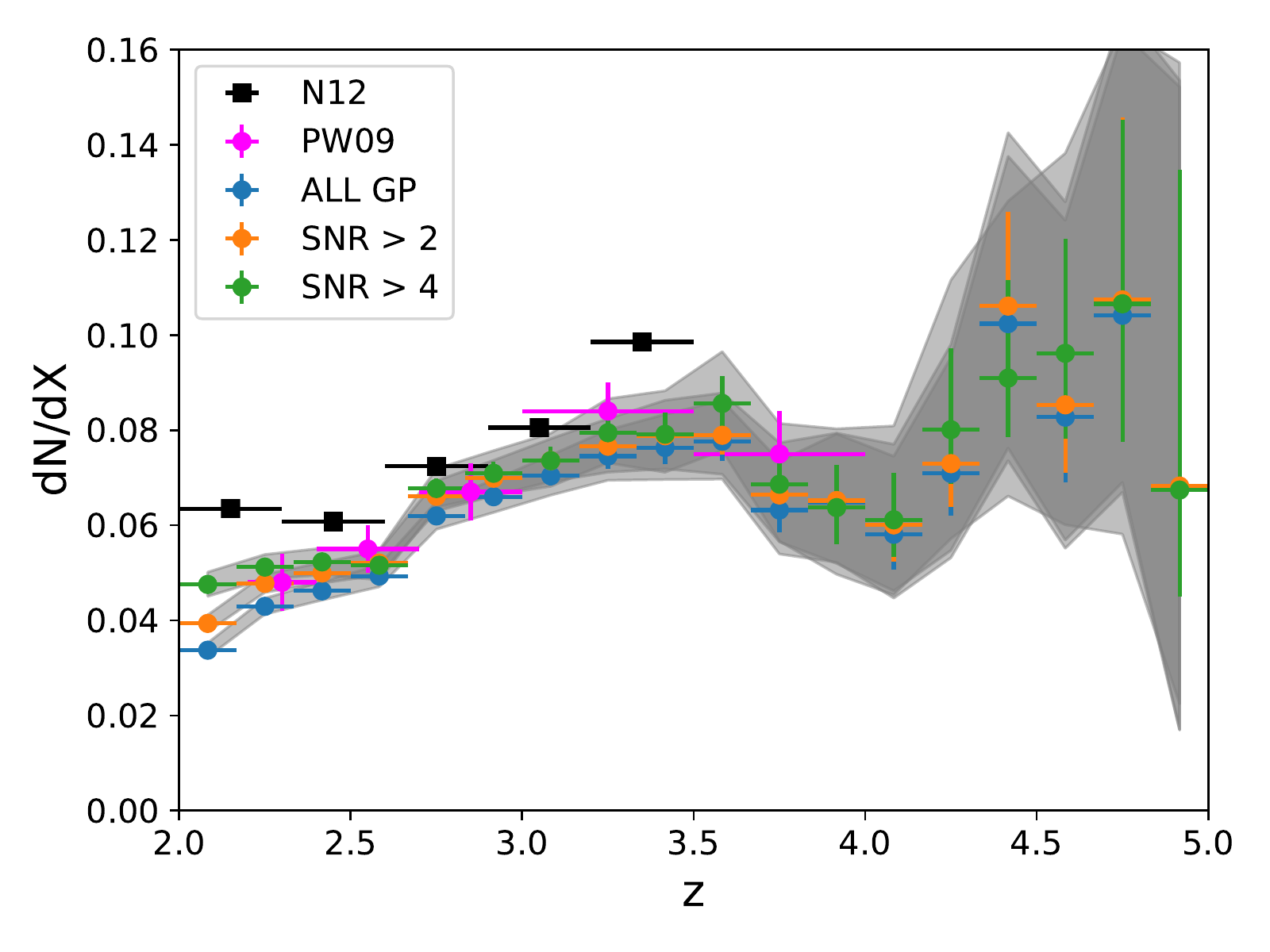}
   \includegraphics[width=\columnwidth]{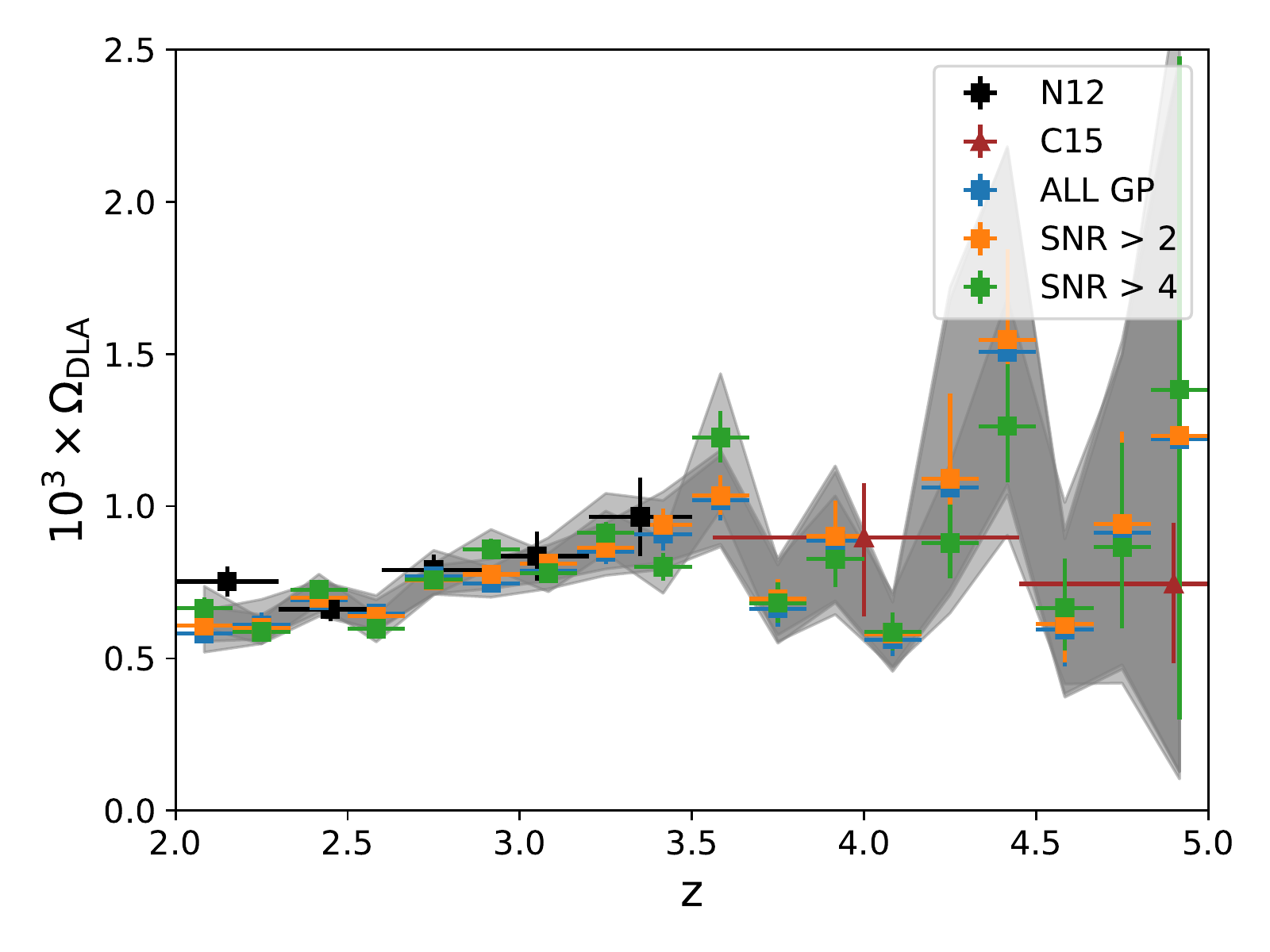}
   \caption{The line density \textbf{(left)} and total {$\nhi$} mass \textbf{(right)} in DLAs as a function of absorber redshift
   from subsets of samples with different minimal SNRs.
   SNR $> 2$ (orange) excludes $20\%$ of the noisiest spectra, and SNR $> 4$ (green) excludes $54\%$ of the spectra.
   }
   \label{fig:snrs_dndx_omegadla}
\end{figure*}

Figure~\ref{fig:snrs_cddf} and Figure~\ref{fig:snrs_dndx_omegadla} show the abundance of DLAs from subsets of our catalogue with various signal-to-noise cuts (SNR), SNR $>2$ and $>4$.
We define our SNR as the median of the ratio between the flux and the instrumental noise within the quasar spectrum redwards of the {\Lya} emission peak.
This specific choice is to avoid introducing correlations between the detected DLAs and the SNR.
With this definition,
$80\%$ of the quasar spectra have SNR $> 2$, and
$46\%$ of the spectra have SNR $> 4$.

We verify that, in Figure~\ref{fig:snrs_cddf},
the CDDF is not sensitive to the SNR when $\nhi < 10^{22} \cm^{-2}$.
However, we note that the highest non-zero column density at 95\% confident limits changed from $\nhi < 3\times 10^{22} \cm^{-2}$ to $\nhi \lesssim 10^{22} \cm^{-2}$ for samples with SNR $> 4$. This is likely because there are too few high column density absorbers to constrain the CDDF sufficiently at the high end in the smaller high SNR sample.

We find that our $\omegadla$ measurement exhibits no systematic correlation with the SNR cuts.
We notice a dependence of SNR on $\dd N/ \dd X$ at $z \in [2.0, 2.5]$,
which is due to the difficulty of finding DLAs in short and noisy spectra. 
As discussed in Section~\ref{subsection:occams}, it is hard to find features in these spectra, and the observing window cannot fully cover a high-$\nhi$ DLA profile with damping wings. To secure our samples' purity,
we use an Occam’s razor penalty which may also introduce this SNR dependence at $z \in [2, 2.5]$.

As mentioned in \cite{Krogager:2019}, the colour and magnitude criteria used in SDSS for quasar target selection is biased against dusty DLAs, which harbour a certain amount of cold neutral gas. \cite{Krogager:2019} showed that in SDSS DR7 this caused $\omegadla$ to be underestimated by $10 - 50\%$ at $z \sim 3$. Also, redder quasars containing metal rich dusty DLAs will have lower SNR in the blue part of the spectrum and thus may be excluded from the sample of \cite{Noterdaeme:2009}, who enforced $\mathrm{CNR} > 3$. This effect is likely to be substantially reduced in our sample, if present at all, as we use all quasars irrespective of SNR. We also define SNR using the region redwards of the \Ly{$\alpha$} emission peak specifically to avoid this kind of selection effect, and we are using DR16, which has a different and more complex selection function. More quantitatively, the XQ-100 targets in \cite{Berg:2019} use only radio-selected quasars, or quasars previously found by other techniques, and so avoids any SDSS colour selection bias. Figure~\ref{fig:omegadla} shows that our $\omegadla$ mostly agrees with \cite{Berg:2019}, implying that colour effects in our sample are smaller than those in DR7.

\subsection{Effect of quasar redshifts}
\label{subsec:z_qsos}

\begin{figure}
   \includegraphics[width=\columnwidth]{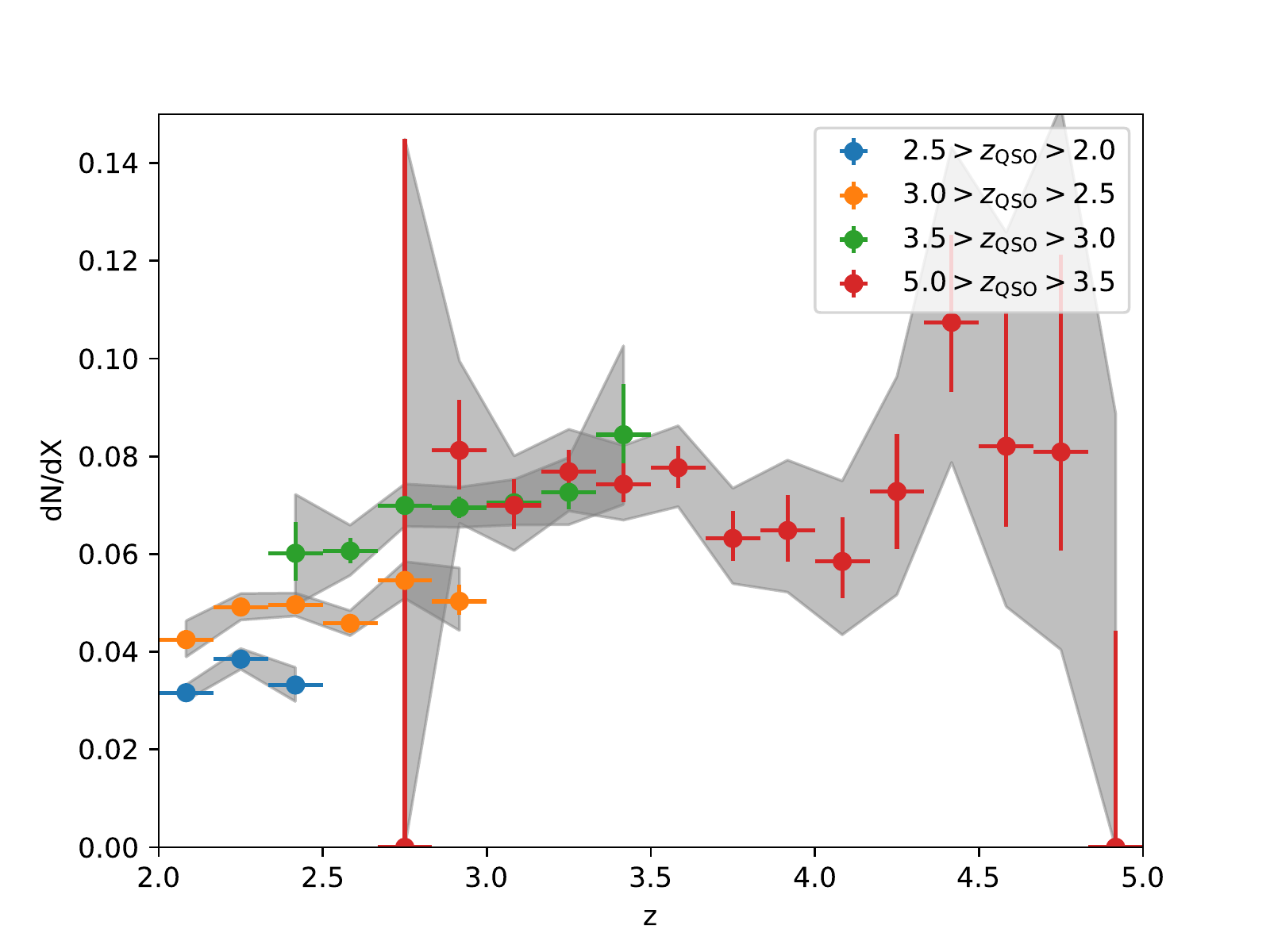}
   \caption{The redshift evolution of the incident rate of DLAs, cutting with different quasar redshift intervals.
   Any correlation between the absorber properties and the background quasars redshifts might indicate systematics.}
   \label{fig:zqsos_dndx_omegadla}
\end{figure}

In Figure~\ref{fig:zqsos_dndx_omegadla},
we test our measured $\dd N/\dd X$ with different quasar redshift bins.
In a perfect scenario without systematics, we expect that the absorber properties be uncorrelated with the background quasars, as they are widely separated in physical space.
However, Figure~\ref{fig:zqsos_dndx_omegadla},
shows some residual correlation between absorber properties and the redshifts of the background quasars for DLAs in spectra with $\zqso < 3$.

\begin{figure}
   \includegraphics[width=\columnwidth]{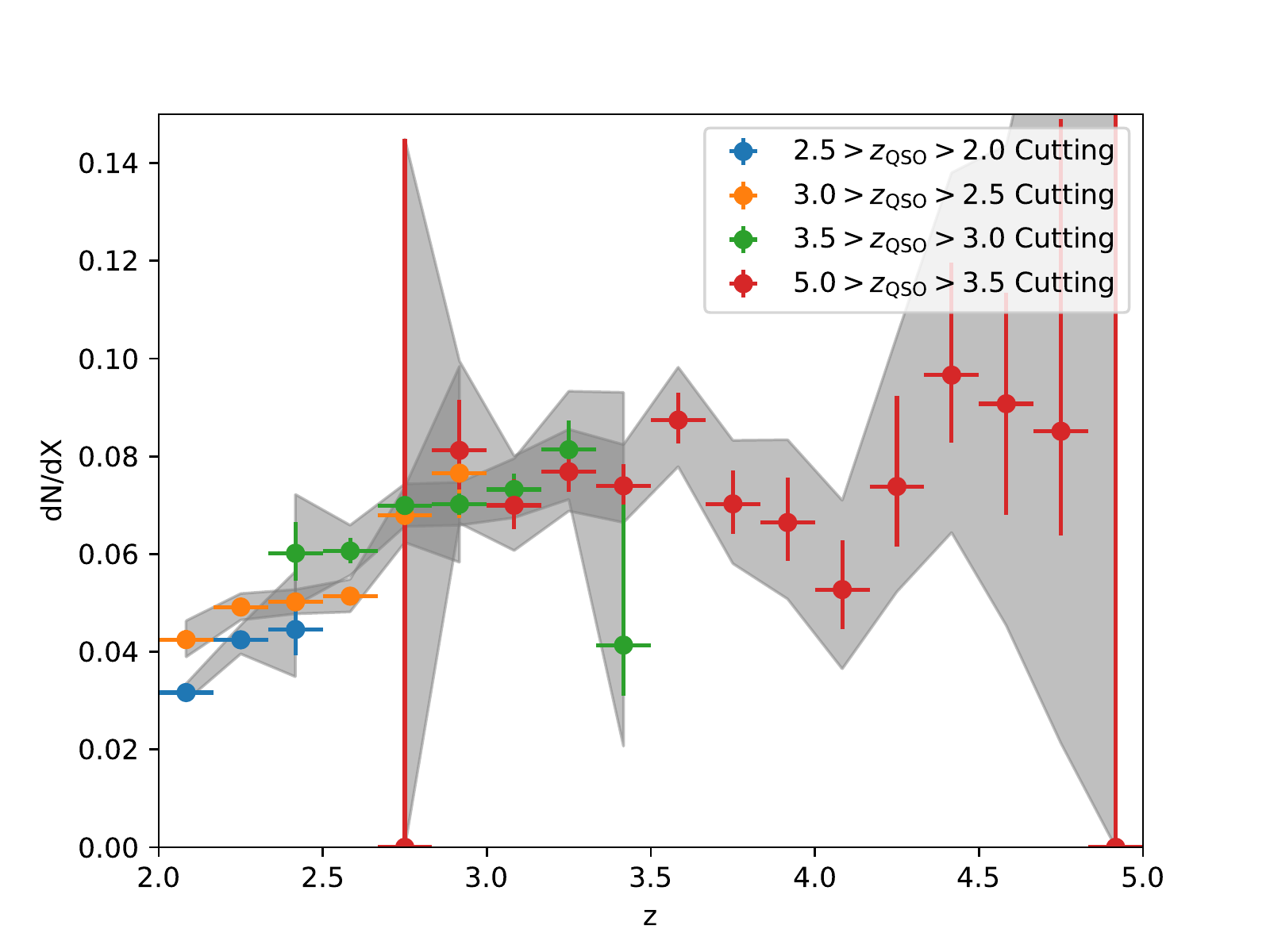}
   \caption{The redshift evolution of the incident rate of DLAs , cutting with different quasar redshift intervals.
   Unlike Figure~\ref{fig:zqsos_dndx_omegadla}, we remove the putative absorbers near the {\Lya} emission line with $|\zqso - \zdla | < 30\,000 \kms$.
   }
   \label{fig:zqso_omegadla_dndx_cutting}
\end{figure}

In Figure~\ref{fig:zqso_omegadla_dndx_cutting},
we have investigated removing the sampling range near the quasar redshift, $|\zqso - z| < 30\,000 \kms$.
We found removing the putative absorbers near the {\Lya} emission is sufficient to remove the correlation between quasar redshifts and DLA properties at $z < 3$.
A small tension still exists for the $z = 2$ bin within $2.5 > \zqso > 2.0$ for $\dd N/ \dd X$, which may be due to the effect discussed above for SNR, as these very short spectra are often also noisy.


\subsection{Additional noise test}

To understand the implication of applying Occam's razor to the model posteriors,
we conduct a test based on adding noise to a DLA spectrum.
We choose a quasar spectrum that we are very confident contains a DLA and add additional Gaussian noise with zero mean and standard deviation $\sigma$ to the flux and noise variance.

We then examine changes in the DLA model posterior $p(\{\mdla\} \mid \Data)$.
This test will mimic the effect of SNR on the model's ability to detect the underlying DLAs.
For Occam's razor $N = 30\,000$, the model posterior is $p(\{\mdla\} \mid \Data) \simeq 0.9$ for $\sigma \leq 1.5$, which corresponds to SNR $\simeq 0.9$.
On the other hand, for a model without Occam's razor, the model posterior is $p(\{\mdla\} \mid \Data) \simeq 0.9$ for $\sigma \leq 3$, which means SNR $\simeq 0.5$.
A strong Occam's razor thus introduces false negatives in very noisy spectra.
However, by visually inspecting the flux with $\sigma = 3$ we determined that it is almost impossible for humans to identify the underlying DLA.
Therefore, we choose to follow the value ($N = 1\,000$) we determined in Section~\ref{subsection:occams}.

We were unable to quantify the number of false positives, as our simple assumption of Gaussian noise rarely produces correlated structures that resemble DLAs. In practice, false positives are likely caused by oscillatory structure embedded in the noise, present when the SNR is extremely low.


\section{Results with DLAs in the Lyman $\beta$ region}
\label{sec:lymanlimit}

We have shown the CDDF, $\dd N/ \dd X$, and $\omegadla$ of our GP model in Section~\ref{sec:results}.
In this section, instead of using a sampling range from Ly$\beta$ to Ly$\alpha$,
we only compute the population statistics of DLAs detected \textit{within the Ly$\beta$ forest region}.
We set the sampling range to be Lyman limit $+ 30\,000 \kms$ to \Lyb.
We cut off a wider velocity width at the blue end to avoid counting DLAs detected right on the edge of the Lyman break.

\begin{figure}
   \includegraphics[width=\columnwidth]{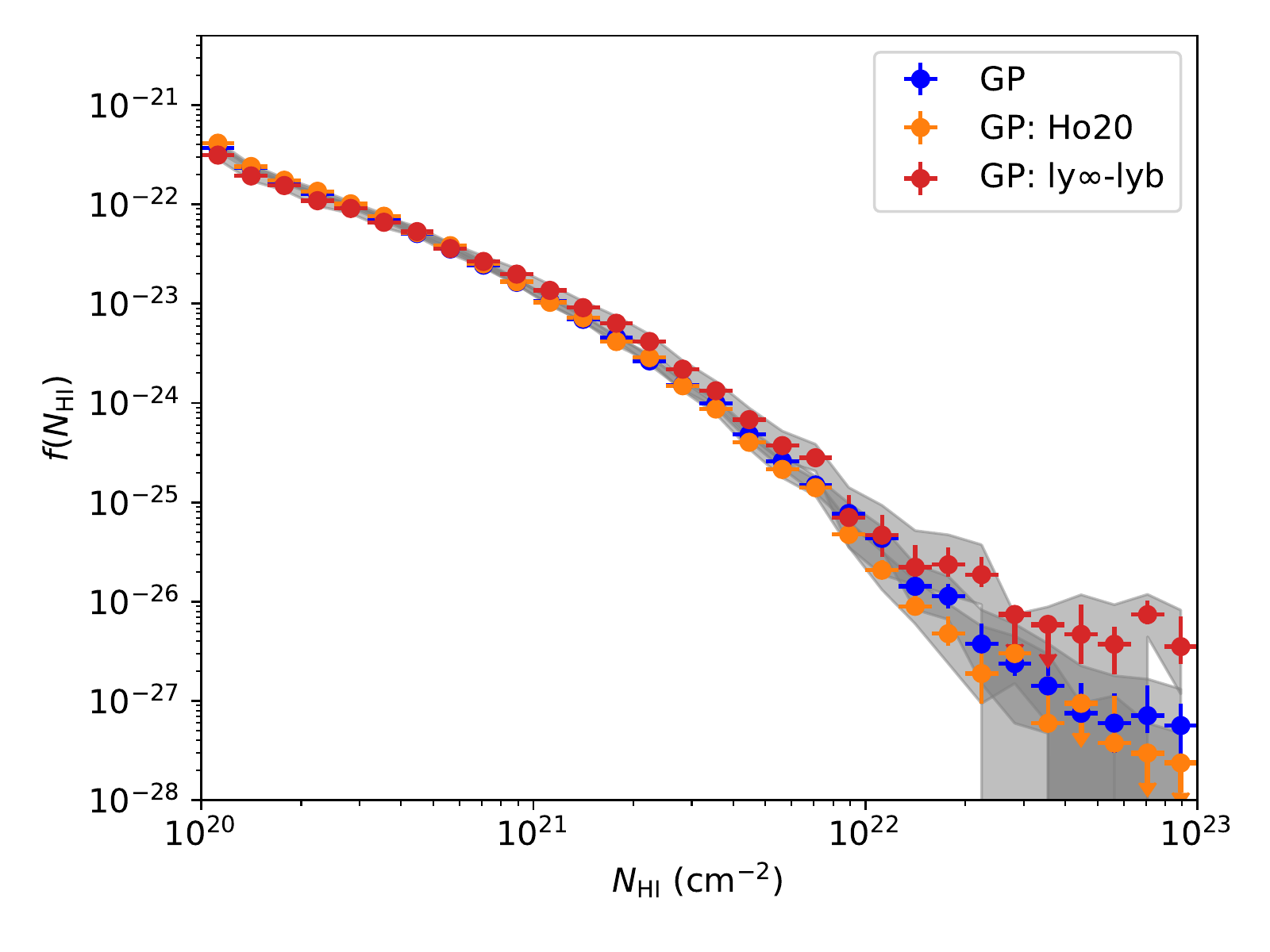}
   \caption{
   Comparing the CDDFs between the sampling range from Ly$\beta$-Ly$\alpha$ (Blue; GP) and \textbf{Ly$\infty$-Ly$\beta$} (Red; GP: ly$\infty$-lyb).
   The error bars are $68\%$ confidence limits, and the shaded areas are $95\%$ confidence limits.
   \protect\cite{Ho:2020} (GP: Ho20; Orange) also used a sampling range from Ly$\beta$-Ly$\alpha$.
   }
   \label{fig:cddf_limit}
\end{figure}

Figure~\ref{fig:cddf_limit} shows the CDDF for DLAs in the Lyman $\beta$ region.
As we can see from the figure,
it is mostly consistent with the CDDF from Ly$\beta$-Ly$\alpha$ for $\nhi < 10^{21}$,
and it starts to diverge for $\nhi > 20^{22}$.
We visually inspected those spectra and found that they are mostly due to fitting large DLAs on the spectra's noisy left edges.
This may indicate that additional regularisation is still needed to avoid spurious detections at the blue end of high redshift spectra. In particular, if the redshift measurement is slightly inaccurate, parts of the Lyman break move into our modelling window.

\begin{figure*}
   \includegraphics[width=\columnwidth]{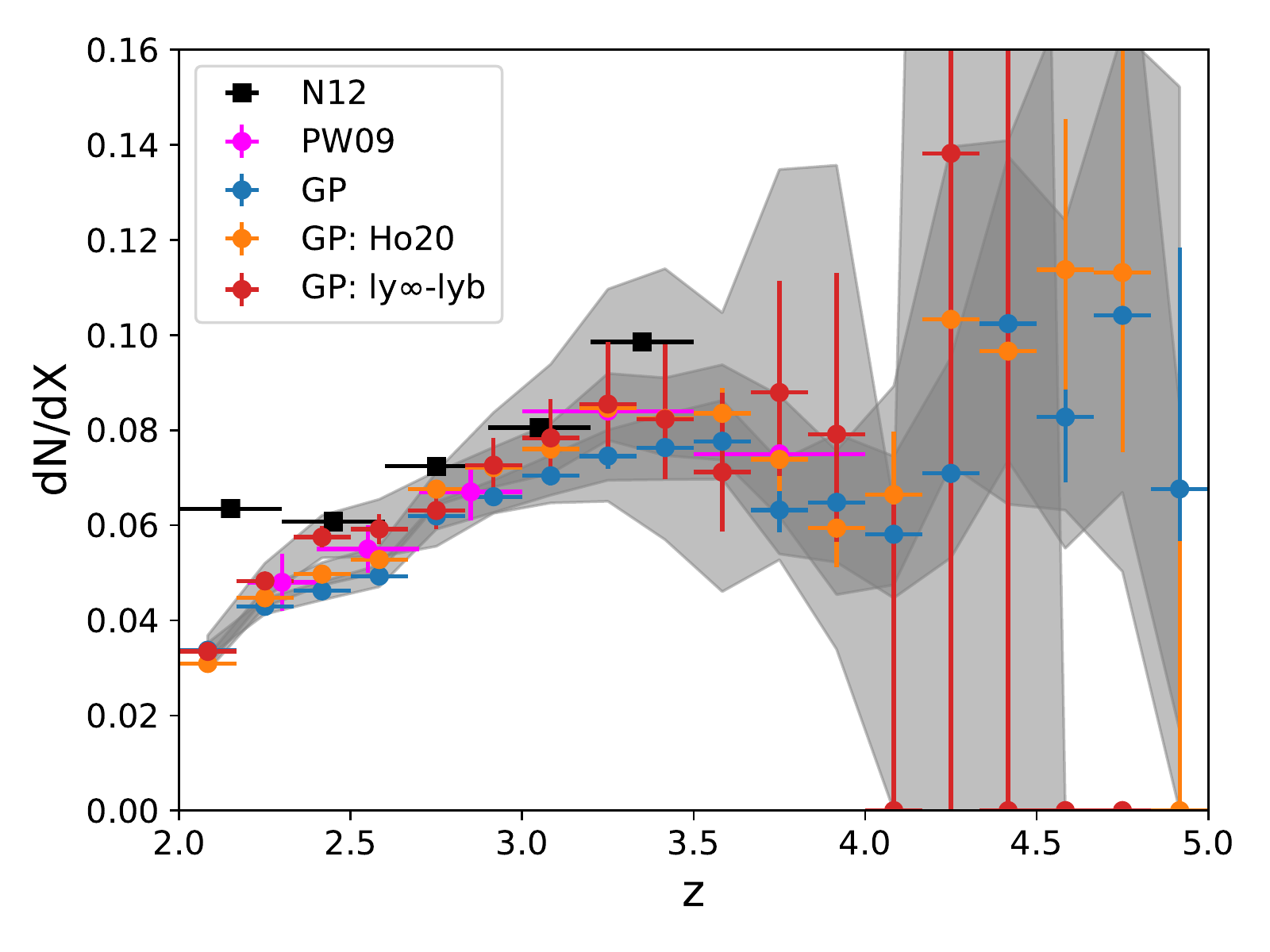}
   \includegraphics[width=\columnwidth]{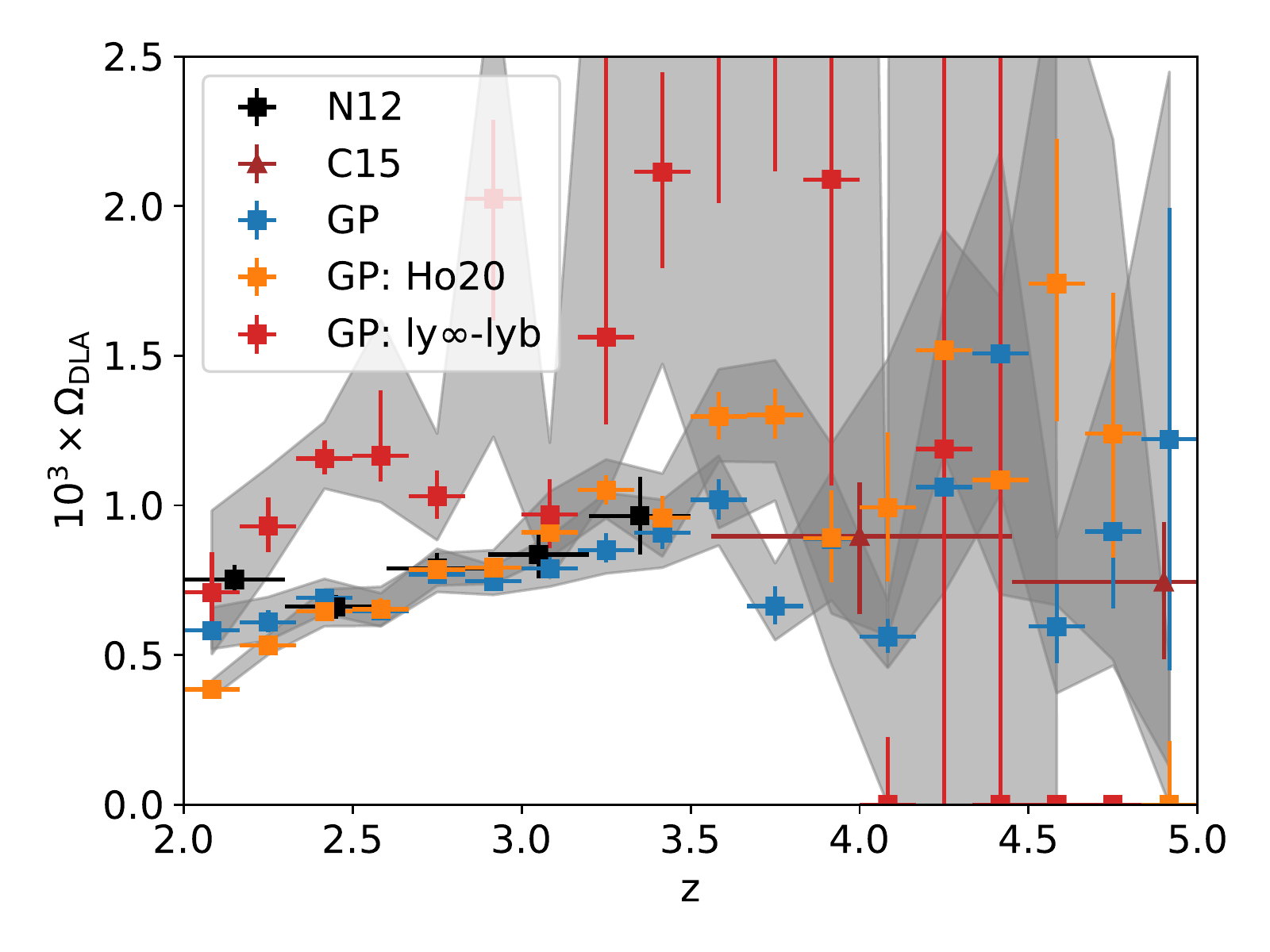}
   \caption{
   \textbf{(Left)}
   The comparison of $\dd N / \dd X$ with different sampling ranges, Ly$\beta$-Ly$\alpha$ (blue) and Ly$\infty$-Ly$\beta$ (red).
   Other plot settings are the same as Figure~\ref{fig:dndx}.
   \textbf{(Right)}
   The comparison of $\omegadla$ with difference sampling ranges, Ly$\beta$-Ly$\alpha$ (blue) and Ly$\infty$-Ly$\beta$ (red).
   Other plot settings are the same as Figure~\ref{fig:omegadla}.}
   \label{fig:dndx_omegadla_limit}
\end{figure*}

We also show the $\dd N/\dd X$ and $\omegadla$ for DLAs in the {\Lyb} forest region in Figure~\ref{fig:dndx_omegadla_limit}.
$\dd N/ \dd X$ in the {\Lyb} region is broadly consistent with other measurements,
with the detection consistent with zero at $\zdla > 4$.

For $\omegadla$, in the right panel of Figure~\ref{fig:dndx_omegadla_limit},
we observe our measurement is biased high and highly uncertain for $\zdla > 3.5$.
This may be because our current model can only poorly estimate the column density from the {\Lyb} region from high-redshift quasar spectra, perhaps due to the high level of absorption from the {\Lyb} and {\Lya} forests at these redshifts.
Alternatively, it could again reflect that the mean flux measure is not certain at these redshifts,
so the degeneracy between large DLAs and the effective {\Lya}/{\Lyb} absorption is not fully broken by sampling $(\taumf, \betamf)$.

\section{Comparison to the CNN model}
\label{sec:comparison}

SDSS DR16Q includes DLA measurements using the convolutional neural network (CNN) model of \cite{Parks18}.
The DLAs from the CNN model are recorded as \texttt{CONF\_DLA}, \texttt{Z\_DLA}, and \texttt{NHI\_DLA} columns in the DR16Q catalogue.\footnote{This column is the log column density of the given DLA.}

To compare our model and the CNN model,
we restrict the $\zdla$ sampling range of the CNN DLAs to be the same as our GP DLAs.
Table~\ref{table:confusion_matrix} shows the confusion matrix.
On the existence of DLAs, which means the binary classification of having at least one DLA or no DLA,
the GP model is $\sim 94.8 \%$ in agreement with the CNN model.
If we only consider only spectra with SNR $> 6$, the rate of agreement
climbs to $\sim 96.5 \%$.

\begin{table*}
   \caption{The confusion matrix for multi-DLAs detections between the GP and the CNN model \citep{Parks18}.
   Note we require both the model posteriors of our GP model and DLA confidence in Parks to be larger than $0.98$.
   We also require $\lognhi > 20.3$. The maximum number of DLAs is fixed to three, and everything larger than three is considered three.}
   \begin{tabular}{r | r r r r r}
      CNN & 0 DLA & 1 DLA & 2 DLAs & 3 DLAs             \\
      GP with Multi-DLAs & & & & & \\
      \hline
      0 DLA  & {\bf 142759} & 5686  & 93    & 2          \\
      1 DLA  & 2397   & {\bf 8007}  & 208    & 1          \\
      2 DLAs & 117    & 234   & {\bf 333}    & 5         \\
      3 DLAs & 8     & 6    & 11     & {\bf 4}
   \end{tabular}
   \label{table:confusion_matrix}
\end{table*}

\begin{figure*}
   \includegraphics[width=\columnwidth]{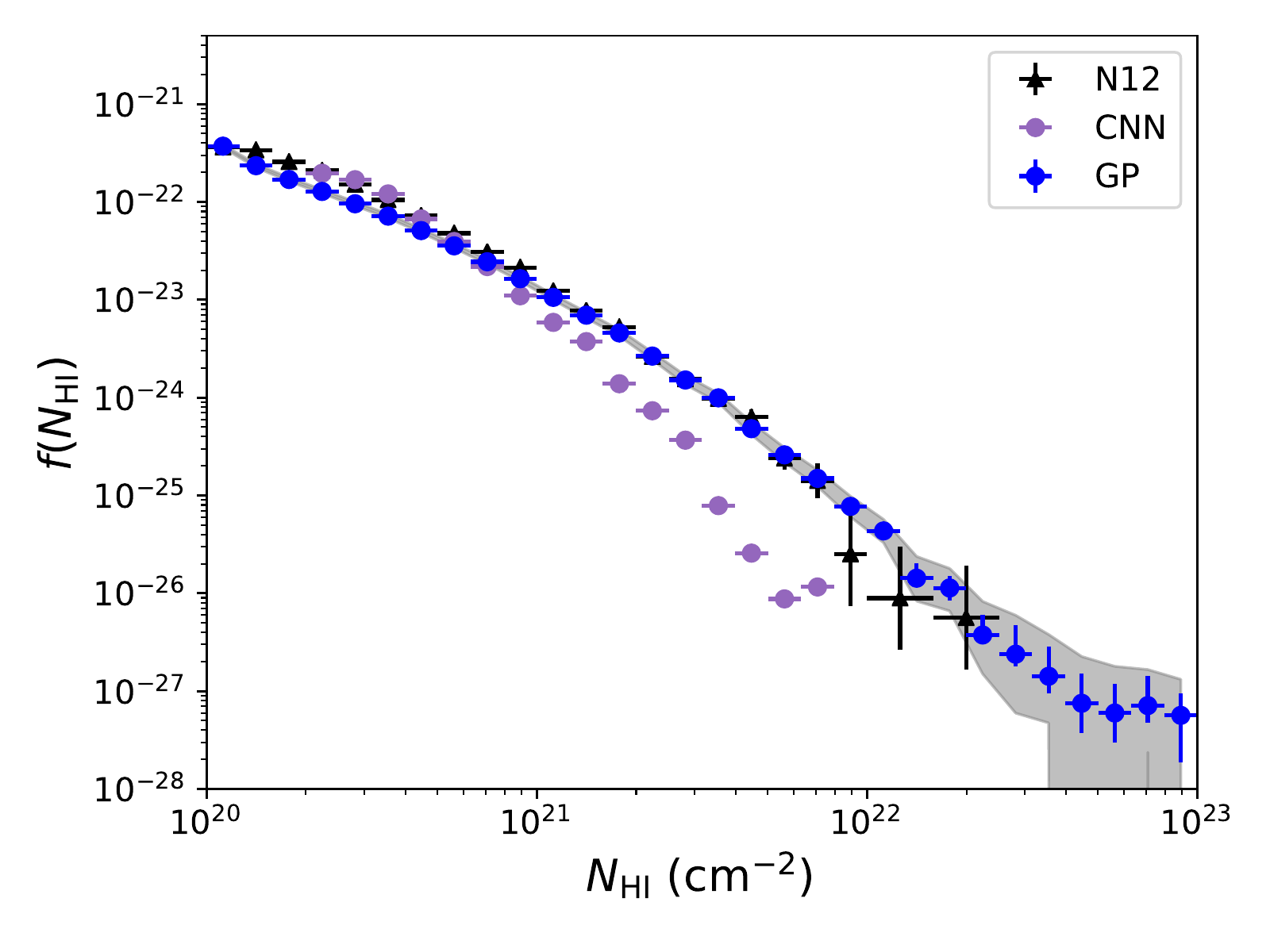}
   \includegraphics[width=\columnwidth]{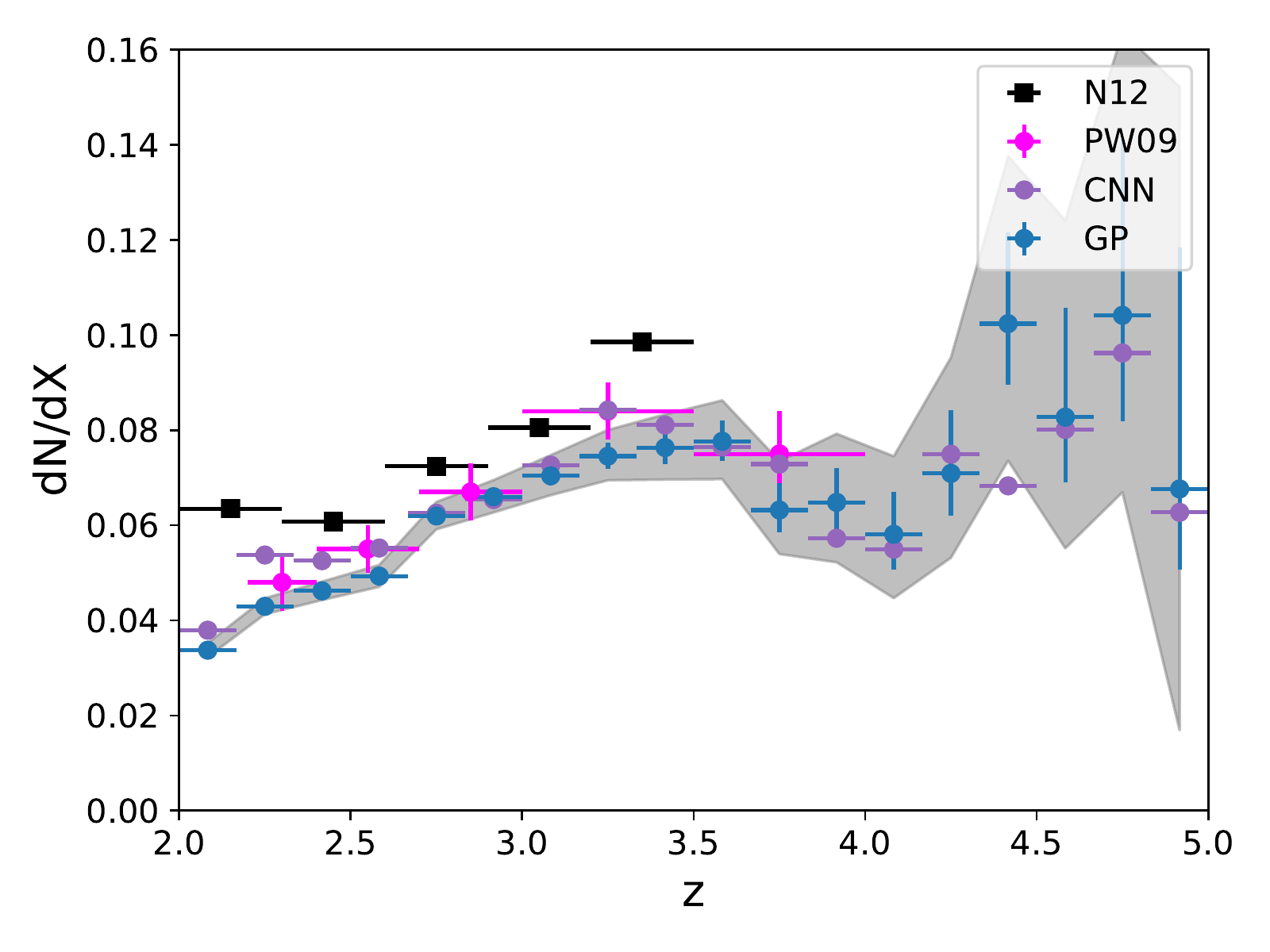}
   \caption{
      \textbf{(Left)} The CDDF of the DLAs detected by the CNN model presented in \protect\cite{Parks18}.
      The $\zdla$ and $\lognhi$ values are taken from the SDSS DR16Q catalogue in column \texttt{Z\_DLA} and \texttt{NHI\_DLA}.
      We require the confidence of DLAs to be larger than $0.98$ and set the search range of the CNN DLAs to be the same as our search range, which is {\Lyb} $+ 3\,000 \kms$ to $\zqso - 3\,000 \kms$.
      \textbf{(Right)} The line density of the DLAs detected by the CNN model.
      All three measurements, GP, PW09, and CNN are consistent on the line density.
   }
   \label{fig:cddf_dndx_parks}
\end{figure*}

We have also checked the CDDF of the CNN DLAs,
as shown in Figure~\ref{fig:cddf_dndx_parks}.
The sampling range is restricted to be the same as ours, and we only count the DLAs with \texttt{CONF\_DLA} larger than $0.98$.
The CDDF of the CNN model under-detects DLAs with $\nhi > 7 \times 10^{20}$, compared to N12.
We have discussed this issue in Figure 19 of \cite{Ho:2020}.
The CDDF of the CNN model in the DR16Q catalogue shows improvements in detecting more high column density systems comparing to \cite{Parks18},
but it is still an order of magnitude lower than N12 for $\nhi > 2 \times 10^{21}$.
Thus the lack of high column density systems in the CNN DLAs, as identified in \cite{Ho:2020}, is still present in the latest catalogue.

The $\dd N/ \dd X$ of the CNN model, in contrast, mostly agree with our GP measurements.
Bins with $z > 4.5$ are even consistent at the 1-$\sigma$ level.
Since $\dd N/ \dd X$ is sensitive to low column density systems, it shows these two codes find consistent small DLAs, but differ in their column density estimates.

\begin{figure*}
   \includegraphics[width=2\columnwidth]{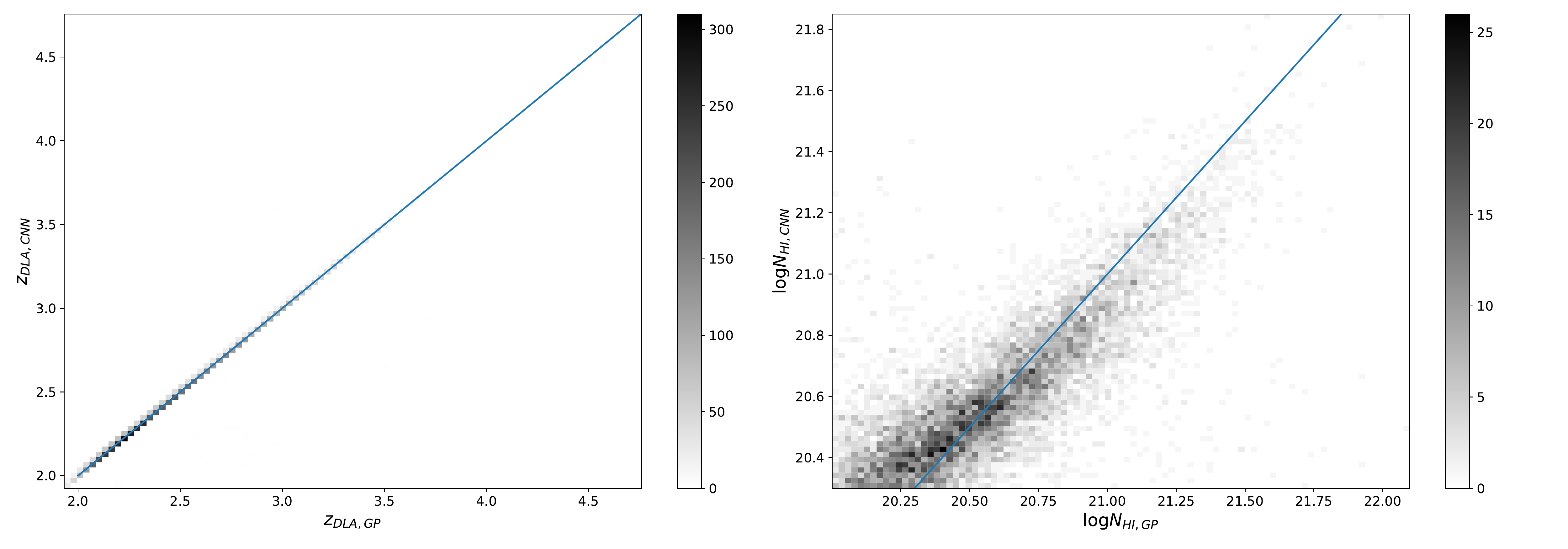}
   \caption{The 2D histograms for $\zdla$ \textbf{(left)} and $\lognhi$ \textbf{(right)} estimated by the GP code and the CNN.
   We use the maximum a posteriori (MAP) estimate for parameter estimation for the GP code.
   The colourbars indicate the number of DLAs within the bin.
   The blue line is a straight line that shows the diagonal line of the 2D histogram.
   }
   \label{fig:map_cnn_gp}
\end{figure*}

We compare DLAs detected by the CNN and GP codes on a spectrum-by-spectrum basis in Figure~\ref{fig:map_cnn_gp}.
As anticipated, the CNN and the GP code have a perfect agreement in $\zdla$,
but the CNN predicts slightly lower $\lognhi$ than the GP code, consistent with the CDDF plot in Figure~\ref{fig:cddf_dndx_parks}.

We visually inspected $319$ quasar spectra, where the CNN code strongly disagrees with the GP code's detections.
As expected, most cases are spectra with low SNRs, where even human experts will have difficulty identifying DLAs.
Besides those low-SNR cases,
in general, the CNN code has false negatives on DLAs overlapping with sub-DLAs or DLAs very close to each other.
There are $24$ out of $319$ cases which show a clear pattern where the CNN missed the DLAs when multiple absorption systems are overlapping or nearby.\footnote{We put the figures for these $24$ spectra in here \url{http://tiny.cc/overlapping\_dlas} for future investigators.}
Some of these are ambiguous detections, but $9$ out of $24$ have apparent damping wings on the absorber.

\begin{figure*}
   \includegraphics[width=2\columnwidth]{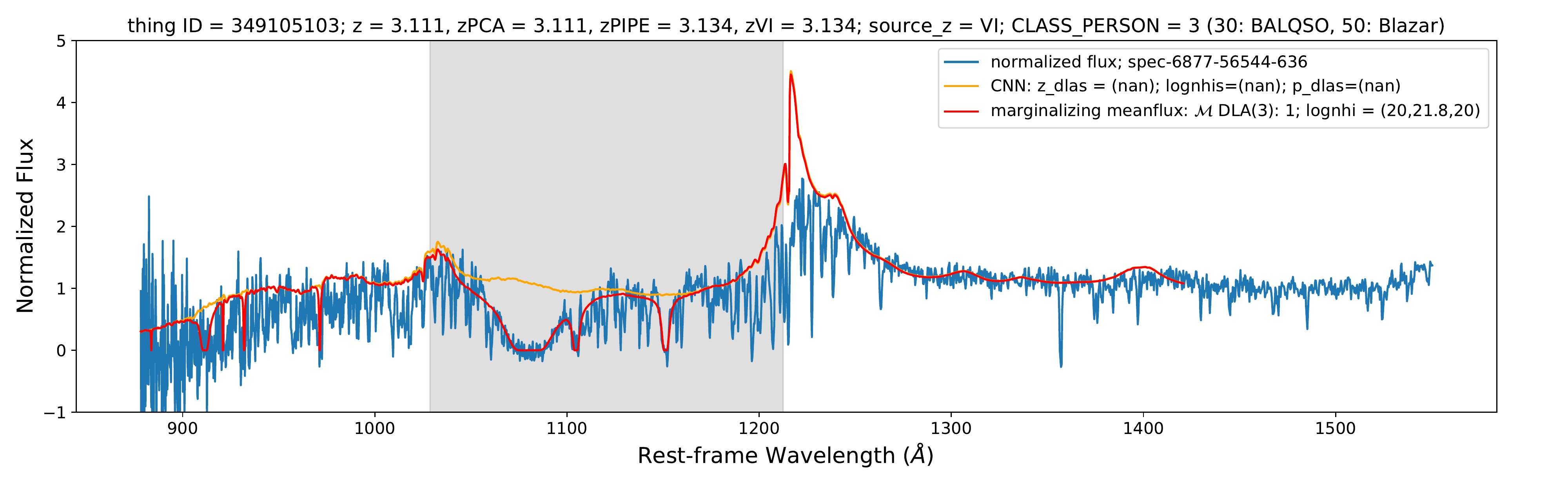}
   \includegraphics[width=2\columnwidth]{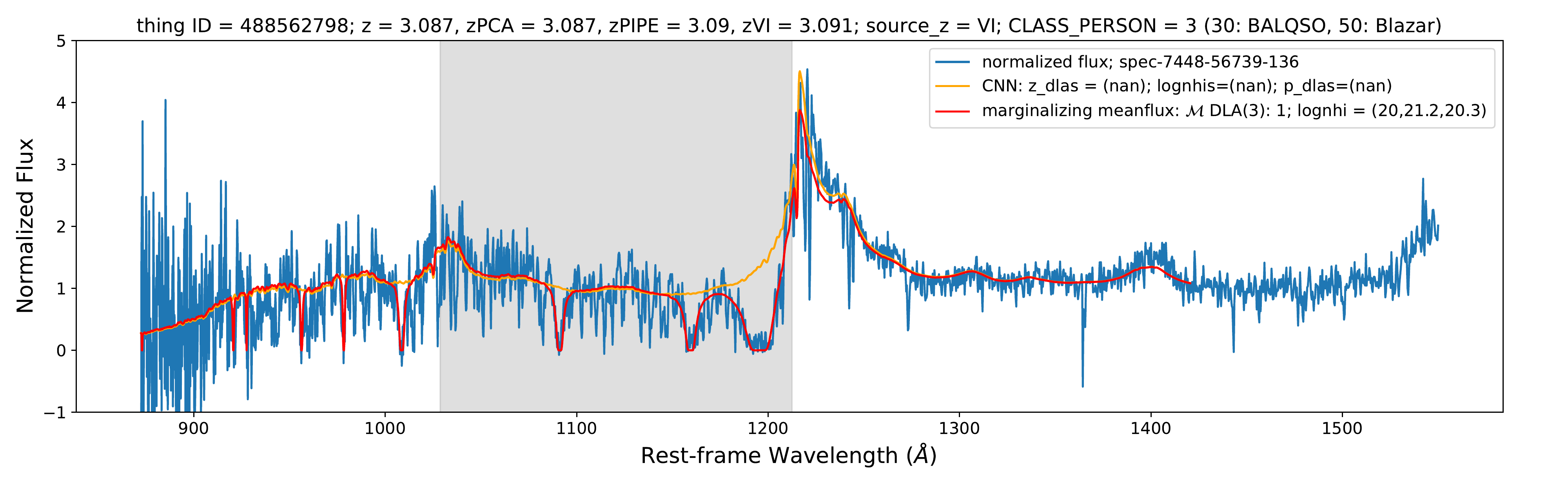}
   \caption{Examples showing \textbf{(top)} the case of a sub-DLA overlapping a DLA and \textbf{(bottom)} the case of a DLA near to another DLA.
   The red line indicates the GP code predictions, and we describe the $\lognhi$ in the legend.
   We intervene the DLAs from the CNN model in the DR16Q catalogue onto our null model in the orange line.
   Both spectra have high enough SNR: the upper one has SNR $= 3.45$ while the bottom one has SNR $= 7.52$.
   The damping wings and the {\Lyb} absorption lines of the DLAs are visible in the plots.}
   \label{fig:overlapping}
\end{figure*}

We show two examples in Figure~\ref{fig:overlapping}.
The first one shows a sub-DLA intervening on the right of the DLA damping wings.
Though the damping wings are disturbed by the sub-DLA,
the pattern of a DLA is still visible.
The second example shows two DLAs close to each other, but not close enough to overlap.
We suspect these non-detections for the CNN code are due to the lack of training data for multiple absorption systems (sub-DLAs or DLAs) close to each other.
Since these overlapping cases are rare in the real dataset,
we think one might need to implement simulated DLAs/sub-DLAs to augment the CNN training set.

\section{Conclusion}
\label{sec:conclusion}

We have presented a new estimate of the abundance of DLAs from $z = 2$ to $z = 5$ and a DLA catalogue built from SDSS DR16Q spectra \citep{SDSSDR16Q:2020} using our Gaussian process model \cite{Garnett17,Ho:2020}.
We verify our results are in good agreement with previous measurements from \cite{Noterdaeme12}, \cite{Prochaska2009}, and \cite{Crighton2015}.
We newly integrate out the uncertainty in the measured mean flux, which improves our modelling of DLA detection uncertainties for $\zqso > 4$ without biasing towards high $\nhi$ detections.

We note, nevertheless, that there is a residual dependence on low-redshift spectra with $\zqso < 2.5$.
This could be due to unmodelled systematics or simply because the low-redshift optical spectra are incomplete in the Lyman series range, so we can not securely detect DLAs in low $\zqso$.
Incorporating spectra with shorter observed wavelengths could potentially verify these detections at $\zqso < 2.5$.

Our measurement shows the abundance of DLAs and neural hydrogen increases moderately over $2 < z < 4$,
while the trend beyond $z = 4$ is unclear due to statistical uncertainties.
Larger datasets and better mean flux measurements are needed to give more robust constraints for DLA detections at $z > 4$.

\section*{Data Availability}

Our DLA catalogue is publicly available at \url{http://tiny.cc/gp_dla_dr16q},
including both MATLAB catalogue and JSON catalogue.
A sub-DLA candidate catalogue is available in JSON format.
README files are included to describe the data formats of both catalogues.
The data files for DLA population statistics are also included, including CDDF, $\dd N/ \dd X$, and $\omegadla$ with or without SNR cuts.
A tutorial for manipulating the MATLAB catalogue is publicly available at \url{https://github.com/jibanCat/gp_dla_detection_dr16q_public/tree/master/notebooks} as a notebook file.
Our GP code is also publicly available at \url{https://github.com/jibanCat/gp_dla_detection_dr16q_public/}.

\section*{Acknowledgements}

We thank the anonymous referee for the constructive comments and suggestions.
We thank Reza Monadi and Bryan Scott for useful discussions and comments. SB was supported by NSF AST-1817256.
RG was supported by the NSF under award numbers IIS–1939677, OAC–1940224, and IIS–1845434.
SB and RG were supported by an Amazon.com Machine Learning
Research Award, which also provided computing time.
We hope the world can recover from the COVID pandemic soon.

\bibliography{sample}

\begin{thebibliography}{}
\makeatletter
\relax
\def\mn@urlcharsother{\let\do\@makeother \do\$\do\&\do\#\do\^\do\_\do\%\do\~}
\def\mn@doi{\begingroup\mn@urlcharsother \@ifnextchar [ {\mn@doi@}
  {\mn@doi@[]}}
\def\mn@doi@[#1]#2{\def\@tempa{#1}\ifx\@tempa\@empty \href
  {http://dx.doi.org/#2} {doi:#2}\else \href {http://dx.doi.org/#2} {#1}\fi
  \endgroup}
\def\mn@eprint#1#2{\mn@eprint@#1:#2::\@nil}
\def\mn@eprint@arXiv#1{\href {http://arxiv.org/abs/#1} {{\tt arXiv:#1}}}
\def\mn@eprint@dblp#1{\href {http://dblp.uni-trier.de/rec/bibtex/#1.xml}
  {dblp:#1}}
\def\mn@eprint@#1:#2:#3:#4\@nil{\def\@tempa {#1}\def\@tempb {#2}\def\@tempc
  {#3}\ifx \@tempc \@empty \let \@tempc \@tempb \let \@tempb \@tempa \fi \ifx
  \@tempb \@empty \def\@tempb {arXiv}\fi \@ifundefined
  {mn@eprint@\@tempb}{\@tempb:\@tempc}{\expandafter \expandafter \csname
  mn@eprint@\@tempb\endcsname \expandafter{\@tempc}}}

\bibitem[\protect\citeauthoryear{{Alam} et~al.,}{{Alam}
  et~al.}{2015}]{Alam:2015}
{Alam} S.,  et~al., 2015, \mn@doi [\apjs] {10.1088/0067-0049/219/1/12}, \href
  {https://ui.adsabs.harvard.edu/abs/2015ApJS..219...12A} {219, 12} (\mn@eprint
  {arXiv} {1501.00963})

\bibitem[\protect\citeauthoryear{{Alonso}, {Colosimo}, {Font-Ribera}  \&
  {Slosar}}{{Alonso} et~al.}{2018}]{Alonso:2018}
{Alonso} D.,  {Colosimo} J.,  {Font-Ribera} A.,   {Slosar} A.,  2018, \mn@doi
  [\jcap] {10.1088/1475-7516/2018/04/053}, \href
  {https://ui.adsabs.harvard.edu/abs/2018JCAP...04..053A} {2018, 053}
  (\mn@eprint {arXiv} {1712.02738})

\bibitem[\protect\citeauthoryear{{Berg} et~al.,}{{Berg}
  et~al.}{2019}]{Berg:2019}
{Berg} T. A.~M.,  et~al., 2019, \mn@doi [\mnras] {10.1093/mnras/stz2012}, \href
  {https://ui.adsabs.harvard.edu/abs/2019MNRAS.488.4356B} {488, 4356}
  (\mn@eprint {arXiv} {1907.07703})

\bibitem[\protect\citeauthoryear{Bird, Haehnelt, Neeleman, Genel, Vogelsberger
  \& Hernquist}{Bird et~al.}{2015}]{Bird2015}
Bird S.,  Haehnelt M.,  Neeleman M.,  Genel S.,  Vogelsberger M.,   Hernquist
  L.,  2015, \mn@doi [\mnras] {10.1093/mnras/stu2542}, 447, 1834

\bibitem[\protect\citeauthoryear{Bird, Garnett  \& Ho}{Bird
  et~al.}{2017}]{Bird17}
Bird S.,  Garnett R.,   Ho S.,  2017, \mn@doi [\mnras] {10.1093/mnras/stw3246},
  466, 2111

\bibitem[\protect\citeauthoryear{{Busca} \& {Balland}}{{Busca} \&
  {Balland}}{2018}]{QuasarNET:2018}
{Busca} N.,  {Balland} C.,  2018, arXiv e-prints, \href
  {https://ui.adsabs.harvard.edu/abs/2018arXiv180809955B} {p. arXiv:1808.09955}
  (\mn@eprint {arXiv} {1808.09955})

\bibitem[\protect\citeauthoryear{Carithers}{Carithers}{2012}]{Concordance2012}
Carithers W.,  2012, Published internally to SDSS

\bibitem[\protect\citeauthoryear{Cen}{Cen}{2012}]{Cen2012}
Cen R.,  2012, \mn@doi [\apj] {10.1088/0004-637x/748/2/121}, 748, 121

\bibitem[\protect\citeauthoryear{{Chabanier} et~al.,}{{Chabanier}
  et~al.}{2019}]{Chabanier:2019}
{Chabanier} S.,  et~al., 2019, \mn@doi [\jcap] {10.1088/1475-7516/2019/07/017},
  \href {https://ui.adsabs.harvard.edu/abs/2019JCAP...07..017C} {2019, 017}
  (\mn@eprint {arXiv} {1812.03554})

\bibitem[\protect\citeauthoryear{Crighton et~al.,}{Crighton
  et~al.}{2015}]{Crighton2015}
Crighton N. H.~M.,  et~al., 2015, \mn@doi [\mnras] {10.1093/mnras/stv1182},
  452, 217

\bibitem[\protect\citeauthoryear{{Croft}, {Weinberg}, {Katz}  \&
  {Hernquist}}{{Croft} et~al.}{1998}]{Croft:1998}
{Croft} R. A.~C.,  {Weinberg} D.~H.,  {Katz} N.,   {Hernquist} L.,  1998,
  \mn@doi [\apj] {10.1086/305289}, \href
  {https://ui.adsabs.harvard.edu/abs/1998ApJ...495...44C} {495, 44} (\mn@eprint
  {arXiv} {astro-ph/9708018})

\bibitem[\protect\citeauthoryear{{Cuceu}, {Font-Ribera}  \& {Joachimi}}{{Cuceu}
  et~al.}{2020}]{Cuceu:2020}
{Cuceu} A.,  {Font-Ribera} A.,   {Joachimi} B.,  2020, \mn@doi [\jcap]
  {10.1088/1475-7516/2020/07/035}, \href
  {https://ui.adsabs.harvard.edu/abs/2020JCAP...07..035C} {2020, 035}
  (\mn@eprint {arXiv} {2004.02761})

\bibitem[\protect\citeauthoryear{{Dawson} et~al.,}{{Dawson}
  et~al.}{2013}]{Dawson:2013}
{Dawson} K.~S.,  et~al., 2013, \mn@doi [\aj] {10.1088/0004-6256/145/1/10},
  \href {https://ui.adsabs.harvard.edu/abs/2013AJ....145...10D} {145, 10}
  (\mn@eprint {arXiv} {1208.0022})

\bibitem[\protect\citeauthoryear{{Dawson} et~al.,}{{Dawson}
  et~al.}{2016}]{Dawson:2016}
{Dawson} K.~S.,  et~al., 2016, \mn@doi [\aj] {10.3847/0004-6256/151/2/44},
  \href {https://ui.adsabs.harvard.edu/abs/2016AJ....151...44D} {151, 44}
  (\mn@eprint {arXiv} {1508.04473})

\bibitem[\protect\citeauthoryear{{Eisenstein} et~al.,}{{Eisenstein}
  et~al.}{2011}]{Eisenstein:2011}
{Eisenstein} D.~J.,  et~al., 2011, \mn@doi [\aj] {10.1088/0004-6256/142/3/72},
  \href {https://ui.adsabs.harvard.edu/abs/2011AJ....142...72E} {142, 72}
  (\mn@eprint {arXiv} {1101.1529})

\bibitem[\protect\citeauthoryear{{Fauber}, {Ho}, {Bird}, {Shelton}, {Garnett}
  \& {Korde}}{{Fauber} et~al.}{2020}]{Fauber:2020}
{Fauber} L.,  {Ho} M.-F.,  {Bird} S.,  {Shelton} C.~R.,  {Garnett} R.,
  {Korde} I.,  2020, \mn@doi [\mnras] {10.1093/mnras/staa2826}, \href
  {https://ui.adsabs.harvard.edu/abs/2020MNRAS.498.5227F} {498, 5227}
  (\mn@eprint {arXiv} {2006.07343})

\bibitem[\protect\citeauthoryear{{Font-Ribera} et~al.,}{{Font-Ribera}
  et~al.}{2012}]{FontRibera:2012}
{Font-Ribera} A.,  et~al., 2012, \mn@doi [\jcap]
  {10.1088/1475-7516/2012/11/059}, \href
  {https://ui.adsabs.harvard.edu/abs/2012JCAP...11..059F} {2012, 059}
  (\mn@eprint {arXiv} {1209.4596})

\bibitem[\protect\citeauthoryear{{Fumagalli}, {O'Meara}, {Prochaska}  \&
  {Worseck}}{{Fumagalli} et~al.}{2013}]{Fumagalli:2013}
{Fumagalli} M.,  {O'Meara} J.~M.,  {Prochaska} J.~X.,   {Worseck} G.,  2013,
  \mn@doi [\apj] {10.1088/0004-637X/775/1/78}, \href
  {http://adsabs.harvard.edu/abs/2013ApJ...775...78F} {775, 78} (\mn@eprint
  {arXiv} {1308.1101})

\bibitem[\protect\citeauthoryear{Gardner, Katz, Weinberg  \& Hernquist}{Gardner
  et~al.}{1997}]{Gardner1997}
Gardner J.~P.,  Katz N.,  Weinberg D.~H.,   Hernquist L.,  1997, \mn@doi [\apj]
  {10.1086/304526}, 486, 42

\bibitem[\protect\citeauthoryear{Garnett, Ho, Bird  \& Schneider}{Garnett
  et~al.}{2017}]{Garnett17}
Garnett R.,  Ho S.,  Bird S.,   Schneider J.,  2017, \mn@doi [\mnras]
  {10.1093/mnras/stx1958}, 472, 1850

\bibitem[\protect\citeauthoryear{{Guo} \& {Martini}}{{Guo} \&
  {Martini}}{2019}]{Guo:2019}
{Guo} Z.,  {Martini} P.,  2019, \mn@doi [\apj] {10.3847/1538-4357/ab2590},
  \href {https://ui.adsabs.harvard.edu/abs/2019ApJ...879...72G} {879, 72}
  (\mn@eprint {arXiv} {1901.04506})

\bibitem[\protect\citeauthoryear{Haehnelt, Steinmetz  \& Rauch}{Haehnelt
  et~al.}{1998}]{Haehnelt1998}
Haehnelt M.~G.,  Steinmetz M.,   Rauch M.,  1998, \mn@doi [\apj]
  {10.1086/305323}, 495, 647

\bibitem[\protect\citeauthoryear{{Hassan}, {Finlator}, {Dav{\'e}}, {Churchill}
  \& {Prochaska}}{{Hassan} et~al.}{2020}]{Hassan:2020}
{Hassan} S.,  {Finlator} K.,  {Dav{\'e}} R.,  {Churchill} C.~W.,   {Prochaska}
  J.~X.,  2020, \mn@doi [\mnras] {10.1093/mnras/staa056}, \href
  {https://ui.adsabs.harvard.edu/abs/2020MNRAS.492.2835H} {492, 2835}
  (\mn@eprint {arXiv} {1910.07541})

\bibitem[\protect\citeauthoryear{{Ho}, {Bird}  \& {Garnett}}{{Ho}
  et~al.}{2020}]{Ho:2020}
{Ho} M.-F.,  {Bird} S.,   {Garnett} R.,  2020, \mn@doi [\mnras]
  {10.1093/mnras/staa1806}, \href
  {https://ui.adsabs.harvard.edu/abs/2020MNRAS.496.5436H} {496, 5436}
  (\mn@eprint {arXiv} {2003.11036})

\bibitem[\protect\citeauthoryear{{Ir{\v{s}}i{\v{c}}}
  et~al.,}{{Ir{\v{s}}i{\v{c}}} et~al.}{2017}]{Irvic:2017}
{Ir{\v{s}}i{\v{c}}} V.,  et~al., 2017, \mn@doi [\mnras]
  {10.1093/mnras/stw3372}, \href
  {https://ui.adsabs.harvard.edu/abs/2017MNRAS.466.4332I} {466, 4332}
  (\mn@eprint {arXiv} {1702.01761})

\bibitem[\protect\citeauthoryear{{Kamble}, {Dawson}, {du Mas des Bourboux},
  {Bautista}  \& {Scheinder}}{{Kamble} et~al.}{2020}]{Kamble:2020}
{Kamble} V.,  {Dawson} K.,  {du Mas des Bourboux} H.,  {Bautista} J.,
  {Scheinder} D.~P.,  2020, \mn@doi [\apj] {10.3847/1538-4357/ab76bd}, \href
  {https://ui.adsabs.harvard.edu/abs/2020ApJ...892...70K} {892, 70} (\mn@eprint
  {arXiv} {1904.01110})

\bibitem[\protect\citeauthoryear{Kim, Bolton, Viel, Haehnelt  \& Carswell}{Kim
  et~al.}{2007}]{Kim07}
Kim T.-S.,  Bolton J.~S.,  Viel M.,  Haehnelt M.~G.,   Carswell R.~F.,  2007,
  \mn@doi [\mnras] {10.1111/j.1365-2966.2007.12406.x}, 382, 1657

\bibitem[\protect\citeauthoryear{{Krogager}, {Fynbo}, {M{\o}ller},
  {Noterdaeme}, {Heintz}  \& {Pettini}}{{Krogager}
  et~al.}{2019}]{Krogager:2019}
{Krogager} J.-K.,  {Fynbo} J. P.~U.,  {M{\o}ller} P.,  {Noterdaeme} P.,
  {Heintz} K.~E.,   {Pettini} M.,  2019, \mn@doi [\mnras]
  {10.1093/mnras/stz1120}, \href
  {https://ui.adsabs.harvard.edu/abs/2019MNRAS.486.4377K} {486, 4377}
  (\mn@eprint {arXiv} {1904.06966})

\bibitem[\protect\citeauthoryear{Lee et~al.,}{Lee et~al.}{2013}]{Lee2013}
Lee K.-G.,  et~al., 2013, \mn@doi [\aj] {10.1088/0004-6256/145/3/69}, 145, 69

\bibitem[\protect\citeauthoryear{{Lin}, {Cai}, {Li}, {Krolewski}  \&
  {Ferraro}}{{Lin} et~al.}{2020}]{Lin:2020}
{Lin} X.,  {Cai} Z.,  {Li} Y.,  {Krolewski} A.,   {Ferraro} S.,  2020, arXiv
  e-prints, \href {https://ui.adsabs.harvard.edu/abs/2020arXiv201101234L} {p.
  arXiv:2011.01234} (\mn@eprint {arXiv} {2011.01234})

\bibitem[\protect\citeauthoryear{{Lyke} et~al.,}{{Lyke}
  et~al.}{2020}]{SDSSDR16Q:2020}
{Lyke} B.~W.,  et~al., 2020, \mn@doi [\apjs] {10.3847/1538-4365/aba623}, \href
  {https://ui.adsabs.harvard.edu/abs/2020ApJS..250....8L} {250, 8} (\mn@eprint
  {arXiv} {2007.09001})

\bibitem[\protect\citeauthoryear{{McDonald}, {Miralda-Escud{\'e}}, {Rauch},
  {Sargent}, {Barlow}, {Cen}  \& {Ostriker}}{{McDonald}
  et~al.}{2000}]{McDonald:2000}
{McDonald} P.,  {Miralda-Escud{\'e}} J.,  {Rauch} M.,  {Sargent} W. L.~W.,
  {Barlow} T.~A.,  {Cen} R.,   {Ostriker} J.~P.,  2000, \mn@doi [\apj]
  {10.1086/317079}, \href
  {https://ui.adsabs.harvard.edu/abs/2000ApJ...543....1M} {543, 1} (\mn@eprint
  {arXiv} {astro-ph/9911196})

\bibitem[\protect\citeauthoryear{{McDonald}, {Seljak}, {Cen}, {Bode}  \&
  {Ostriker}}{{McDonald} et~al.}{2005a}]{McDonald:2005a}
{McDonald} P.,  {Seljak} U.,  {Cen} R.,  {Bode} P.,   {Ostriker} J.~P.,  2005a,
  \mn@doi [\mnras] {10.1111/j.1365-2966.2005.09141.x}, \href
  {https://ui.adsabs.harvard.edu/abs/2005MNRAS.360.1471M} {360, 1471}
  (\mn@eprint {arXiv} {astro-ph/0407378})

\bibitem[\protect\citeauthoryear{{McDonald} et~al.,}{{McDonald}
  et~al.}{2005b}]{McDonald:2005b}
{McDonald} P.,  et~al., 2005b, \mn@doi [\apj] {10.1086/497563}, \href
  {https://ui.adsabs.harvard.edu/abs/2005ApJ...635..761M} {635, 761}
  (\mn@eprint {arXiv} {astro-ph/0407377})

\bibitem[\protect\citeauthoryear{{Noterdaeme}, {Petitjean}, {Ledoux}  \&
  {Srianand}}{{Noterdaeme} et~al.}{2009}]{Noterdaeme:2009}
{Noterdaeme} P.,  {Petitjean} P.,  {Ledoux} C.,   {Srianand} R.,  2009, \mn@doi
  [\aap] {10.1051/0004-6361/200912768}, \href
  {https://ui.adsabs.harvard.edu/abs/2009A&A...505.1087N} {505, 1087}
  (\mn@eprint {arXiv} {0908.1574})

\bibitem[\protect\citeauthoryear{{Noterdaeme} et~al.,}{{Noterdaeme}
  et~al.}{2012}]{Noterdaeme12}
{Noterdaeme} et~al., 2012, \mn@doi [A\&A] {10.1051/0004-6361/201220259}, 547,
  L1

\bibitem[\protect\citeauthoryear{{P\^aris, Isabelle} et~al.,}{{P\^aris,
  Isabelle} et~al.}{2018}]{Paris2018}
{P\^aris, Isabelle} et~al., 2018, \mn@doi [A\&A] {10.1051/0004-6361/201732445},
  613, A51

\bibitem[\protect\citeauthoryear{Parks, Prochaska, Dong  \& Cai}{Parks
  et~al.}{2018}]{Parks18}
Parks D.,  Prochaska J.~X.,  Dong S.,   Cai Z.,  2018, \mn@doi [\mnras]
  {10.1093/mnras/sty196}, 476, 1151

\bibitem[\protect\citeauthoryear{{P{\'e}rez-R{\`a}fols}
  et~al.,}{{P{\'e}rez-R{\`a}fols} et~al.}{2018}]{PerezRafols:2018}
{P{\'e}rez-R{\`a}fols} I.,  et~al., 2018, \mn@doi [\mnras]
  {10.1093/mnras/stx2525}, \href
  {https://ui.adsabs.harvard.edu/abs/2018MNRAS.473.3019P} {473, 3019}
  (\mn@eprint {arXiv} {1709.00889})

\bibitem[\protect\citeauthoryear{Pontzen et~al.,}{Pontzen
  et~al.}{2008}]{Pontzen2008}
Pontzen A.,  et~al., 2008, \mn@doi [\mnras] {10.1111/j.1365-2966.2008.13782.x},
  390, 1349

\bibitem[\protect\citeauthoryear{{Prochaska} \& {Wolfe}}{{Prochaska} \&
  {Wolfe}}{1997}]{Prochaska1997}
{Prochaska} J.~X.,  {Wolfe} A.~M.,  1997, \mn@doi [\apj] {10.1086/304591}, 487,
  73

\bibitem[\protect\citeauthoryear{{Prochaska} \& {Wolfe}}{{Prochaska} \&
  {Wolfe}}{2009}]{Prochaska2009}
{Prochaska} J.~X.,  {Wolfe} A.~M.,  2009, \mn@doi [\apj]
  {10.1088/0004-637X/696/2/1543}, 696, 1543

\bibitem[\protect\citeauthoryear{Prochaska, Herbert-Fort  \& Wolfe}{Prochaska
  et~al.}{2005}]{Prochaska05}
Prochaska J.~X.,  Herbert-Fort S.,   Wolfe A.~M.,  2005, \apj, 635, 123

\bibitem[\protect\citeauthoryear{{Rahmati} \& {Schaye}}{{Rahmati} \&
  {Schaye}}{2014}]{Rahmati:2014}
{Rahmati} A.,  {Schaye} J.,  2014, \mn@doi [\mnras] {10.1093/mnras/stt2235},
  \href {https://ui.adsabs.harvard.edu/abs/2014MNRAS.438..529R} {438, 529}
  (\mn@eprint {arXiv} {1310.3317})

\bibitem[\protect\citeauthoryear{Rasmussen \& Williams}{Rasmussen \&
  Williams}{2005}]{Rasmussen05}
Rasmussen C.~E.,  Williams C. K.~I.,  2005, Gaussian Processes for Machine
  Learning (Adaptive Computation and Machine Learning).
The MIT Press

\bibitem[\protect\citeauthoryear{{Rogers}, {Bird}, {Peiris}, {Pontzen},
  {Font-Ribera}  \& {Leistedt}}{{Rogers} et~al.}{2018a}]{Rogers:2018b}
{Rogers} K.~K.,  {Bird} S.,  {Peiris} H.~V.,  {Pontzen} A.,  {Font-Ribera} A.,
   {Leistedt} B.,  2018a, \mn@doi [\mnras] {10.1093/mnras/stx2942}, \href
  {https://ui.adsabs.harvard.edu/abs/2018MNRAS.474.3032R} {474, 3032}
  (\mn@eprint {arXiv} {1706.08532})

\bibitem[\protect\citeauthoryear{{Rogers}, {Bird}, {Peiris}, {Pontzen},
  {Font-Ribera}  \& {Leistedt}}{{Rogers} et~al.}{2018b}]{Rogers:2018a}
{Rogers} K.~K.,  {Bird} S.,  {Peiris} H.~V.,  {Pontzen} A.,  {Font-Ribera} A.,
   {Leistedt} B.,  2018b, \mn@doi [\mnras] {10.1093/mnras/sty603}, \href
  {https://ui.adsabs.harvard.edu/abs/2018MNRAS.476.3716R} {476, 3716}
  (\mn@eprint {arXiv} {1711.06275})

\bibitem[\protect\citeauthoryear{{Schaye}}{{Schaye}}{2001}]{Schaye:2001}
{Schaye} J.,  2001, \mn@doi [\apjl] {10.1086/338106}, \href
  {https://ui.adsabs.harvard.edu/abs/2001ApJ...562L..95S} {562, L95}
  (\mn@eprint {arXiv} {astro-ph/0109280})

\bibitem[\protect\citeauthoryear{Slosar et~al.,}{Slosar
  et~al.}{2011}]{Slosar11}
Slosar A.,  et~al., 2011, \jcap, 2011, 001

\bibitem[\protect\citeauthoryear{{Viel}, {Haehnelt}, {Carswell}  \&
  {Kim}}{{Viel} et~al.}{2004}]{Viel:2004}
{Viel} M.,  {Haehnelt} M.~G.,  {Carswell} R.~F.,   {Kim} T.~S.,  2004, \mn@doi
  [\mnras] {10.1111/j.1365-2966.2004.07753.x}, \href
  {https://ui.adsabs.harvard.edu/abs/2004MNRAS.349L..33V} {349, L33}
  (\mn@eprint {arXiv} {astro-ph/0308078})

\bibitem[\protect\citeauthoryear{{Wolfe}, {Turnshek}, {Smith}  \&
  {Cohen}}{{Wolfe} et~al.}{1986}]{Wolfe1986}
{Wolfe} A.~M.,  {Turnshek} D.~A.,  {Smith} H.~E.,   {Cohen} R.~D.,  1986,
  \mn@doi [ApJS] {10.1086/191114}, 61, 249

\bibitem[\protect\citeauthoryear{{Zafar, T.}, {P\'eroux, C.}, {Popping, A.},
  {Milliard, B.}, {Deharveng, J.-M.}  \& {Frank, S.}}{{Zafar, T.}
  et~al.}{2013}]{Zafar2013}
{Zafar, T.} {P\'eroux, C.} {Popping, A.} {Milliard, B.} {Deharveng, J.-M.}
  {Frank, S.} 2013, \mn@doi [A\&A] {10.1051/0004-6361/201321154}, 556, A141

\makeatother
\end{thebibliography}

\appendix



\section{Tables of the measurements}

\begin{table*}
   \centering
   \caption{Table of $\dd N/\dd X$ values, integrated over all putative absorbers with $\nhi > 10^{20.3}$ in our catalogue.}
  \begin{tabular}{cccccc}
  \hline
  $z$ & dN/dX & $68$\% limits & $95$\% limits \\
   \hline
   $ 2.00 - 2.17 $ & $ 0.0337 $ & $ 0.0330 - 0.0345 $ & $ 0.0323 - 0.0352 $  \\
   $ 2.17 - 2.33 $ & $ 0.0429 $ & $ 0.0421 - 0.0438 $ & $ 0.0413 - 0.0446 $  \\
   $ 2.33 - 2.50 $ & $ 0.0462 $ & $ 0.0452 - 0.0472 $ & $ 0.0443 - 0.0481 $  \\
   $ 2.50 - 2.67 $ & $ 0.0493 $ & $ 0.0482 - 0.0505 $ & $ 0.0471 - 0.0516 $  \\
   $ 2.67 - 2.83 $ & $ 0.0620 $ & $ 0.0606 - 0.0634 $ & $ 0.0592 - 0.0649 $  \\
   $ 2.83 - 3.00 $ & $ 0.0660 $ & $ 0.0643 - 0.0678 $ & $ 0.0627 - 0.0695 $  \\
   $ 3.00 - 3.17 $ & $ 0.0704 $ & $ 0.0683 - 0.0726 $ & $ 0.0663 - 0.0747 $  \\
   $ 3.17 - 3.33 $ & $ 0.0745 $ & $ 0.0719 - 0.0774 $ & $ 0.0695 - 0.0800 $  \\
   $ 3.33 - 3.50 $ & $ 0.0763 $ & $ 0.0729 - 0.0800 $ & $ 0.0696 - 0.0833 $  \\
   $ 3.50 - 3.67 $ & $ 0.0777 $ & $ 0.0735 - 0.0821 $ & $ 0.0697 - 0.0862 $  \\
   $ 3.67 - 3.83 $ & $ 0.0632 $ & $ 0.0586 - 0.0688 $ & $ 0.0539 - 0.0735 $  \\
   $ 3.83 - 4.00 $ & $ 0.0648 $ & $ 0.0585 - 0.0720 $ & $ 0.0522 - 0.0792 $  \\
   $ 4.00 - 4.17 $ & $ 0.0581 $ & $ 0.0507 - 0.0670 $ & $ 0.0447 - 0.0745 $  \\
   $ 4.17 - 4.33 $ & $ 0.0709 $ & $ 0.0620 - 0.0842 $ & $ 0.0532 - 0.0953 $  \\
   $ 4.33 - 4.50 $ & $ 0.1024 $ & $ 0.0896 - 0.1216 $ & $ 0.0736 - 0.1376 $  \\
   $ 4.50 - 4.67 $ & $ 0.0827 $ & $ 0.0689 - 0.1057 $ & $ 0.0552 - 0.1241 $  \\
   $ 4.67 - 4.83 $ & $ 0.1041 $ & $ 0.0818 - 0.1413 $ & $ 0.0669 - 0.1636 $  \\
   $ 4.83 - 5.00 $ & $ 0.0676 $ & $ 0.0507 - 0.1184 $ & $ 0.0169 - 0.1522 $  \\
     \hline
    \end{tabular}
   \label{tab:CDDF_analysis/dr16q_full_int_lyb_occam_zqso7_1_30_delta_z_0_1/dndx_all.txt}
\end{table*}
\begin{table*}
   \centering
   \caption{$\Omega_\mathrm{DLA}$ values,
   integrated over all putative absorbers with $\nhi > 10^{20.3}$ in our catalogue.}
  \begin{tabular}{cccccc}
  \hline
  $z$ & $\Omega_\mathrm{DLA} (10^{-3}) $ & $68$\% limits & $95$\% limits \\
   \hline
   $ 2.00 - 2.17 $ & $ 0.582 $ & $ 0.550 - 0.619 $ & $ 0.520 - 0.659 $  \\
   $ 2.17 - 2.33 $ & $ 0.610 $ & $ 0.576 - 0.651 $ & $ 0.548 - 0.694 $  \\
   $ 2.33 - 2.50 $ & $ 0.691 $ & $ 0.664 - 0.722 $ & $ 0.638 - 0.755 $  \\
   $ 2.50 - 2.67 $ & $ 0.647 $ & $ 0.621 - 0.676 $ & $ 0.596 - 0.706 $  \\
   $ 2.67 - 2.83 $ & $ 0.770 $ & $ 0.738 - 0.809 $ & $ 0.711 - 0.855 $  \\
   $ 2.83 - 3.00 $ & $ 0.747 $ & $ 0.723 - 0.773 $ & $ 0.701 - 0.799 $  \\
   $ 3.00 - 3.17 $ & $ 0.789 $ & $ 0.758 - 0.829 $ & $ 0.729 - 0.896 $  \\
   $ 3.17 - 3.33 $ & $ 0.850 $ & $ 0.810 - 0.909 $ & $ 0.773 - 1.042 $  \\
   $ 3.33 - 3.50 $ & $ 0.908 $ & $ 0.855 - 0.962 $ & $ 0.792 - 1.019 $  \\
   $ 3.50 - 3.67 $ & $ 1.019 $ & $ 0.953 - 1.087 $ & $ 0.866 - 1.166 $  \\
   $ 3.67 - 3.83 $ & $ 0.664 $ & $ 0.604 - 0.731 $ & $ 0.550 - 0.806 $  \\
   $ 3.83 - 4.00 $ & $ 0.887 $ & $ 0.781 - 1.000 $ & $ 0.683 - 1.112 $  \\
   $ 4.00 - 4.17 $ & $ 0.562 $ & $ 0.508 - 0.622 $ & $ 0.457 - 0.684 $  \\
   $ 4.17 - 4.33 $ & $ 1.061 $ & $ 0.843 - 1.337 $ & $ 0.708 - 1.675 $  \\
   $ 4.33 - 4.50 $ & $ 1.507 $ & $ 1.252 - 1.810 $ & $ 1.038 - 2.182 $  \\
   $ 4.50 - 4.67 $ & $ 0.595 $ & $ 0.473 - 0.737 $ & $ 0.373 - 0.892 $  \\
   $ 4.67 - 4.83 $ & $ 0.913 $ & $ 0.657 - 1.208 $ & $ 0.465 - 1.498 $  \\
   $ 4.83 - 5.00 $ & $ 1.221 $ & $ 0.449 - 1.995 $ & $ 0.127 - 2.449 $  \\
     \hline
    \end{tabular}
   \label{tab:CDDF_analysis/dr16q_full_int_lyb_occam_zqso7_1_30_delta_z_0_1/omega_dla_all.txt}
\end{table*}
\begin{table*}
   \centering
   \caption{The column density distribution function integrated over all spectral lengths within $2 < z < 5$.}
  \begin{tabular}{cccccc}
  \hline
  $\log_{10} \mathrm{N}_\mathrm{HI}$ & $f(N_\mathrm{HI})$  $( 10^{ -21 } )$ & $68$\% limits $( 10^{ -21 } )$ & $95$\% limits $( 10^{ -21 } )$ \\
   \hline
   $ 20.0 - 20.1 $ & $ 0.371 $ & $ 0.365 - 0.378 $ & $ 0.358 - 0.385 $  \\
   $ 20.1 - 20.2 $ & $ 0.235 $ & $ 0.230 - 0.240 $ & $ 0.225 - 0.244 $  \\
   $ 20.2 - 20.3 $ & $ 0.170 $ & $ 0.166 - 0.173 $ & $ 0.162 - 0.177 $  \\
   $ 20.3 - 20.4 $ & $ 0.128 $ & $ 0.125 - 0.131 $ & $ 0.122 - 0.134 $  \\
   $ 20.4 - 20.5 $ & $ 9.58 \times 10^{ -2 }$ & $ [9.36  - 9.80 ]\times 10^{ -2 }$ & $ [9.15  - 10.02 ]\times 10^{ -2 }$  \\
   $ 20.5 - 20.6 $ & $ 7.16 \times 10^{ -2 }$ & $ [6.99  - 7.33 ]\times 10^{ -2 }$ & $ [6.83  - 7.50 ]\times 10^{ -2 }$  \\
   $ 20.6 - 20.7 $ & $ 5.09 \times 10^{ -2 }$ & $ [4.97  - 5.23 ]\times 10^{ -2 }$ & $ [4.85  - 5.35 ]\times 10^{ -2 }$  \\
   $ 20.7 - 20.8 $ & $ 3.56 \times 10^{ -2 }$ & $ [3.47  - 3.66 ]\times 10^{ -2 }$ & $ [3.38  - 3.75 ]\times 10^{ -2 }$  \\
   $ 20.8 - 20.9 $ & $ 2.45 \times 10^{ -2 }$ & $ [2.38  - 2.52 ]\times 10^{ -2 }$ & $ [2.31  - 2.59 ]\times 10^{ -2 }$  \\
   $ 20.9 - 21.0 $ & $ 1.64 \times 10^{ -2 }$ & $ [1.59  - 1.69 ]\times 10^{ -2 }$ & $ [1.55  - 1.74 ]\times 10^{ -2 }$  \\
   $ 21.0 - 21.1 $ & $ 1.06 \times 10^{ -2 }$ & $ [1.02  - 1.09 ]\times 10^{ -2 }$ & $ [9.92  - 11.29 ]\times 10^{ -3 }$  \\
   $ 21.1 - 21.2 $ & $ 6.96 \times 10^{ -3 }$ & $ [6.72  - 7.22 ]\times 10^{ -3 }$ & $ [6.48  - 7.47 ]\times 10^{ -3 }$  \\
   $ 21.2 - 21.3 $ & $ 4.58 \times 10^{ -3 }$ & $ [4.41  - 4.77 ]\times 10^{ -3 }$ & $ [4.25  - 4.94 ]\times 10^{ -3 }$  \\
   $ 21.3 - 21.4 $ & $ 2.66 \times 10^{ -3 }$ & $ [2.55  - 2.79 ]\times 10^{ -3 }$ & $ [2.43  - 2.91 ]\times 10^{ -3 }$  \\
   $ 21.4 - 21.5 $ & $ 1.51 \times 10^{ -3 }$ & $ [1.44  - 1.60 ]\times 10^{ -3 }$ & $ [1.36  - 1.68 ]\times 10^{ -3 }$  \\
   $ 21.5 - 21.6 $ & $ 9.95 \times 10^{ -4 }$ & $ [9.43  - 10.56 ]\times 10^{ -4 }$ & $ [8.91  - 11.08 ]\times 10^{ -4 }$  \\
   $ 21.6 - 21.7 $ & $ 4.82 \times 10^{ -4 }$ & $ [4.52  - 5.23 ]\times 10^{ -4 }$ & $ [4.18  - 5.57 ]\times 10^{ -4 }$  \\
   $ 21.7 - 21.8 $ & $ 2.60 \times 10^{ -4 }$ & $ [2.39  - 2.84 ]\times 10^{ -4 }$ & $ [2.18  - 3.08 ]\times 10^{ -4 }$  \\
   $ 21.8 - 21.9 $ & $ 1.50 \times 10^{ -4 }$ & $ [1.35  - 1.69 ]\times 10^{ -4 }$ & $ [1.23  - 1.83 ]\times 10^{ -4 }$  \\
   $ 21.9 - 22.0 $ & $ 7.73 \times 10^{ -5 }$ & $ [6.98  - 8.86 ]\times 10^{ -5 }$ & $ [6.03  - 9.81 ]\times 10^{ -5 }$  \\
   $ 22.0 - 22.1 $ & $ 4.34 \times 10^{ -5 }$ & $ [3.74  - 5.09 ]\times 10^{ -5 }$ & $ [3.30  - 5.69 ]\times 10^{ -5 }$  \\
   $ 22.1 - 22.2 $ & $ 1.43 \times 10^{ -5 }$ & $ [1.19  - 2.02 ]\times 10^{ -5 }$ & $ [8.33  - 23.80 ]\times 10^{ -6 }$  \\
   $ 22.2 - 22.3 $ & $ 1.13 \times 10^{ -5 }$ & $ [8.51  - 15.12 ]\times 10^{ -6 }$ & $ [6.62  - 17.96 ]\times 10^{ -6 }$  \\
   $ 22.3 - 22.4 $ & $ 3.75 \times 10^{ -6 }$ & $ [3.00  - 6.01 ]\times 10^{ -6 }$ & $ [1.50  - 8.26 ]\times 10^{ -6 }$  \\
   $ 22.4 - 22.5 $ & $ 2.39 \times 10^{ -6 }$ & $ [1.79  - 4.77 ]\times 10^{ -6 }$ & $ [5.96  - 59.63 ]\times 10^{ -7 }$  \\
   $ 22.5 - 22.6 $ & $ 1.42 \times 10^{ -6 }$ & $ [9.47  - 28.42 ]\times 10^{ -7 }$ & $ [4.74  - 37.90 ]\times 10^{ -7 }$  \\
   $ 22.6 - 22.7 $ & $ 7.53 \times 10^{ -7 }$ & $ [3.76  - 15.05 ]\times 10^{ -7 }$ & $0 -  2.26 \times 10^{ -6 }$  \\
   $ 22.7 - 22.8 $ & $ 5.98 \times 10^{ -7 }$ & $ [2.99  - 11.96 ]\times 10^{ -7 }$ & $0 -  1.79 \times 10^{ -6 }$  \\
   $ 22.8 - 22.9 $ & $ 7.12 \times 10^{ -7 }$ & $ [4.75  - 14.24 ]\times 10^{ -7 }$ & $ [2.37  - 16.62 ]\times 10^{ -7 }$  \\
   $ 22.9 - 23.0 $ & $ 5.66 \times 10^{ -7 }$ & $ [1.89  - 9.43 ]\times 10^{ -7 }$ & $0 -  1.32 \times 10^{ -6 }$  \\
     \hline
    \end{tabular}
   \label{tab:CDDF_analysis/dr16q_full_int_lyb_occam_zqso7_1_30_delta_z_0_1/cddf_all.txt}
\end{table*}
\begin{table*}
   \centering
   \caption{The column density distribution function integrated over spectral lengths within $2 < z < 2.5$.}
  \begin{tabular}{cccccc}
  \hline
  $\log_{10} \mathrm{N}_\mathrm{HI}$ & $f(N_\mathrm{HI})$  $( 10^{ -21 } )$ & $68$\% limits $( 10^{ -21 } )$ & $95$\% limits $( 10^{ -21 } )$ \\
   \hline
   $ 20.0 - 20.1 $ & $ 0.286 $ & $ 0.278 - 0.295 $ & $ 0.271 - 0.303 $  \\
   $ 20.1 - 20.2 $ & $ 0.190 $ & $ 0.184 - 0.196 $ & $ 0.179 - 0.202 $  \\
   $ 20.2 - 20.3 $ & $ 0.135 $ & $ 0.131 - 0.140 $ & $ 0.127 - 0.145 $  \\
   $ 20.3 - 20.4 $ & $ 0.103 $ & $ [9.99  - 10.71 ]\times 10^{ -2 }$ & $ [9.66  - 11.05 ]\times 10^{ -2 }$  \\
   $ 20.4 - 20.5 $ & $ 7.94 \times 10^{ -2 }$ & $ [7.66  - 8.22 ]\times 10^{ -2 }$ & $ [7.40  - 8.49 ]\times 10^{ -2 }$  \\
   $ 20.5 - 20.6 $ & $ 6.00 \times 10^{ -2 }$ & $ [5.79  - 6.21 ]\times 10^{ -2 }$ & $ [5.59  - 6.42 ]\times 10^{ -2 }$  \\
   $ 20.6 - 20.7 $ & $ 4.23 \times 10^{ -2 }$ & $ [4.07  - 4.40 ]\times 10^{ -2 }$ & $ [3.92  - 4.55 ]\times 10^{ -2 }$  \\
   $ 20.7 - 20.8 $ & $ 2.97 \times 10^{ -2 }$ & $ [2.86  - 3.10 ]\times 10^{ -2 }$ & $ [2.75  - 3.21 ]\times 10^{ -2 }$  \\
   $ 20.8 - 20.9 $ & $ 2.01 \times 10^{ -2 }$ & $ [1.93  - 2.10 ]\times 10^{ -2 }$ & $ [1.85  - 2.19 ]\times 10^{ -2 }$  \\
   $ 20.9 - 21.0 $ & $ 1.37 \times 10^{ -2 }$ & $ [1.31  - 1.44 ]\times 10^{ -2 }$ & $ [1.25  - 1.50 ]\times 10^{ -2 }$  \\
   $ 21.0 - 21.1 $ & $ 9.51 \times 10^{ -3 }$ & $ [9.06  - 9.98 ]\times 10^{ -3 }$ & $ [8.61  - 10.46 ]\times 10^{ -3 }$  \\
   $ 21.1 - 21.2 $ & $ 6.16 \times 10^{ -3 }$ & $ [5.85  - 6.52 ]\times 10^{ -3 }$ & $ [5.53  - 6.86 ]\times 10^{ -3 }$  \\
   $ 21.2 - 21.3 $ & $ 4.14 \times 10^{ -3 }$ & $ [3.91  - 4.41 ]\times 10^{ -3 }$ & $ [3.69  - 4.64 ]\times 10^{ -3 }$  \\
   $ 21.3 - 21.4 $ & $ 2.36 \times 10^{ -3 }$ & $ [2.20  - 2.53 ]\times 10^{ -3 }$ & $ [2.06  - 2.71 ]\times 10^{ -3 }$  \\
   $ 21.4 - 21.5 $ & $ 1.47 \times 10^{ -3 }$ & $ [1.35  - 1.58 ]\times 10^{ -3 }$ & $ [1.24  - 1.70 ]\times 10^{ -3 }$  \\
   $ 21.5 - 21.6 $ & $ 9.55 \times 10^{ -4 }$ & $ [8.79  - 10.47 ]\times 10^{ -4 }$ & $ [8.12  - 11.22 ]\times 10^{ -4 }$  \\
   $ 21.6 - 21.7 $ & $ 4.86 \times 10^{ -4 }$ & $ [4.39  - 5.39 ]\times 10^{ -4 }$ & $ [3.92  - 5.85 ]\times 10^{ -4 }$  \\
   $ 21.7 - 21.8 $ & $ 2.43 \times 10^{ -4 }$ & $ [2.11  - 2.80 ]\times 10^{ -4 }$ & $ [1.85  - 3.12 ]\times 10^{ -4 }$  \\
   $ 21.8 - 21.9 $ & $ 1.43 \times 10^{ -4 }$ & $ [1.22  - 1.68 ]\times 10^{ -4 }$ & $ [1.01  - 1.89 ]\times 10^{ -4 }$  \\
   $ 21.9 - 22.0 $ & $ 8.33 \times 10^{ -5 }$ & $ [7.00  - 10.00 ]\times 10^{ -5 }$ & $ [6.00  - 11.33 ]\times 10^{ -5 }$  \\
   $ 22.0 - 22.1 $ & $ 3.71 \times 10^{ -5 }$ & $ [3.18  - 5.03 ]\times 10^{ -5 }$ & $ [2.38  - 5.83 ]\times 10^{ -5 }$  \\
   $ 22.1 - 22.2 $ & $ 1.89 \times 10^{ -5 }$ & $ [1.47  - 2.73 ]\times 10^{ -5 }$ & $ [1.05  - 3.15 ]\times 10^{ -5 }$  \\
   $ 22.2 - 22.3 $ & $ 1.34 \times 10^{ -5 }$ & $ [8.35  - 18.38 ]\times 10^{ -6 }$ & $ [6.68  - 21.72 ]\times 10^{ -6 }$  \\
   $ 22.3 - 22.4 $ & $ 3.98 \times 10^{ -6 }$ & $ [2.65  - 7.96 ]\times 10^{ -6 }$ & $ [1.33  - 10.62 ]\times 10^{ -6 }$  \\
   $ 22.4 - 22.5 $ & $ 2.11 \times 10^{ -6 }$ & $ [1.05  - 5.27 ]\times 10^{ -6 }$ & $0 -  7.38 \times 10^{ -6 }$  \\
   $ 22.5 - 22.6 $ & $ 1.67 \times 10^{ -6 }$ & $0 -  3.35 \times 10^{ -6 }$ & $0 -  5.02 \times 10^{ -6 }$  \\
   $ 22.6 - 22.7 $ & $ 6.65 \times 10^{ -7 }$ & $0 -  2.00 \times 10^{ -6 }$ & $0 -  3.33 \times 10^{ -6 }$  \\
   $ 22.7 - 22.8 $ & $ 5.28 \times 10^{ -7 }$ & $0 -  2.11 \times 10^{ -6 }$ & $0 -  2.64 \times 10^{ -6 }$  \\
   $ 22.8 - 22.9 $ & $ 8.39 \times 10^{ -7 }$ & $ [4.20  - 20.98 ]\times 10^{ -7 }$ & $ [4.20  - 25.18 ]\times 10^{ -7 }$  \\
   $ 22.9 - 23.0 $ & $ 6.67 \times 10^{ -7 }$ & $ [3.33  - 13.33 ]\times 10^{ -7 }$ & $0 -  2.00 \times 10^{ -6 }$  \\
     \hline
    \end{tabular}
   \label{tab:CDDF_analysis/dr16q_full_int_lyb_occam_zqso7_1_30_delta_z_0_1/cddf_z225.txt}
\end{table*}
\begin{table*}
   \centering
   \caption{The column density distribution function integrated over spectral lengths within $2.5 < z < 3$}
  \begin{tabular}{cccccc}
  \hline
  $\log_{10} \mathrm{N}_\mathrm{HI}$ & $f(N_\mathrm{HI})$  $( 10^{ -21 } )$ & $68$\% limits $( 10^{ -21 } )$ & $95$\% limits $( 10^{ -21 } )$ \\
   \hline
   $ 20.0 - 20.1 $ & $ 0.465 $ & $ 0.450 - 0.480 $ & $ 0.437 - 0.494 $  \\
   $ 20.1 - 20.2 $ & $ 0.281 $ & $ 0.271 - 0.292 $ & $ 0.261 - 0.302 $  \\
   $ 20.2 - 20.3 $ & $ 0.211 $ & $ 0.204 - 0.220 $ & $ 0.196 - 0.228 $  \\
   $ 20.3 - 20.4 $ & $ 0.162 $ & $ 0.156 - 0.168 $ & $ 0.150 - 0.175 $  \\
   $ 20.4 - 20.5 $ & $ 0.117 $ & $ 0.113 - 0.122 $ & $ 0.108 - 0.127 $  \\
   $ 20.5 - 20.6 $ & $ 8.65 \times 10^{ -2 }$ & $ [8.31  - 9.03 ]\times 10^{ -2 }$ & $ [7.97  - 9.39 ]\times 10^{ -2 }$  \\
   $ 20.6 - 20.7 $ & $ 6.37 \times 10^{ -2 }$ & $ [6.11  - 6.66 ]\times 10^{ -2 }$ & $ [5.86  - 6.93 ]\times 10^{ -2 }$  \\
   $ 20.7 - 20.8 $ & $ 4.36 \times 10^{ -2 }$ & $ [4.16  - 4.57 ]\times 10^{ -2 }$ & $ [3.98  - 4.77 ]\times 10^{ -2 }$  \\
   $ 20.8 - 20.9 $ & $ 2.99 \times 10^{ -2 }$ & $ [2.86  - 3.15 ]\times 10^{ -2 }$ & $ [2.72  - 3.29 ]\times 10^{ -2 }$  \\
   $ 20.9 - 21.0 $ & $ 1.99 \times 10^{ -2 }$ & $ [1.88  - 2.10 ]\times 10^{ -2 }$ & $ [1.79  - 2.20 ]\times 10^{ -2 }$  \\
   $ 21.0 - 21.1 $ & $ 1.26 \times 10^{ -2 }$ & $ [1.19  - 1.34 ]\times 10^{ -2 }$ & $ [1.12  - 1.40 ]\times 10^{ -2 }$  \\
   $ 21.1 - 21.2 $ & $ 7.91 \times 10^{ -3 }$ & $ [7.41  - 8.49 ]\times 10^{ -3 }$ & $ [6.96  - 8.99 ]\times 10^{ -3 }$  \\
   $ 21.2 - 21.3 $ & $ 5.28 \times 10^{ -3 }$ & $ [4.95  - 5.67 ]\times 10^{ -3 }$ & $ [4.63  - 5.99 ]\times 10^{ -3 }$  \\
   $ 21.3 - 21.4 $ & $ 3.16 \times 10^{ -3 }$ & $ [2.94  - 3.42 ]\times 10^{ -3 }$ & $ [2.71  - 3.68 ]\times 10^{ -3 }$  \\
   $ 21.4 - 21.5 $ & $ 1.61 \times 10^{ -3 }$ & $ [1.45  - 1.79 ]\times 10^{ -3 }$ & $ [1.31  - 1.95 ]\times 10^{ -3 }$  \\
   $ 21.5 - 21.6 $ & $ 1.12 \times 10^{ -3 }$ & $ [1.01  - 1.24 ]\times 10^{ -3 }$ & $ [9.17  - 13.31 ]\times 10^{ -4 }$  \\
   $ 21.6 - 21.7 $ & $ 4.43 \times 10^{ -4 }$ & $ [3.86  - 5.14 ]\times 10^{ -4 }$ & $ [3.29  - 5.86 ]\times 10^{ -4 }$  \\
   $ 21.7 - 21.8 $ & $ 2.95 \times 10^{ -4 }$ & $ [2.50  - 3.40 ]\times 10^{ -4 }$ & $ [2.16  - 3.86 ]\times 10^{ -4 }$  \\
   $ 21.8 - 21.9 $ & $ 1.62 \times 10^{ -4 }$ & $ [1.35  - 1.98 ]\times 10^{ -4 }$ & $ [1.17  - 2.25 ]\times 10^{ -4 }$  \\
   $ 21.9 - 22.0 $ & $ 7.16 \times 10^{ -5 }$ & $ [5.73  - 9.31 ]\times 10^{ -5 }$ & $ [4.30  - 11.46 ]\times 10^{ -5 }$  \\
   $ 22.0 - 22.1 $ & $ 4.55 \times 10^{ -5 }$ & $ [3.98  - 6.26 ]\times 10^{ -5 }$ & $ [2.84  - 7.39 ]\times 10^{ -5 }$  \\
   $ 22.1 - 22.2 $ & $ 4.52 \times 10^{ -6 }$ & $ [4.52  - 13.55 ]\times 10^{ -6 }$ & $0 -  2.26 \times 10^{ -5 }$  \\
   $ 22.2 - 22.3 $ & $ 7.18 \times 10^{ -6 }$ & $ [3.59  - 14.35 ]\times 10^{ -6 }$ & $0 -  1.79 \times 10^{ -5 }$  \\
   $ 22.3 - 22.4 $ & $ 2.85 \times 10^{ -6 }$ & $0 -  5.70 \times 10^{ -6 }$ & $0 -  1.14 \times 10^{ -5 }$  \\
   $ 22.4 - 22.5 $ & $ 2.26 \times 10^{ -6 }$ & $0 -  6.79 \times 10^{ -6 }$ & $0 -  9.06 \times 10^{ -6 }$  \\
   $ 22.5 - 22.6 $ & $0$ & $0 -  1.80 \times 10^{ -6 }$ & $0 -  3.60 \times 10^{ -6 }$  \\
   $ 22.6 - 22.7 $ & $0$ & $0 -  1.43 \times 10^{ -6 }$ & $0 -  2.86 \times 10^{ -6 }$  \\
   $ 22.7 - 22.8 $ & $0$ & $0 -  1.13 \times 10^{ -6 }$ & $0 -  1.13 \times 10^{ -6 }$  \\
   $ 22.8 - 22.9 $ & $0$ & $0 -  1.80 \times 10^{ -6 }$ & $0 -  1.80 \times 10^{ -6 }$  \\
   $ 22.9 - 23.0 $ & $0$ & $0 -  7.16 \times 10^{ -7 }$ & $0 -  1.43 \times 10^{ -6 }$  \\
     \hline
    \end{tabular}
   \label{tab:CDDF_analysis/dr16q_full_int_lyb_occam_zqso7_1_30_delta_z_0_1/cddf_z253.txt}
\end{table*}
\begin{table*}
   \centering
   \caption{The column density distribution function integrated over spectral lengths within $3 < z < 4$.}
  \begin{tabular}{cccccc}
  \hline
  $\log_{10} \mathrm{N}_\mathrm{HI}$ & $f(N_\mathrm{HI})$  $( 10^{ -21 } )$ & $68$\% limits $( 10^{ -21 } )$ & $95$\% limits $( 10^{ -21 } )$ \\
   \hline
   $ 20.0 - 20.1 $ & $ 0.700 $ & $ 0.673 - 0.728 $ & $ 0.647 - 0.756 $  \\
   $ 20.1 - 20.2 $ & $ 0.407 $ & $ 0.389 - 0.428 $ & $ 0.371 - 0.447 $  \\
   $ 20.2 - 20.3 $ & $ 0.290 $ & $ 0.276 - 0.306 $ & $ 0.263 - 0.319 $  \\
   $ 20.3 - 20.4 $ & $ 0.205 $ & $ 0.195 - 0.217 $ & $ 0.185 - 0.228 $  \\
   $ 20.4 - 20.5 $ & $ 0.155 $ & $ 0.147 - 0.164 $ & $ 0.139 - 0.172 $  \\
   $ 20.5 - 20.6 $ & $ 0.116 $ & $ 0.110 - 0.123 $ & $ 0.105 - 0.129 $  \\
   $ 20.6 - 20.7 $ & $ 7.87 \times 10^{ -2 }$ & $ [7.46  - 8.35 ]\times 10^{ -2 }$ & $ [7.05  - 8.79 ]\times 10^{ -2 }$  \\
   $ 20.7 - 20.8 $ & $ 5.52 \times 10^{ -2 }$ & $ [5.20  - 5.87 ]\times 10^{ -2 }$ & $ [4.90  - 6.20 ]\times 10^{ -2 }$  \\
   $ 20.8 - 20.9 $ & $ 4.04 \times 10^{ -2 }$ & $ [3.81  - 4.30 ]\times 10^{ -2 }$ & $ [3.57  - 4.54 ]\times 10^{ -2 }$  \\
   $ 20.9 - 21.0 $ & $ 2.51 \times 10^{ -2 }$ & $ [2.34  - 2.68 ]\times 10^{ -2 }$ & $ [2.19  - 2.85 ]\times 10^{ -2 }$  \\
   $ 21.0 - 21.1 $ & $ 1.41 \times 10^{ -2 }$ & $ [1.30  - 1.53 ]\times 10^{ -2 }$ & $ [1.21  - 1.64 ]\times 10^{ -2 }$  \\
   $ 21.1 - 21.2 $ & $ 1.00 \times 10^{ -2 }$ & $ [9.27  - 10.88 ]\times 10^{ -3 }$ & $ [8.51  - 11.64 ]\times 10^{ -3 }$  \\
   $ 21.2 - 21.3 $ & $ 6.33 \times 10^{ -3 }$ & $ [5.91  - 6.93 ]\times 10^{ -3 }$ & $ [5.39  - 7.45 ]\times 10^{ -3 }$  \\
   $ 21.3 - 21.4 $ & $ 3.67 \times 10^{ -3 }$ & $ [3.33  - 4.08 ]\times 10^{ -3 }$ & $ [3.06  - 4.42 ]\times 10^{ -3 }$  \\
   $ 21.4 - 21.5 $ & $ 1.89 \times 10^{ -3 }$ & $ [1.67  - 2.16 ]\times 10^{ -3 }$ & $ [1.51  - 2.43 ]\times 10^{ -3 }$  \\
   $ 21.5 - 21.6 $ & $ 1.03 \times 10^{ -3 }$ & $ [9.01  - 12.01 ]\times 10^{ -4 }$ & $ [7.72  - 13.30 ]\times 10^{ -4 }$  \\
   $ 21.6 - 21.7 $ & $ 6.47 \times 10^{ -4 }$ & $ [5.45  - 7.50 ]\times 10^{ -4 }$ & $ [4.77  - 8.52 ]\times 10^{ -4 }$  \\
   $ 21.7 - 21.8 $ & $ 2.98 \times 10^{ -4 }$ & $ [2.44  - 3.79 ]\times 10^{ -4 }$ & $ [2.17  - 4.33 ]\times 10^{ -4 }$  \\
   $ 21.8 - 21.9 $ & $ 1.50 \times 10^{ -4 }$ & $ [1.29  - 1.93 ]\times 10^{ -4 }$ & $ [8.60  - 23.65 ]\times 10^{ -5 }$  \\
   $ 21.9 - 22.0 $ & $ 5.12 \times 10^{ -5 }$ & $ [3.42  - 8.54 ]\times 10^{ -5 }$ & $ [1.71  - 10.25 ]\times 10^{ -5 }$  \\
   $ 22.0 - 22.1 $ & $ 4.07 \times 10^{ -5 }$ & $ [4.07  - 6.78 ]\times 10^{ -5 }$ & $ [2.71  - 8.14 ]\times 10^{ -5 }$  \\
   $ 22.1 - 22.2 $ & $ 1.08 \times 10^{ -5 }$ & $0 -  2.15 \times 10^{ -5 }$ & $0 -  3.23 \times 10^{ -5 }$  \\
   $ 22.2 - 22.3 $ & $0$ & $0 -  8.56 \times 10^{ -6 }$ & $0 -  1.71 \times 10^{ -5 }$  \\
   $ 22.3 - 22.4 $ & $ 6.80 \times 10^{ -6 }$ & $ [6.80  - 20.39 ]\times 10^{ -6 }$ & $0 -  2.04 \times 10^{ -5 }$  \\
   $ 22.4 - 22.5 $ & $0$ & $0 -  5.40 \times 10^{ -6 }$ & $0 -  5.40 \times 10^{ -6 }$  \\
   $ 22.5 - 22.6 $ & $ 4.29 \times 10^{ -6 }$ & $ [4.29  - 8.58 ]\times 10^{ -6 }$ & $0 -  1.29 \times 10^{ -5 }$  \\
   $ 22.6 - 22.7 $ & $0$ & $0 -  3.41 \times 10^{ -6 }$ & $0 -  3.41 \times 10^{ -6 }$  \\
   $ 22.7 - 22.8 $ & $0$ & $0 -  2.71 \times 10^{ -6 }$ & $0 -  2.71 \times 10^{ -6 }$  \\
   $ 22.8 - 22.9 $ & $0$ & $0 -  2.15 \times 10^{ -6 }$ & $0 -  4.30 \times 10^{ -6 }$  \\
   $ 22.9 - 23.0 $ & $0$ & $0 -  1.71 \times 10^{ -6 }$ & $0 -  3.42 \times 10^{ -6 }$  \\
     \hline
    \end{tabular}
   \label{tab:CDDF_analysis/dr16q_full_int_lyb_occam_zqso7_1_30_delta_z_0_1/cddf_z34.txt}
\end{table*}
\begin{table*}
   \centering
   \caption{The column density distribution function integrated over spectral lengths within $4 < z < 5$.}
  \begin{tabular}{cccccc}
  \hline
  $\log_{10} \mathrm{N}_\mathrm{HI}$ & $f(N_\mathrm{HI})$  $( 10^{ -21 } )$ & $68$\% limits $( 10^{ -21 } )$ & $95$\% limits $( 10^{ -21 } )$ \\
   \hline
   $ 20.0 - 20.1 $ & $ 0.628 $ & $ 0.523 - 0.753 $ & $ 0.439 - 0.858 $  \\
   $ 20.1 - 20.2 $ & $ 0.515 $ & $ 0.432 - 0.598 $ & $ 0.366 - 0.681 $  \\
   $ 20.2 - 20.3 $ & $ 0.317 $ & $ 0.264 - 0.396 $ & $ 0.211 - 0.449 $  \\
   $ 20.3 - 20.4 $ & $ 0.252 $ & $ 0.210 - 0.315 $ & $ 0.168 - 0.356 $  \\
   $ 20.4 - 20.5 $ & $ 0.167 $ & $ 0.142 - 0.208 $ & $ 0.108 - 0.250 $  \\
   $ 20.5 - 20.6 $ & $ 0.112 $ & $ [9.26  - 14.55 ]\times 10^{ -2 }$ & $ [6.61  - 17.20 ]\times 10^{ -2 }$  \\
   $ 20.6 - 20.7 $ & $ 8.41 \times 10^{ -2 }$ & $ [6.83  - 11.03 ]\times 10^{ -2 }$ & $ [5.25  - 12.61 ]\times 10^{ -2 }$  \\
   $ 20.7 - 20.8 $ & $ 5.43 \times 10^{ -2 }$ & $ [4.17  - 7.10 ]\times 10^{ -2 }$ & $ [2.92  - 8.35 ]\times 10^{ -2 }$  \\
   $ 20.8 - 20.9 $ & $ 3.32 \times 10^{ -2 }$ & $ [2.65  - 4.64 ]\times 10^{ -2 }$ & $ [1.99  - 5.64 ]\times 10^{ -2 }$  \\
   $ 20.9 - 21.0 $ & $ 2.90 \times 10^{ -2 }$ & $ [2.11  - 3.69 ]\times 10^{ -2 }$ & $ [1.58  - 4.48 ]\times 10^{ -2 }$  \\
   $ 21.0 - 21.1 $ & $ 8.37 \times 10^{ -3 }$ & $ [4.18  - 12.55 ]\times 10^{ -3 }$ & $ [2.09  - 16.73 ]\times 10^{ -3 }$  \\
   $ 21.1 - 21.2 $ & $ 1.16 \times 10^{ -2 }$ & $ [9.97  - 16.62 ]\times 10^{ -3 }$ & $ [6.65  - 19.94 ]\times 10^{ -3 }$  \\
   $ 21.2 - 21.3 $ & $ 3.96 \times 10^{ -3 }$ & $ [2.64  - 7.92 ]\times 10^{ -3 }$ & $ [1.32  - 9.24 ]\times 10^{ -3 }$  \\
   $ 21.3 - 21.4 $ & $ 3.15 \times 10^{ -3 }$ & $ [2.10  - 5.24 ]\times 10^{ -3 }$ & $ [1.05  - 6.29 ]\times 10^{ -3 }$  \\
   $ 21.4 - 21.5 $ & $ 2.50 \times 10^{ -3 }$ & $ [1.67  - 4.16 ]\times 10^{ -3 }$ & $ [8.33  - 58.29 ]\times 10^{ -4 }$  \\
   $ 21.5 - 21.6 $ & $ 2.65 \times 10^{ -3 }$ & $ [1.98  - 3.97 ]\times 10^{ -3 }$ & $ [1.98  - 4.63 ]\times 10^{ -3 }$  \\
   $ 21.6 - 21.7 $ & $ 1.05 \times 10^{ -3 }$ & $ [5.25  - 15.76 ]\times 10^{ -4 }$ & $ [5.25  - 21.02 ]\times 10^{ -4 }$  \\
   $ 21.7 - 21.8 $ & $0$ & $0 -  4.17 \times 10^{ -4 }$ & $0 -  8.35 \times 10^{ -4 }$  \\
   $ 21.8 - 21.9 $ & $ 3.32 \times 10^{ -4 }$ & $ [3.32  - 6.63 ]\times 10^{ -4 }$ & $ [3.32  - 9.95 ]\times 10^{ -4 }$  \\
   $ 21.9 - 22.0 $ & $0$ & $0 -  5.27 \times 10^{ -4 }$ & $0 -  5.27 \times 10^{ -4 }$  \\
   $ 22.0 - 22.1 $ & $0$ & $0 -  4.18 \times 10^{ -4 }$ & $0 -  6.28 \times 10^{ -4 }$  \\
   $ 22.1 - 22.2 $ & $0$ & $0 -  1.66 \times 10^{ -4 }$ & $0 -  1.66 \times 10^{ -4 }$  \\
   $ 22.2 - 22.3 $ & $0$ & $0 -  1.32 \times 10^{ -4 }$ & $0 -  1.32 \times 10^{ -4 }$  \\
     \hline
    \end{tabular}
   \label{tab:CDDF_analysis/dr16q_full_int_lyb_occam_zqso7_1_30_delta_z_0_1/cddf_z45.txt}
\end{table*}

\label{lastpage}

\end{document}